Warsaw University of Technology
FACULTY OF PHYSICS

# Master's diploma thesis

in the field of study Applied Physics
and specialisation Medical Physics

## Focusing of high energy electron beam using crystal lenses for applications in radiotherapy

Marta Monikowska
student record book number 284508

thesis supervisor
Ph.D. Eng. Marcin Patecki

## Abstract

Title of the thesis: Focusing of high energy electron beam using crystal lenses for applications in radiotherapy

The two dominant radiotherapy methods are either simplified in terms of beam generation and handling, which compromises the energy deposition curve in tissues (photon therapy), or require extensive accelerator facilities and complex beam delivery systems to provide a favorable shape of the energy deposition curve (hadron therapy). The advantages of both of these methods, such as the low cost of the apparatus, ease of beam generation, and a suitable shape of the energy deposition curve in tissues, can potentially be achieved by using a very high-energy electron beam (beam energy in the order of a few hundreds of MeV) focused on the area of the tumor lesion. However, focusing of the beam is usually done with the use of quadrupole magnets which makes the beam delivery system complex and challenging from the engineering point of view. In this thesis, the feasibility of an alternative solution is explored, where focusing is performed by a bent silicon crystal with an appropriate shape of its exit face. Such a crystal lens can be a very light object (mass in the order of grams), allowing for much simpler beam delivery systems of radiotherapy facilities.

As a result of this feasibility study, a simulation of a bent silicon crystal with profiled exit was prepared in Geant4. The outcome obtained from the simulation proved the focusing ability of such profiled crystal. However, the focusing strength of the crystal is not strong enough. Technical requirements needed to enable an improvement in focusing efficiency were identified and understood. Moreover, alternative solutions providing stronger focusing while using profiled crystals were proposed.

*Keywords: electron radiotherapy, VHEE beam, beam focusing, charged particle channeling, crystal focusing, bent silicon crystal*

(signature of the thesis supervisor) (signature of the student)

## Streszczenie

Tytuł pracy: Skupianie wysokoenergetycznej wiązki elektronowej przy użyciu kryształów do zastosowań w radioterapii

Dwie dominujące metody radioterapii są albo uproszczone pod względem generowania i obsługi wiązki, co pogarsza krzywą depozycji energii w tkankach (terapia fotonowa), albo wymagają rozbudowanych akceleratorów i złożonych systemów dostarczania wiązki, aby zapewnić korzystny kształt krzywej depozycji energii (terapia hadronowa). Zalety obu tych metod, takie jak, niski koszt aparatury, łatwość generowania wiązki i odpowiedni kształt krzywej depozycji energii w tkankach, można potencjalnie osiągnąć za pomocą wiązki elektronów o bardzo wysokiej energii (energia wiązki rzędu kilkuset MeV), która jest skupiona w obszarze zmiany nowotworowej. Ogniskowanie wiązki jest zwykle wykonywane przy użyciu magnesów kwadrupolowych, co sprawia, że system dostarczania wiązki jest złożony i stanowi wyzwanie z inżynieryjnego punktu widzenia. W niniejszej pracy dyplomowej przeprowadzone zostało studium wykonalności alternatywnego rozwiązania, w którym ogniskowanie odbywa się za pomocą zakrzywionego kryształu krzemu o odpowiednim kształcie jego powierzchni wyjściowej. Taka soczewka krystaliczna może być bardzo lekkim obiektem (masa rzędu gramów), co pozwala na znaczne uproszczenie systemu dostarczania wiązki w miejscach oferujących usługę radioterapii.

W celu przeprowadzenia studium wykonalności przygotowano symulację zakrzywionego kryształu krzemu z wyprofilowanym wyjściem w środowisku programistycznym Geant4. Wynik uzyskany z symulacji wykazał zdolność skupiającą takiego kryształu. Jednak siła ogniskowania kryształu nie jest wystarczająco duża. Zidentyfikowane zostały wymagania techniczne umożliwiające poprawę efektywności soczewek. Zaproponowano również alternatywne rozwiązania, które mogłyby umożliwić silniejsze ogniskowanie przy użyciu badanych kryształów.

(podpis opiekuna naukowego) (podpis dyplomanta)

## Oświadczenie o samodzielności wykonania pracy

Politechnika Warszawska

Warsaw Univeristy of Technology

Marta Monikowska
284508
Fizyka Techniczna
*Applied Physics*

**OŚWIADCZENIE**

**DECLARATION**

Świadoma odpowiedzialności karnej za składanie fałszywych zeznań oświadczam, że niniejsza praca dyplomowa została napisana przeze mnie samodzielnie, pod opieką kierującego pracą dyplomową.

*Under the penalty of perjury, I hereby certify that I wrote my diploma thesis on my own, under the guidance of the thesis supervisor.*

Jednocześnie oświadczam, że:

*I also declare that:*

- niniejsza praca dyplomowa nie narusza praw autorskich w rozumieniu ustawy z dnia 4 lutego 1994 roku o prawie autorskim i prawach pokrewnych (Dz.U. z 2006r. Nr 90, poz. 631 z późn. zm.) oraz dóbr osobistych chronionych prawem cywilnym,
- *this diploma thesis does not constitute infringement of copyright following the act of 4 February 1994 on copyright and related rights (Journal of Acts of 2006 no. 90, item 631 with further amendments) or personal rights protected under the civil law*
- niniejsza praca dyplomowa nie zawiera danych i informacji, które uzyskałem w sposób niedozwolony,
- *the diploma thesis does not contain data or information acquired in an illegal way,*
- niniejsza praca dyplomowa nie była wcześniej podstawą żadnej innej urzędowej procedury związanej z nadawaniem dyplomów lub tytułów zawodowych,
- *the diploma thesis has never been the basis of any other official proceedings leading to the award of diplomas or professional degrees,*
- wszystkie informacje umieszczone w niniejszej pracy, uzyskane ze źródeł pisanych i elektronicznych, zostały udokumentowane w wykazie literatury odpowiednimi odnośnikami,
- *all information included in the diploma thesis, derived from printed and electronic sources, has been documented with relevant references in the literature section,*
- znam regulacje prawne Politechniki Warszawskiej w sprawie zarządzania prawami autorskimi i prawa pokrewnymi, prawami własności przemysłowej oraz zasadami komercjalizacji.
- *I am aware of the regulations at Warsaw University of Technology on management of copyright and related rights, industrial property rights and commercialisation.*

(miejscowość i data) (czytelny podpis dyplomanta)

*(place and date)* *(legible signature of the student)*

## Oświadczenie o udzieleniu Uczelni licencji do pracy

**Politechnika Warszawska**

Warsaw Univeristy of Technology

Marta Monikowska
284508
Fizyka Techniczna
*Applied Physics*

**Oświadczenie studenta w przedmiocie udzielenia licencji Politechnice Warszawskiej**
**Student's statement on granting license to the Warsaw University of Technology**

Oświadczam, że jako autor pracy dyplomowej pt. "Skupianie wysokoenergetycznej wiązki elektronowej przy użyciu kryształów do zastosowań w radioterapii" udzielam Politechnice Warszawskiej nieodpłatnej licencji na niewyłączne, nieograniczone w czasie, umieszczenie pracy dyplomowej w elektronicznych bazach danych oraz udostępnianie pracy dyplomowej w zamkniętym systemie bibliotecznym Politechniki Warszawskiej osobom zainteresowanym.
*I declare that as the author of the thesis titled "Focusing of high energy electron beam using crystal lenses for applications in radiotherapy", I grant the Warsaw University of Technology a royalty-free license for non-exclusive, indefinite placement of the thesis in electronic databases and making the thesis available in the closed library system of the Warsaw University of Technology to interested persons.*

Licencja na udostępnienie pracy dyplomowej nie obejmuje wyrażenia zgody na wykorzystywanie pracy dyplomowej na żadnym innym polu eksploatacji, w szczególności kopiowania pracy dyplomowej w całości lub w części, utrwalania w innej formie czy zwielokrotniania.
*The license to make the thesis available does not include consent to use the thesis in any other field of exploitation, in particular, copying the thesis in whole or in part, fixation in another form, or reproduction.*

(miejscowość i data)
*(place and date)*

(czytelny podpis dyplomanta)
*(legible signature of the student)*

# Contents

# CONTENTS

# 1. Introduction

Cancer is a disease of genetic material. It arises from the transformation of healthy cells in the body into abnormal cells that spread uncontrollably. It is caused by several factors, which can be divided into biological (e.g. age, gender, inherited genetic defects, skin type, etc.), environmental (e.g. natural radioactivity, ultraviolet, cosmic radiation, etc.), occupational (e.g. chemicals, radioactive materials, industrial materials, etc.), and lifestyle-related (e.g. diet, physical condition, etc.). According to the Global Cancer Observatory, it is a global disease that is the second leading cause of death in economically developed countries. [1]

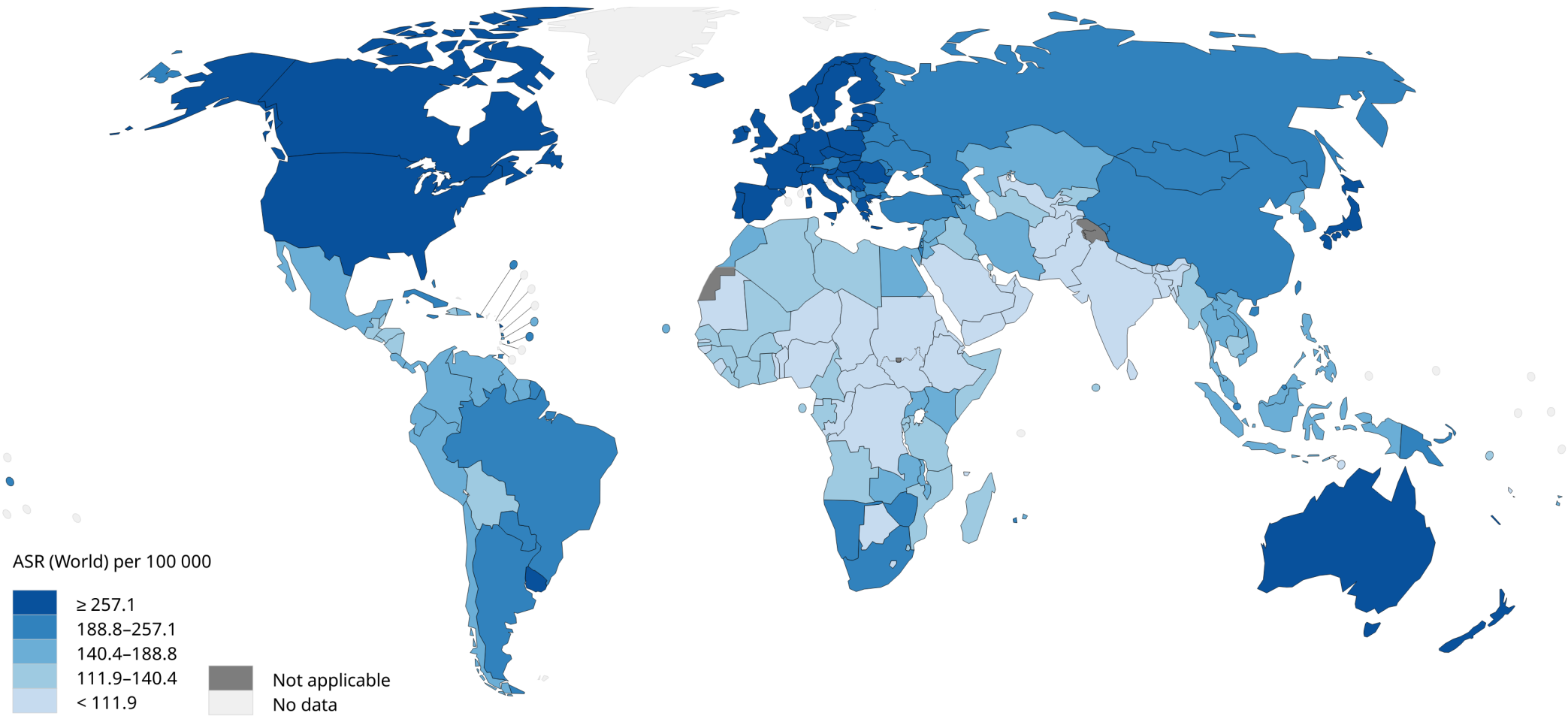

Figure 1.1: Estimated cancer incidence rate, standardized by age in 2020 (graphics from [2]).

The most common cancer treatments are surgery (45%), chemotherapy (28%), and radiation therapy (27%) [3]. These methods are often combined to achieve the best possible result. For example, it means that a patient undergoes both radiotherapy and surgery or chemotherapy at the time of treatment.

Radiation therapy is a method of cancer treatment that uses ionizing radiation. It contributes to a 40% cure rate for cancer [4]. Unfortunately, due to the high cost of radiation therapy, many people worldwide do not have access to it. Innovative radiotherapy techniques are critical to the future provision of high-quality radiotherapy services. To make them broadly available, cost-effective and efficient solutions are required. This research aims to assess the feasibility of enhancing a new radiotherapy technique using very high-energy electron (VHEE) beams characterized by the advantages mentioned above, by utilizing the phenomenon of channeling in crystals.

An overview of the most common radiotherapy methods and a comparison with VHEE-based methods are given in Chapter 2. The theoretical basis of the channeling process is discussed in Chapter 3. Chapter 4 gives a description of computing tools developed to study the feasibility of the proposed idea. In Chapter 5 results obtained from the simulation are presented and discussed. The last Chapter 6 concludes this study.

# 2. Available and prospect radiotherapy methods

The most typical radiation therapy methods are divided into two types: teleradiotherapy and brachytherapy. Teleradiotherapy involves treatment with external beams. Brachytherapy, on the other hand, is a placement of radioactive sources within or adjacent to a cancerous tumor. This thesis focuses on the use of teleradiotherapy.

## 2.1 Energy deposition in tissues

The most important goal of radiotherapy planning is to make sure that the tumor site receives the highest possible dose, while protecting as much as possible the surrounding healthy tissues and critical organs, such as the heart, brain, kidneys, and so on. Absorbed dose is the ratio of the energy imparted by ionizing radiation to the matter in a volume element to the mass of matter contained in that volume element, it can be represented by a formula:

$$D = \frac{dE}{dm}. \tag{2.1}$$

The Bethe-Bloch formula gives the mean rate of energy loss (stopping power) of a heavy charged particle as it passes through a material medium. For electrons, the energy loss is slightly different due to their small mass (requiring relativistic corrections), and since they suffer much larger losses by Bremsstrahlung (high energy electrons lose their energy predominantly through radiation) [5]. The formula was initially derived in the context of non-relativistic particles. However, it can be extended to include relativistic effects by making appropriate modifications. In the relativistic case, the Bethe-Bloch formula takes the form:

$$-\frac{dE}{dx} = K \cdot \frac{Z}{A} \cdot \frac{z^2}{\beta^2} \left[ \frac{1}{2} \ln \frac{2m_e \beta^2 \gamma^2 T_{\max}}{I^2} - \beta^2 - \frac{\delta(\beta\gamma)}{2} \right], \tag{2.2}$$

where:

- $dE/dx$ is the rate of energy loss of the particle per unit path length (stopping power);
- $K$ is a constant related to the properties of the medium;
- $Z$ is the atomic number of the material;
- $A$ is the atomic mass of the material;
- $z$ is the charge of the incident particle (in units of the elementary charge, $e$)
- $\beta = v/c$ is the velocity of the particle relative to the speed of light;
- $\gamma$ is the Lorentz factor $(1/\sqrt{1-\beta^2})$;
- $m_e$ is the rest mass of the electron;
- $T_{max}$ is the maximum kinetic energy transferable from the incident particle to an atomic electron in a single collision;
- $I$ is the mean excitation energy of the material;

- $\delta(\beta\gamma)$ is the density effect correction, which accounts for the fact that at high velocities, the electron density of the medium becomes important.

The formula 2.2 incorporates relativistic effects through the Lorentz factor ($\gamma$) and the velocity ($\beta$) of the particle. As the particle's velocity approaches the speed of light ($c$), the formula approaches the non-relativistic Bethe-Bloch expression [5].

By integration of the Bethe-Bloch formula, a plot of the energy loss inside a material can be achieved. Propagation of the energy loss for radiotherapy purposes is shown in the water phantom because it is one of the closest materials that can imitate human tissues. Depending on the type of beam and its energy, the dose distribution is different, an example is shown in Figure 2.1. The depth and intensity of the energy loss are very important in radiotherapy planning, while cancer cells are usually located at a depth of up to several cm inside the body.

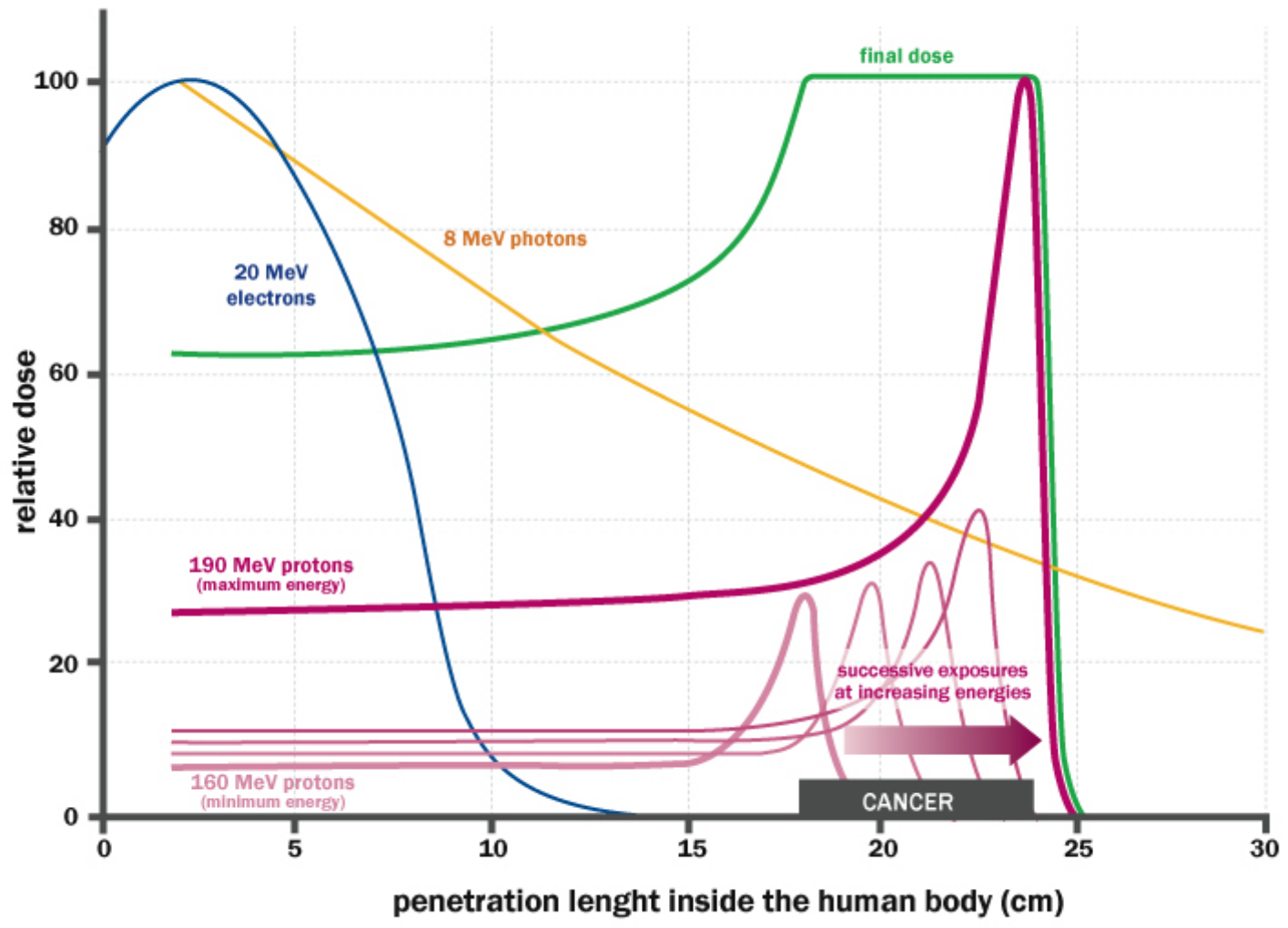

Figure 2.1: Distribution of absorbed dose in a water phantom depending on the type of particle and its energy (graphics from [6]).

## 2.2 The most common radiotherapy methods

Photon, hadron, and electron beams are most frequently used in teleradiotherapy.

Photon radiotherapy is the most widely available. It uses high-energy X-rays or gamma rays generated by high-energy electron beams colliding with a target (e.g. made of tungsten). The machinery is based on a relatively short (usually up to two meters) linear accelerator (e.g. a side-coupled linac) usually integrated within the gantry. Such devices are commercially available on the market, compact, reliable, and easy to operate, so they can be easily integrated into existing hospitals without the special need to build new facilities. The cost of such infrastructure is in the range of 2-3 M$, which

is affordable in most cases. The most commonly used energy range for photon beams in medicine is 4 MeV to 25 MeV. The most important disadvantage is the shape of the energy deposition curve, which irradiates not only the tumor but also healthy tissues. To minimize the impact on healthy cells, irradiation is often done from many directions and in multiple sessions, which is time-consuming and more inconvenient for patients.

On the other hand, hadron radiation therapy uses protons or heavier ions, like e.g. carbon. These are usually produced by a cyclotron or a synchrotron and delivered to the tumor site. The most commonly used hadron beam energy range in medicine is between 70 MeV and 250 MeV. Hadron radiotherapy delivers a highly targeted dose of radiation to the tumor while minimizing exposure to surrounding healthy tissue. Access to this type of radiotherapy is much more limited due to its high cost. Cyclotrons are more compact and more affordable (starting from 30 M$) options for accelerating protons, than synchrotrons (about 150-200 M$). However, they are not suitable for heavier ions, due to beam rigidity. They also require way more space than linacs do. Well-trained and experienced operators or engineers are needed to operate these apparatus. Synchrotrons for medical purposes are very complex machines, that are still considered more like prototypes for which operation methods are still under development.

Electron radiation therapy uses negatively charged particles to deliver the radiation dose. It often uses similar energies and equipment to photon beams. Because electrons have a more shallow penetration range inside tissue than photons, electron radiation therapy is often used to treat tumors located close to the skin surface or during surgeries.

All the described types of available radiotherapy methods can therefore be characterized by the cost and the complexity of the operation and the biological effect which can be derived from the shape of energy deposition curves related to each method, as in Figure 2.1. To sum up, photon and low-energy electron radiotherapy are simplified in terms of beam generation and handling, which compromises the energy deposition curve in tissues. On the other hand, hadron therapy requires extensive accelerator facilities and complex beam delivery systems that provide a favorable shape of the energy deposition curve. Also, the relative biological effectiveness (RBE) of carbon ions is 3, (relative to photon baseline RBE=1). Protons have an RBE similar to photons of around 1.1. Thus, in terms of tumor-killing potential per unit dose, the relation is: photon radiation ≈ proton radiation < carbon ion radiation [7].

## 2.3 VHEE beams in radiotherapy

Very high energy electron (VHEE) beams can also be used for the purpose of radiation therapy. VHEE energies are in the range of 50 MeV to 250 MeV. Figure 2.2 shows a focused electron beam with an energy of 200 MeV in comparison to other beams, it is noticeable that it has a very desirable shape. The absorbed dose reaches the tumor with great accuracy, but the healthy tissues surrounding it are well protected. However, the use of very high-energy electron (VHEE) beams in medicine is still rather rare, but very promising. An example of such treatment is FLASH radiotherapy, which is developed by a collaboration of CERN (European Organization for Nuclear Research),

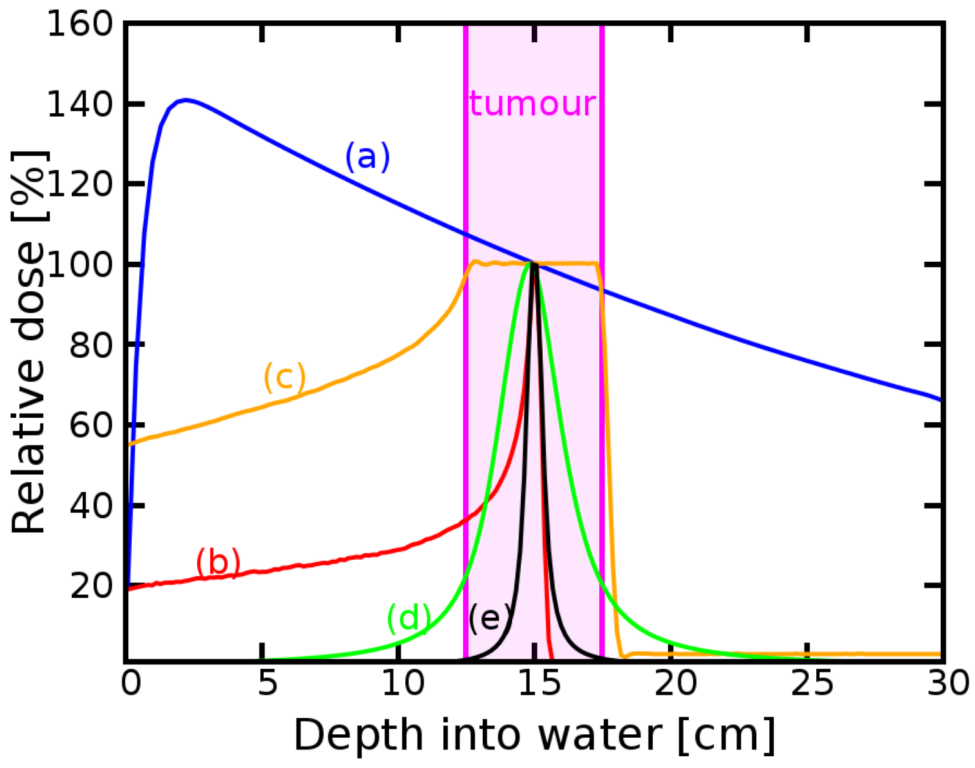

Figure 2.2: Distribution of absorbed dose depending on particle type and energy, where (a) 6 MeV photons; (b) Bragg peak for 148 MeV protons; (c) Bragg peak; (d) 200 MeV electrons; (e) 2 GeV electrons (graphics from [1]).

CHUV Hospital in Lausanne (Switzerland), and THERYQ (ALCEN Group). FLASH radiotherapy is a highly targeted cancer treatment, capable of reaching deep into the body of a patient with fewer side effects than traditional radiotherapy [8]. To obtain the VHEE beam, a linear accelerator is used. The construction of such a machine requires a high accelerating gradient to keep the device compact. Recent advances in the production of high-energy electron beams e.g. the high-performance CLIC linear electron accelerator technology, allow us to construct compact and reliable facilities that could be potentially installed even in existing hospitals.

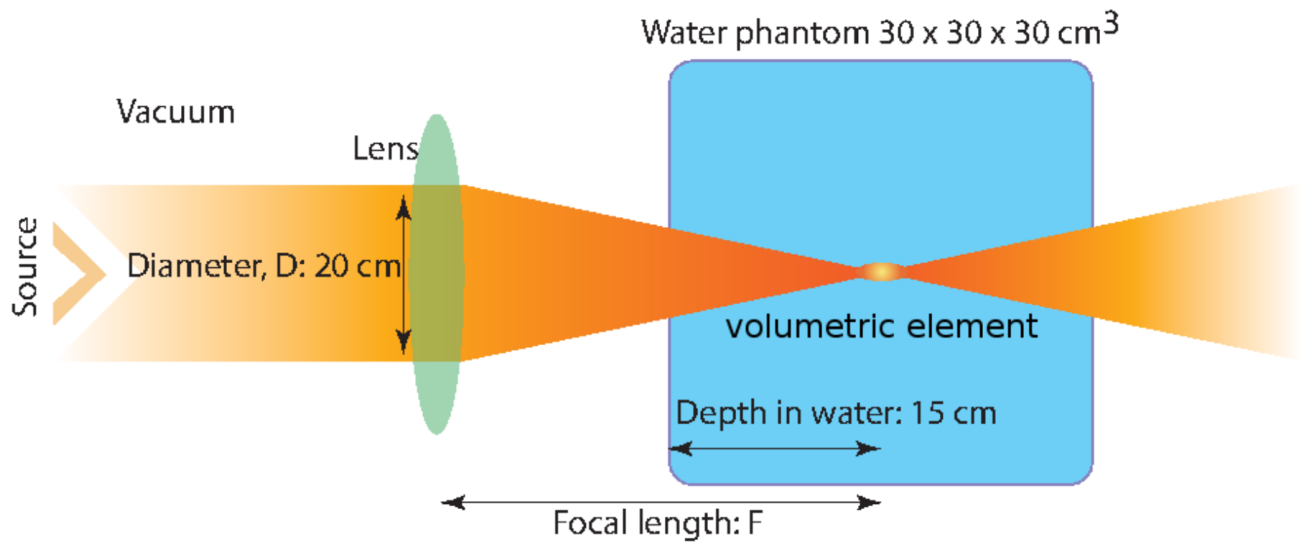

Figure 2.3: The F/D ratio (f-number) is a quantitative representation of the focusing force (graphics from [1]).

Focusing of VHEE beams into the region of a tumor is needed to optimize the shape of the energy deposition curve. A quantitative representation of the focus of a beam is the f-number value, which is the ratio of the focal length to the diameter of the focusing lens. Figure 2.3 shows such a focusing setup.

Figure 2.4 shows the variation in dose propagation for a 200 MeV electron beam, depending on the focusing strength, where f/1.2 is the strongest focusing and f/11.5 is the weakest. Figure 2.5 compares the 200 MeV and f/1.2 electron beam (the turquoise line), with a beam of the same energy but much lower focusing (the gray line), and an electron beam with an energy of 158 MeV, which focusing is f/11.2 (the purple line), and although its energy is lower, the focusing is higher, so ultimately the dose propagation curve better targets the tumor placed at a depth of 15 cm.

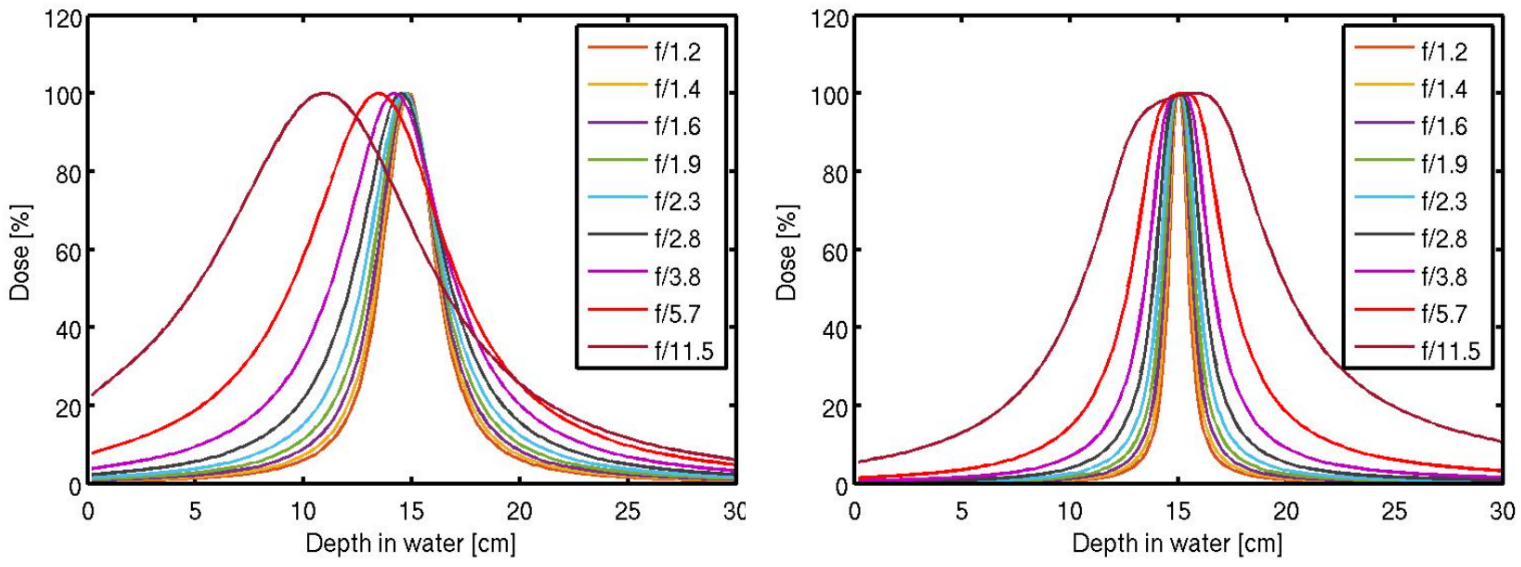

(a) (b)

Figure 2.4: Propagation of absorbed dose in a water phantom for a 200 MeV electron beam focused with f/1.2 - f/11.5 (graphics from [1]).

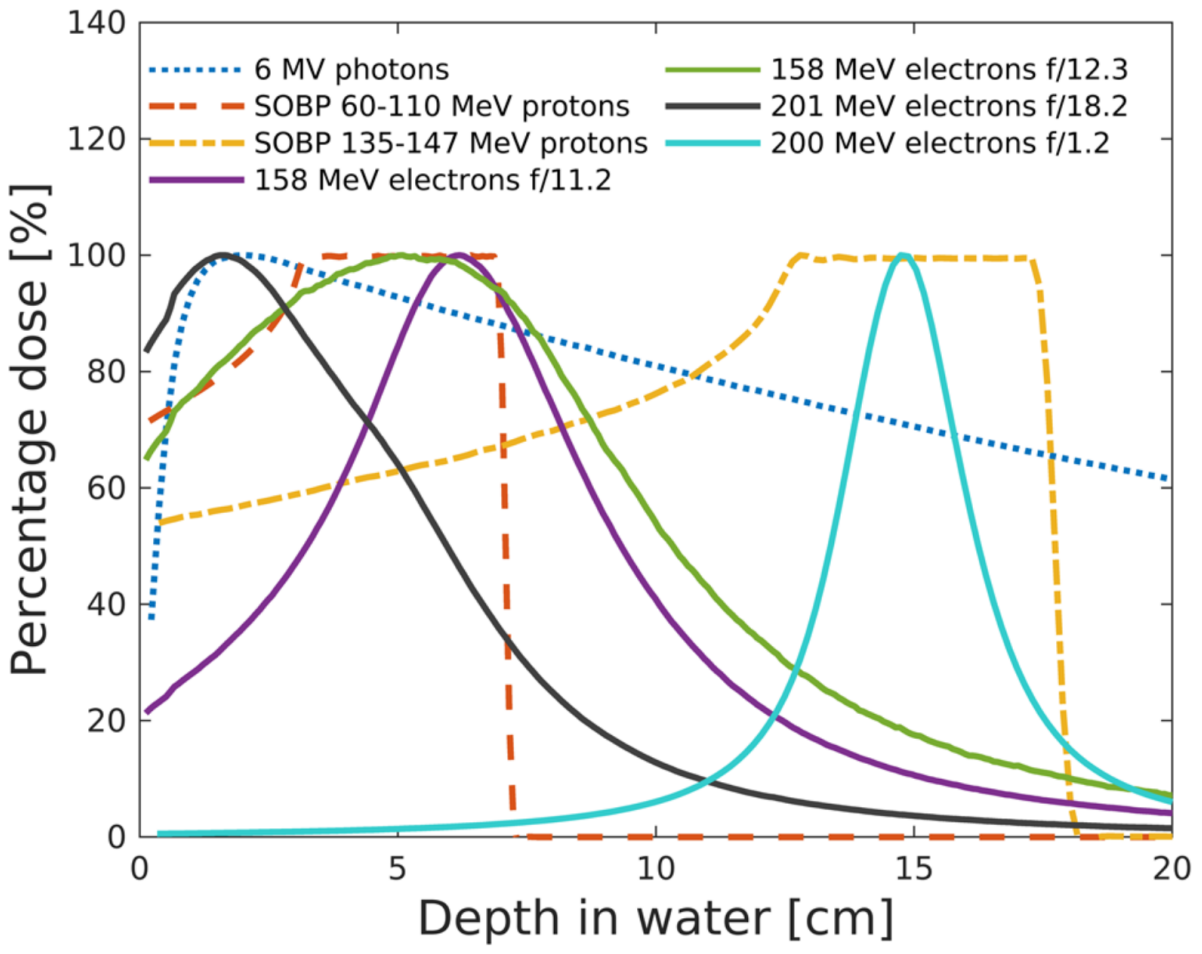

Figure 2.5: Propagation of absorbed dose in a water phantom for different beams (graphics from [9]).

Conventionally, high-energy charged particle beams are focused with the use of electromagnets, which are usually large and heavy. Their weight often ranges from tens to hundreds of kilograms. As a consequence, beam delivery systems can easily become heavy and challenging from a mechanical point of view, especially when a full isocentric gantry is used for beam delivery. An alternative idea to guide the trajectories of charged particles is based on the crystal channeling process. The advantage is that crystals needed for such a purpose are much more compact and weigh up to a few grams. A comparison between a conventional electromagnet and a silicon crystal is given in Figure 2.6 and in Figure 2.7.

Channeling is a phenomenon that may occur when high-energy particles, such as protons or electrons, interact with the internal structure of a crystal. The electric fields originating from the atoms in the crystal's lattice can cause particles to follow some special trajectories within the crystal and therefore make their motion nearly undisturbed. Such an effect is observed for both positively and negatively charged particles, although the trajectories are different depending on the particle charge. More details are given in the next Chapter.

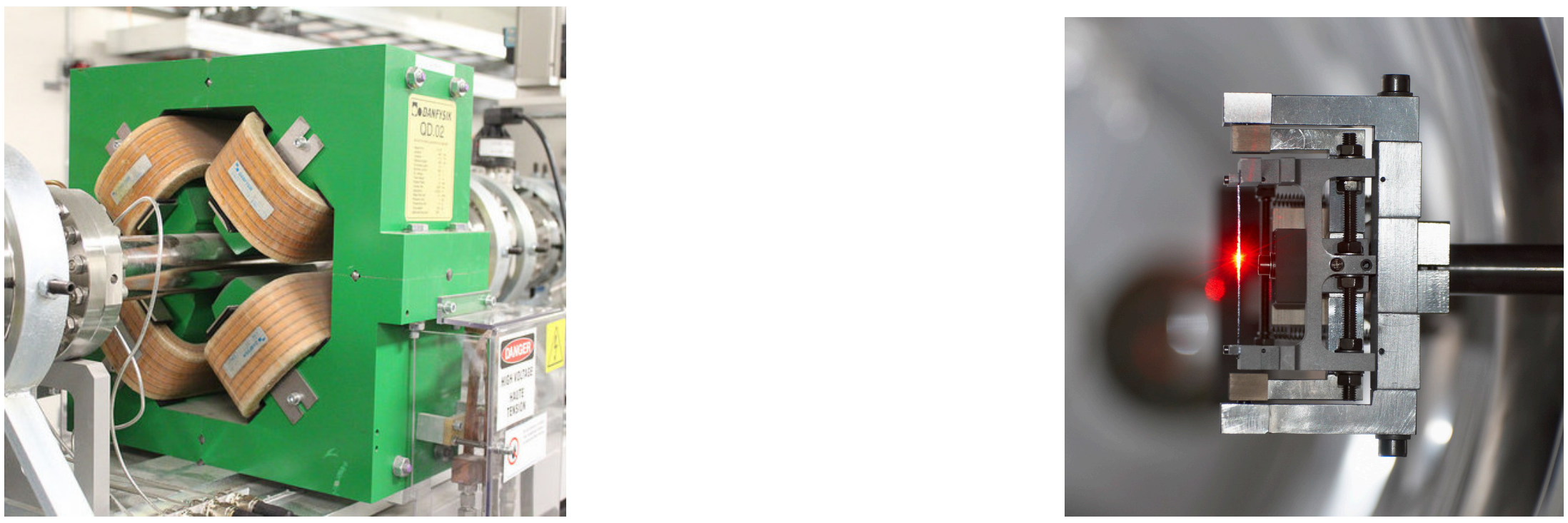

Figure 2.6: An example of a quadrupole magnet used with high-energy electron beams on the left (graphics from [10]), and UA9 bent crystal tested with a laser on the right (graphics from [11]).

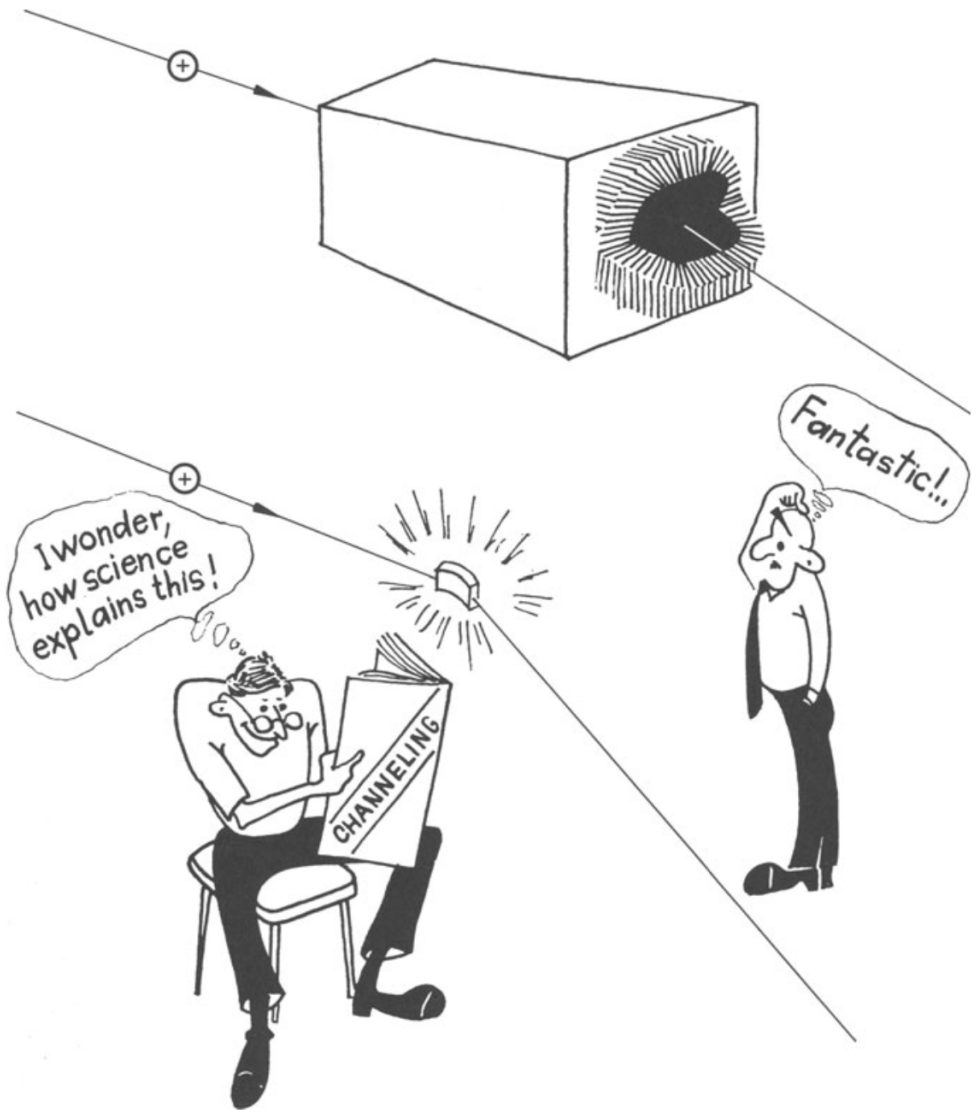

Figure 2.7: An image of a deflected proton beam by a dipol magnet on the top, and by a bent crystal on the bottom (graphics from [12]).

# 3. Crystal channeling

Using crystals to focus high-energy electron beams is potentially possible due to the channeling process that may occur when charged particles enter the crystal. A theoretical overview of the corresponding physics phenomena is discussed in this chapter together with examples of applications of crystals used for high-energy beam steering.

## 3.1 Structure of crystals

Crystals are characterized by an ordered, symmetrical, and periodic arrangement of atoms, referred to as the crystal lattice. A grid of straight lines is often used to illustrate a crystal lattice, with the points of intersection being called the lattice's nodes. In the simplest case, each node corresponds to a single atom. If there are multiple types of atoms, each node in the crystal structure has a group of atoms associated with it; these groups form a base and are identical in terms of position, orientation, and structure [12]. Two examples of some crystal lattices are shown in Figure 3.1.

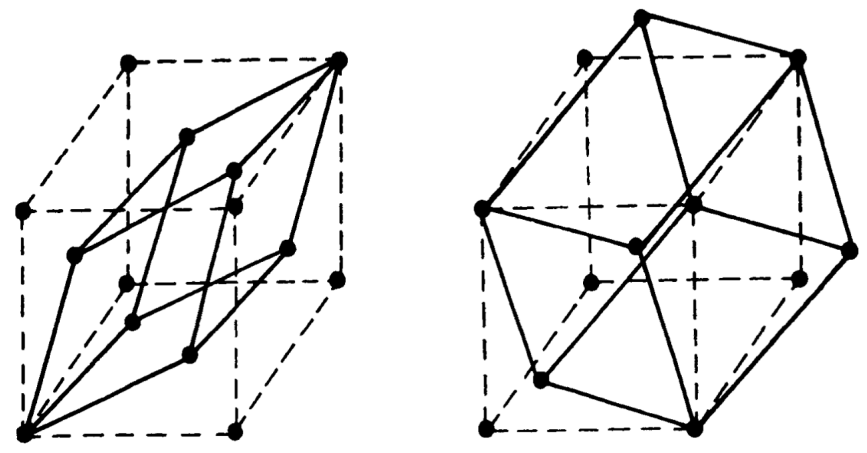

Figure 3.1: The face-centered cubic (FCC) lattice on the left and body-centered cubic (BCC) lattice on the right (graphics from [12]).

The planes of the crystal, passing through an infinite number of lattice nodes, are defined by Miller indices. To describe the direction in the crystal, a straight line passing through the origin of the coordinates is taken. Identical network nodes are placed at equal distances along this line. The position of the line is determined by the coordinates of the nearest nodes n1, n2, and n3, which by definition are three minimum numbers. Let these be h, k, l. In square brackets [hkl] is the axis symbol. This symbol defines a family of parallel axes. The set of axes equivalent to the symmetry sign is denoted by (hkl). For example, in a cubic crystal, the coordinate axes [100], [010], [001] are denoted by (100). The main planes and axes of the cubic crystal are shown in Figure 3.2. They are normal to planes with the same indices. In crystals of other systems, the normality of the axes to planes with the same indices is generally not satisfied. The diagonals of a cube, which are equivalent planes due to the nature of symmetry, can be denoted by (100). The plane passing through the diagonals of opposite faces of the cube has an index (110), and the diagonal plane has an index (111). A strong electromagnetic field occurs between planes and between axes of the crystals [12].

Many elements of the periodic table and their compounds have a cubic structure, especially BCC

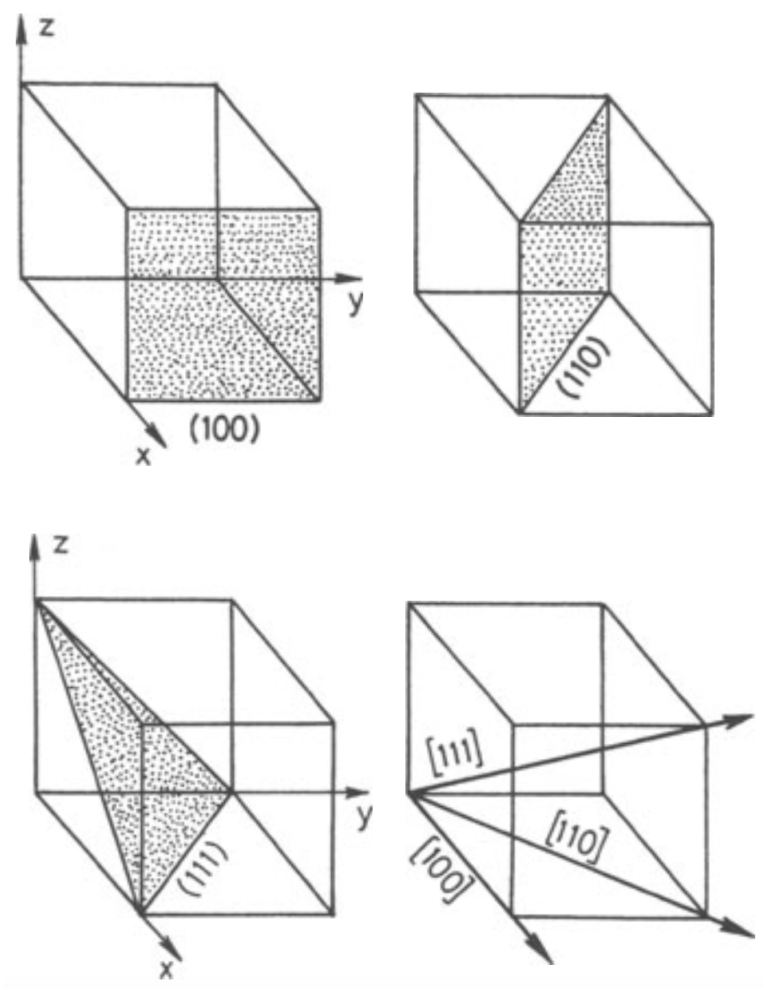

Figure 3.2: The main planes and axes of the simple cubic lattice (graphics from [12]).

and FCC. However, crystals from the diamond group (carbon, silicon, germanium, and tin gray) are used for manipulating trajectories of high-energy beams in particle accelerators [12].

The diamond lattice is cubic, but its cell contains eight atoms instead of four. It can be seen as two identical FCC lattices, pressed one into the other and shifted along the diagonal by a quarter of its length. The cubic lattice structure of the diamond is shown in Figure 3.3, where the full spheres are the atoms of the shifted FCC lattice. The figure shows that each atom has four neighbors located at the vertices of the tetrahedron. This structure is due to the so-called covalent bonding (due to electron exchange) of the atoms. Atoms in the diamond group have four electrons in the outer shell. In order to form a stable eight-electron shell, they lack four electrons, which they compensate for through electron exchange with their four nearest neighbors. Because of the tetragonal covalent bonding, the diamond's lattice is not tightly packed [12].

Monocrystals of silicon and, less frequently, germanium are mainly used for manipulating trajectories of high-energy particle beams. This is due to the high degree of perfection of the crystal lattice in semiconductor materials and the well-developed technology for producing large semiconductor crystals [12].

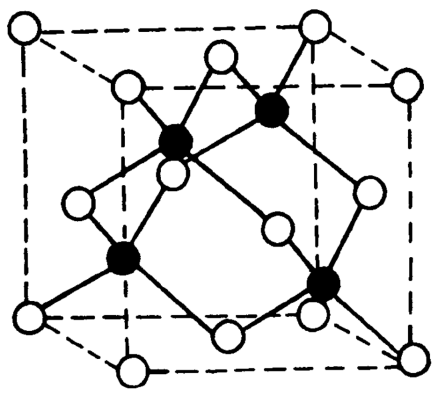

Figure 3.3: Diamond cubic lattice: two identical FCC lattices, inserted one into the other and shifted along the volume diagonal by a quarter of its length. The black spheres are the atoms of the shifted lattice (graphics from [12]).

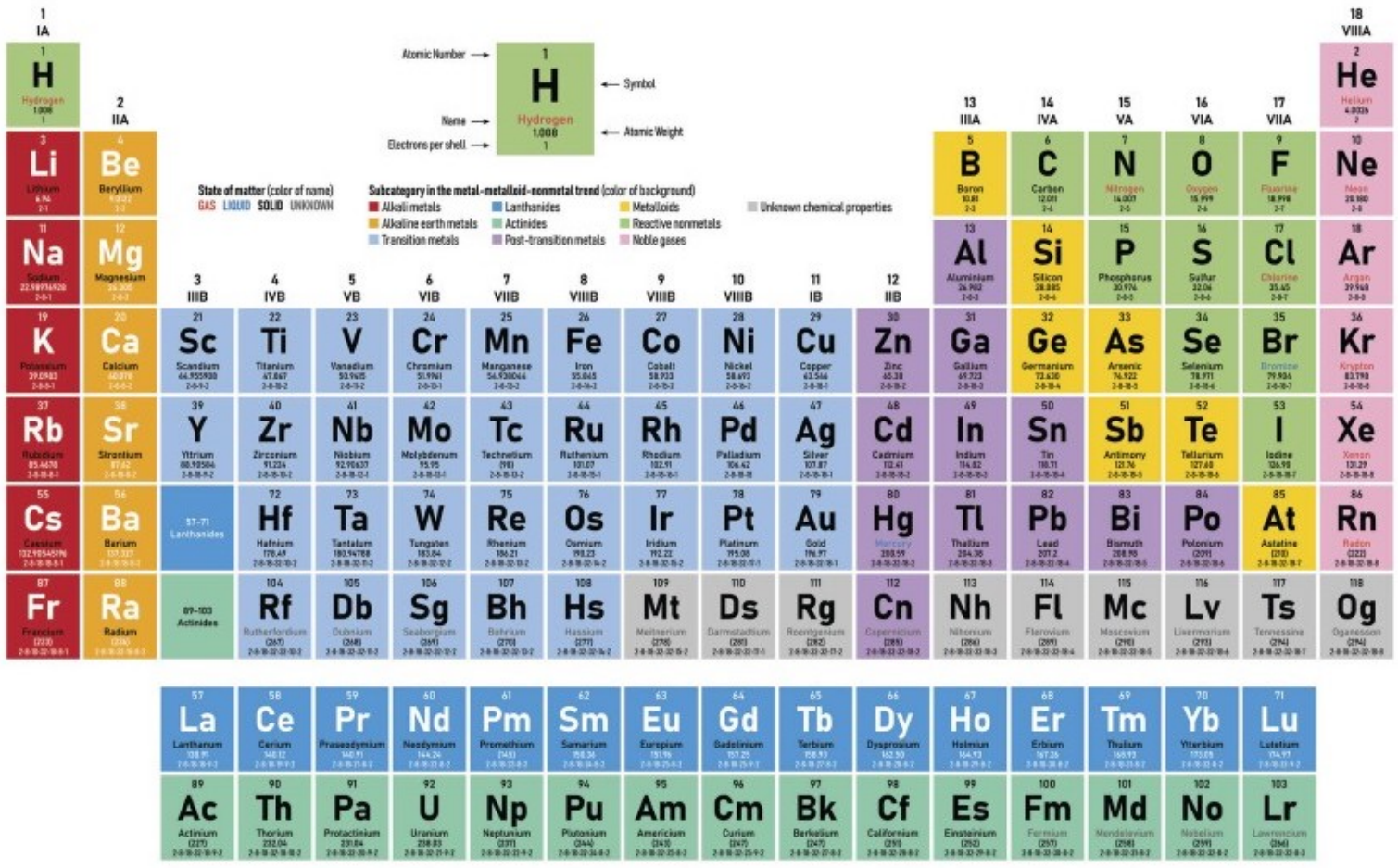

Figure 3.4: Periodic table of the elements, the diamond group is in the 14th column (graphics from [13]).

## 3.2 Channeling

Channeling is the phenomenon of guiding charged particles through a periodic internal structure of a crystal if particle direction is aligned with its planes or axes. The crystal structure forms a collective, electromagnetic potential that captures charged particles as they enter the crystal. Particles become trapped between planes (planar channeling) or axes (axial channeling) which allow them to propagate through the crystal body nearly undisturbed. On the other hand, in amorphous materials charged particles undergo many random processes, for example, ionization and scattering causing a random change of their trajectories and momenta [12].

An electrostatic potential seen by a particle entering the crystal's body can be parameterized based on the Thomas-Fermi model in the following form:

$$V(r) = \frac{Z_i Z e^2}{r} \Phi(\frac{r}{a_{TF}}), \tag{3.1}$$

where $Z$ is the atomic number of the target atom, $Z_i e$ is the charge of the impinging particle, r is the relative distance, and $\Phi(\frac{r}{a_{TF}})$ is a Molière screening, and $a_{TF}$ is a function that takes into account the electronic cloud around the nucleus [14].

According to Lindhard [14], "under the hypothesis of a small impact angle of the striking particle with respect to the crystal plane, we can consider the average potential generated from the entire crystal plane as a continuous potential" that can be written as:

$$U_p(x) = Nd \iint_{-\infty}^{+\infty} V(x, y, z) dy dz, \tag{3.2}$$

where x is the coordinate perpendicular to the plane of the crystal, N is the atomic density, d is the distance from the plane, and V (x, y, z) is the potential in equation 3.1. The potential is determined by the average of this distribution when this motion is assumed to be independent of position within the crystal and using a Gaussian spatial distribution for atoms in the plane. The potential that a positively charged particle would see while looking across the crystal plane is then shown in Figure 3.5 [14].

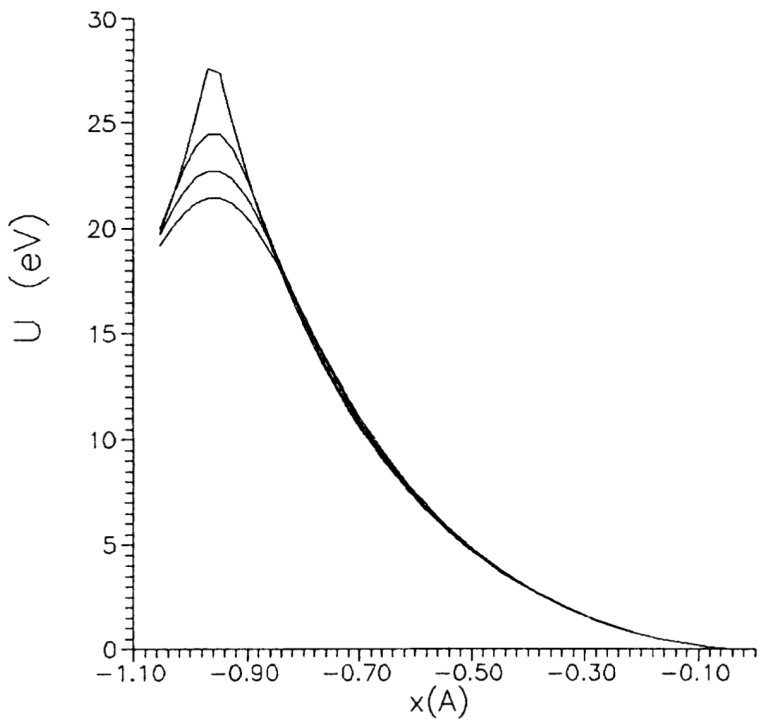

Figure 3.5: Moliere potential of Si(110) planes at different temperatures and static lattice potential. From top to bottom at left edge: static, 77 K, 300 K, 500 K (graphics from [12]).

By superimposing the two planes, the potential (close to the minimum) seen by a particle between them will be a harmonic potential. This follows from the assumption that particles are only affected by the potential of the nearest planes; thus, the entire well of potential affecting particle motion between crystal planes can be approximated as [14]:

$$U(x) \approx U_p(\frac{d_p}{2} - x) + U_p(\frac{d_p}{2} + x) \approx U_{max}(\frac{2x}{d_p})^2. \tag{3.3}$$

Figure 3.6 displays examples of silicon's (110) and (111) planes potential. The analytical estimations frequently use the harmonic approximation $U \approx x^2$, which fits the interplane potential rather adequately and is plotted with a dashed line [12].

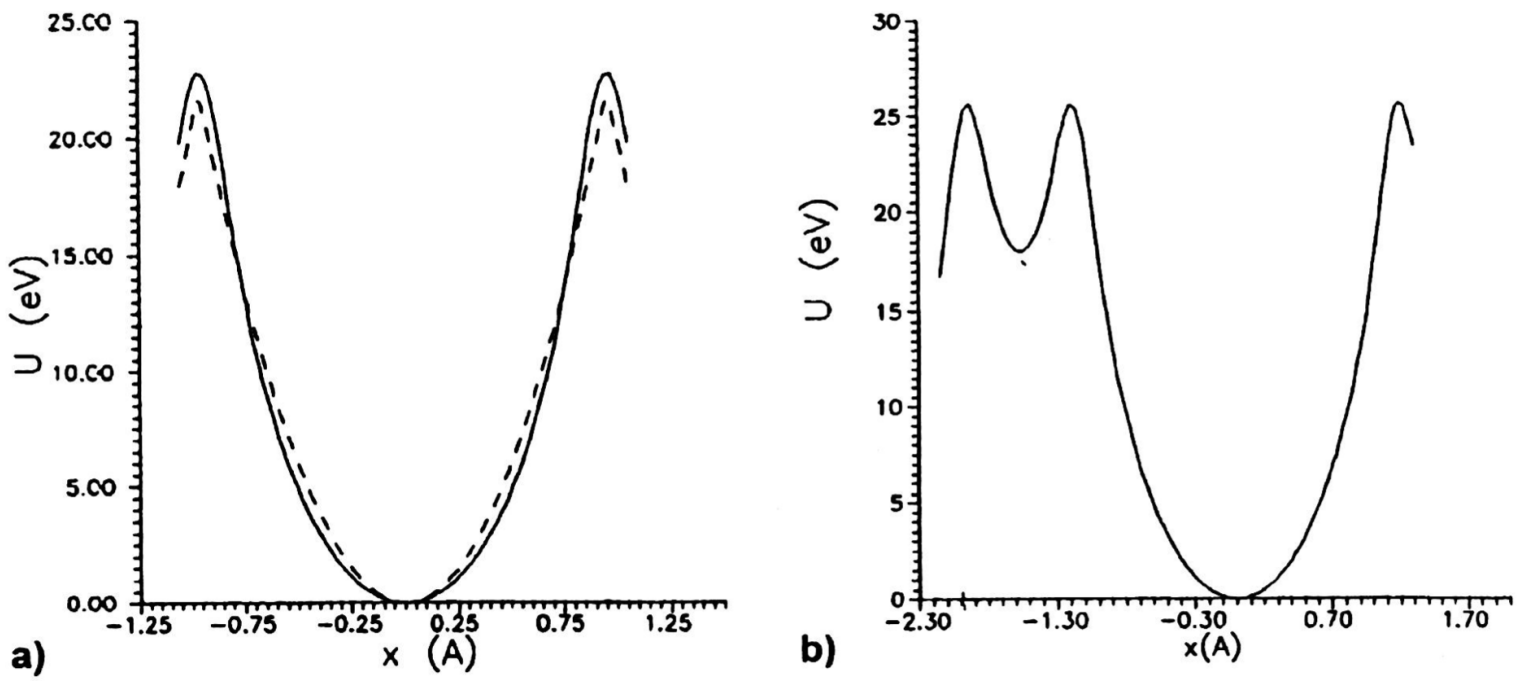

Figure 3.6: The interplanar Molière potential as seen by a proton entering between crystal planes at a small angle; (a) Si(110) plane; (b) Si(111) plane (graphics from [12]).

The silicon potential well has a depth of around $U_{max} \sim 20eV$. In Figure 3.7, the interplanar electric field U'(x) for silicon's planar channels (110) and (111) is depicted [12].

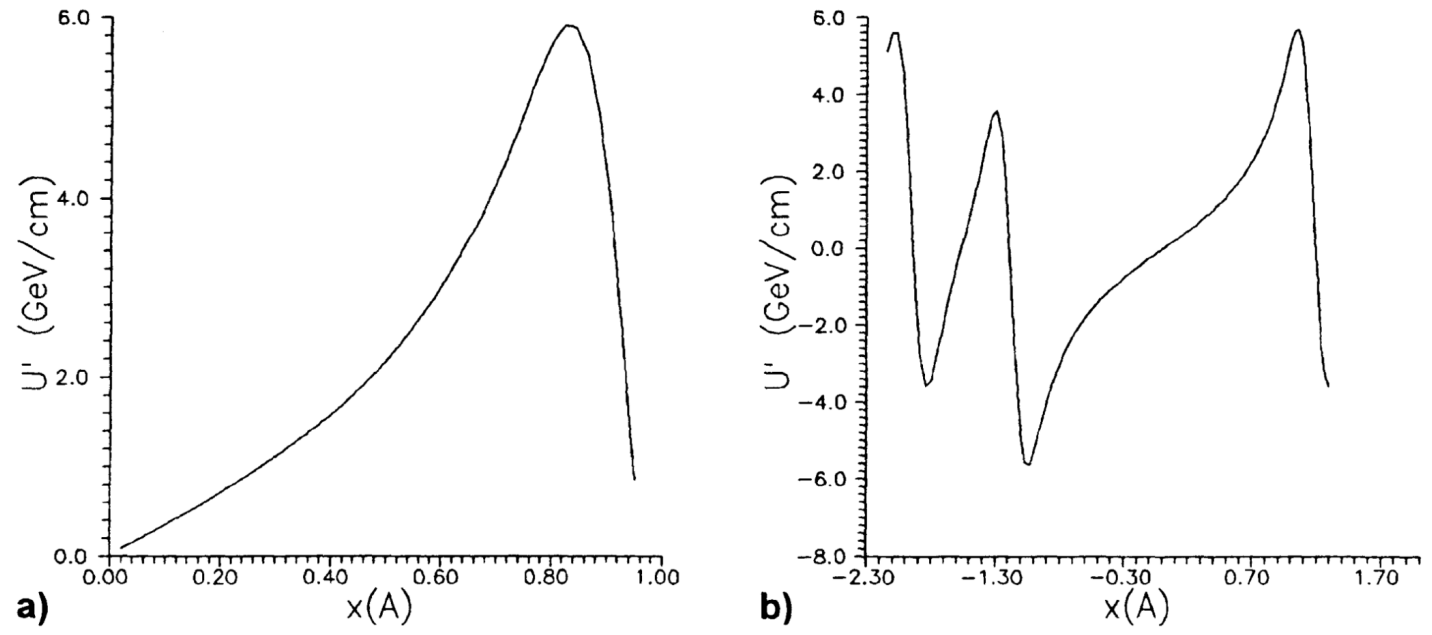

Figure 3.7: The interplanar electric field (Molière) for (a) Si(110) plane; (b) Si(111) plane (graphics from [12]).

To experience trapping, the particles need to have a transverse momentum low enough, such that they stay inside of the potential well. Thus, to undergo planar channeling a particle needs to fulfill the condition [14]:

$$\frac{p^2c^2}{2E}\theta^2 + U(x) \leq U_{max}, \tag{3.4}$$

where $E$ stands for the total particle energy, $p$ is the particle momentum, $c$ is the speed of light. Assuming that the particle enters the center of the channel and using the relation $pc^2 = vE$, where $v$ is the velocity of the particle, the above equation can be simplified to the form [14]:

$$\frac{pv}{2}\theta^2 \leq U_{max}. \tag{3.5}$$

A critical channeling angle $\theta_c$ can be therefore defined, as the maximum angle of incident particle allowing for stable planar channeling. Critical entry angle for a straight crystal [14]:

$$\theta_c = \sqrt{\frac{2U_{max}}{pv}}. \tag{3.6}$$

When the crystal is bent, the behavior of channeled particles is not significantly different from what occurs in straight crystals. The effect of bending of the crystal can be understood as an extra centrifugal motion of a particle which lowers the effective height of the potential well. The deformation of a potential well at different bending radii is shown in Figure 3.8. From this figure, the presence of a critical bending radius $(R_c)$ can be deduced, depending on the energy of the particles, after which the process of planar channeling is no longer possible due to the insufficient depth of the potential well [14]. The critical radius $(R_c)$ can be also understood as the radius for which the centrifugal force is equal to the maximum interplanar field:

$$R_c = \frac{pv}{U'(x_{max})} = \frac{pvx_{max}}{2U_{max}}, \tag{3.7}$$

where $U'(x_{max}) \approx 5.7 GeV/cm$ in silicon crystals, and is calculated in $x_{max} = \frac{d_p}{2} - a_{TF}$ due to the the finite atomic charge distribution [12, 14].

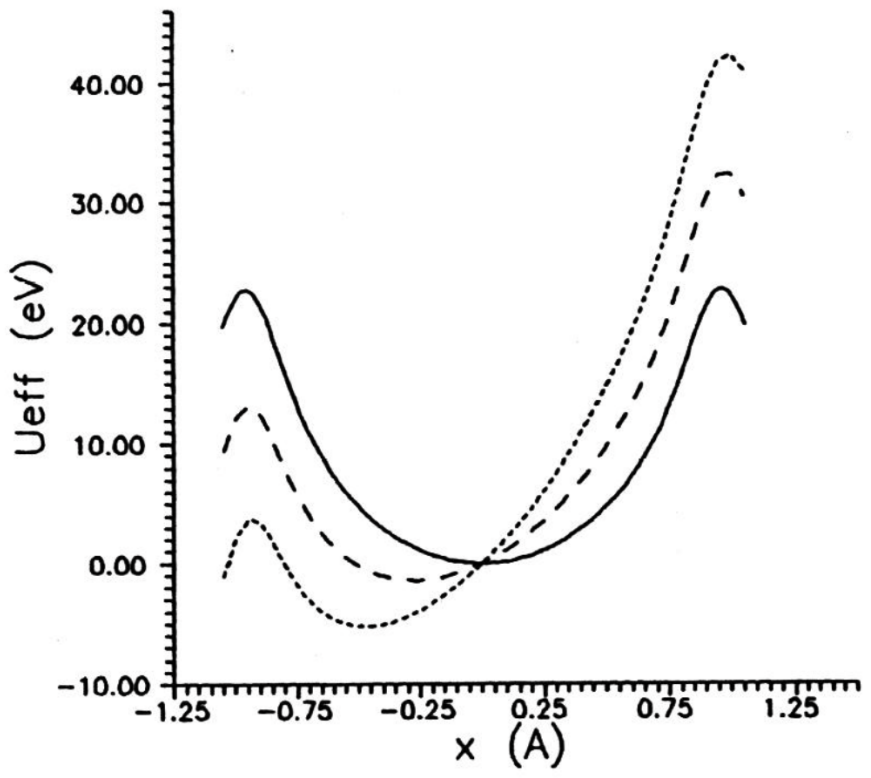

Figure 3.8: Effective potential in Si(110) crystal, where the solid line - straight crystal; dashed line - curved crystal with centrifugal force 1 [GeV/cm]; dotted line - curved crystal with centrifugal force 2 [GeV/cm] (graphics from [14]).

The critical beam incoming angle for a bent crystal is represented by the formula:

$$\theta_c^b = \theta_c \left(1 - \frac{R_c}{R}\right), \tag{3.8}$$

where $R > R_c$.

If the channeling motion is maintained along the entire length of the crystal, the channeled particle is deflected by an angle equal to the geometric bend of the crystal:

$$\theta_b = \frac{l}{R}, \tag{3.9}$$

where $l$ is the crystal length.

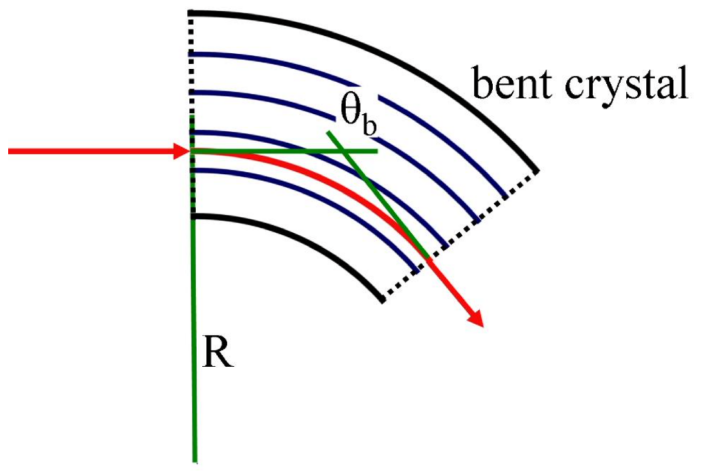

Figure 3.9: Deflection imparted to guided particles along the length of the crystal (graphics from [14]).

Both positively and negatively charged particles can undergo channeling. The shape of the crystal's collective potential determines these particles' trajectory, which varies depending on the charge of the interacting particles. Channeled particles oscillate between or across planes or axes. Under channeling conditions, positive particles can penetrate deeper into the crystal because they are repelled by nuclei, which also have a positive charge. Negatively charged particles, on the other hand, are much more likely to interact with the atomic centers of crystals, which makes them more easily fall out of the channeling process [15].

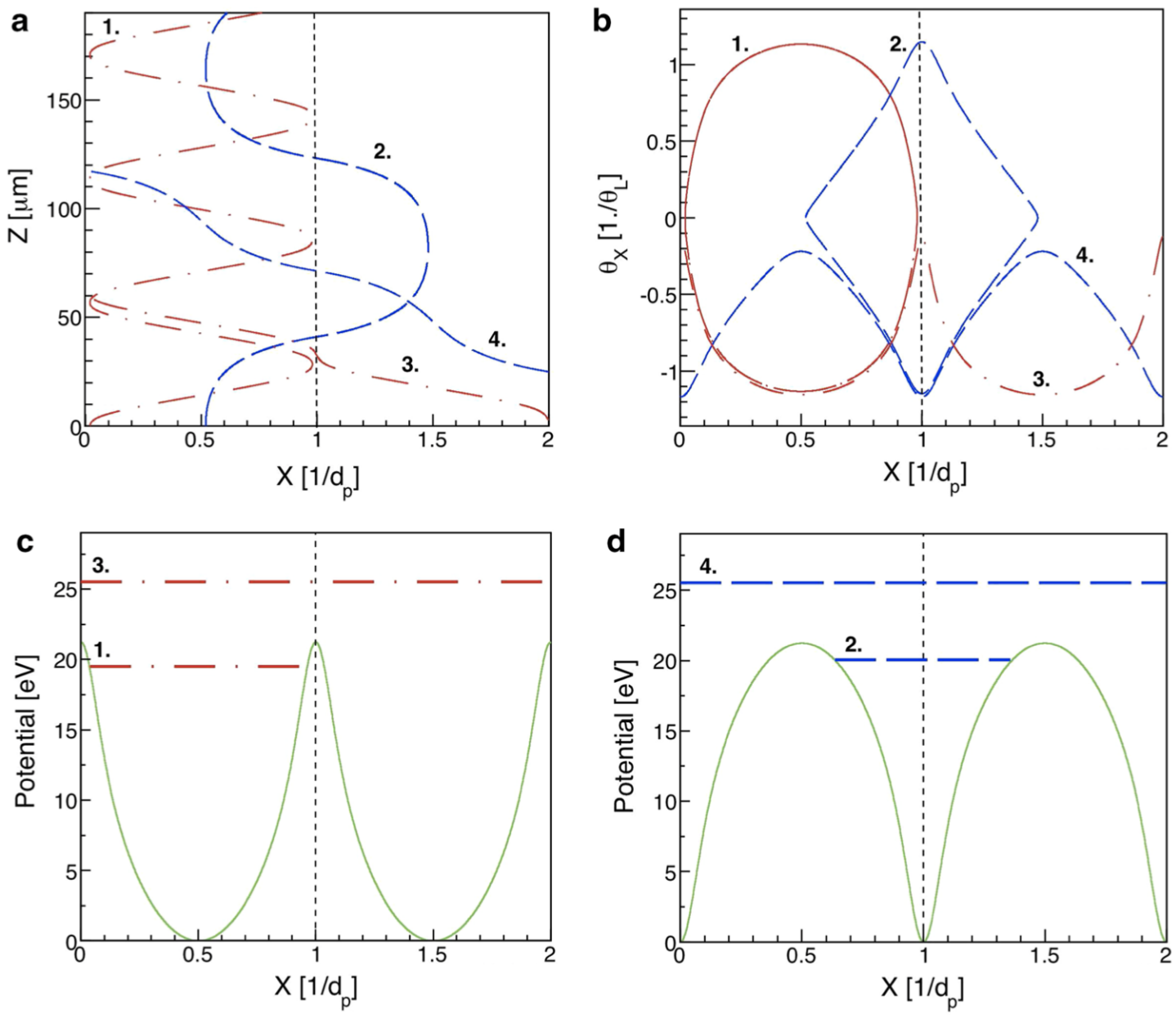

Figure 3.10: 400 GeV/c particles interacting with Si (110) planes (dotted lines). Curves 1 and 2 refer to channeled particles while curves 3 and 4 refer to over-barrier particles. Dashed (dot-dashed) lines represent negative (positive) particles. a) Trajectories as a function of transverse position $(X)$ and penetration depth $(Z)$. b) Trajectories as a function of transverse position $(X)$ and transverse angle $(\theta_X)$. Continuum planar potential (continuous line) and transverse energies for c) positive and d) negative particles (graphics from [15]).

Due to interactions with the crystal structure, not all of the particles entering the crystal are successfully channeled, but there are other possible events that particles can encounter, such as dechanneling, volume capture, volume reflection, and amorphous scattering [14].

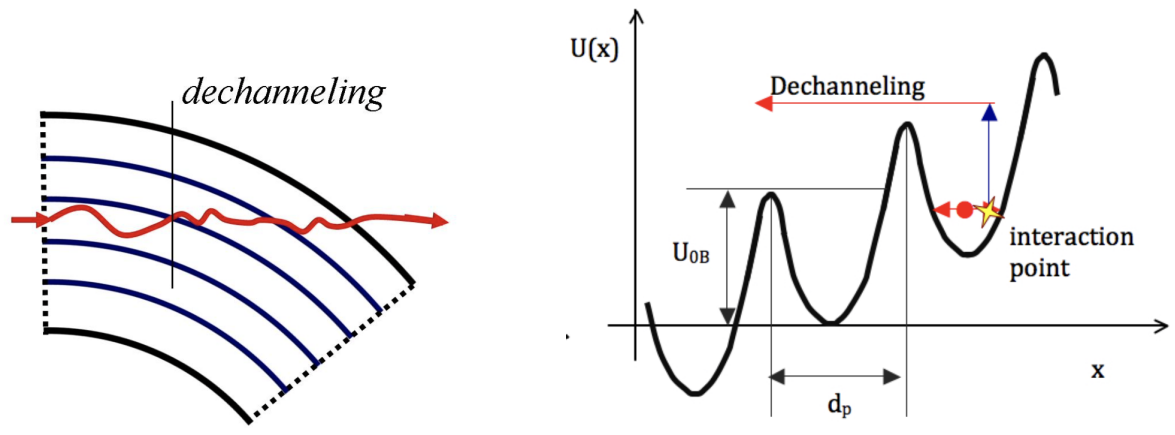

Figure 3.11: Dechanneling (graphics from [14]).

Dechanneling is a phenomenon that occurs when the transverse energy of a particle changes and no longer meets the condition to be channeled and jumps out of the flow. It contributes to decreasing the initial population of channeled particles. On the other hand, volume capture increases the number of channeled particles [14].

Volume capture involves changing the transverse energy of the particle to one that is within the range of the particle channeling condition, and then that particle begins to be channeled, even though it previously was not [14].

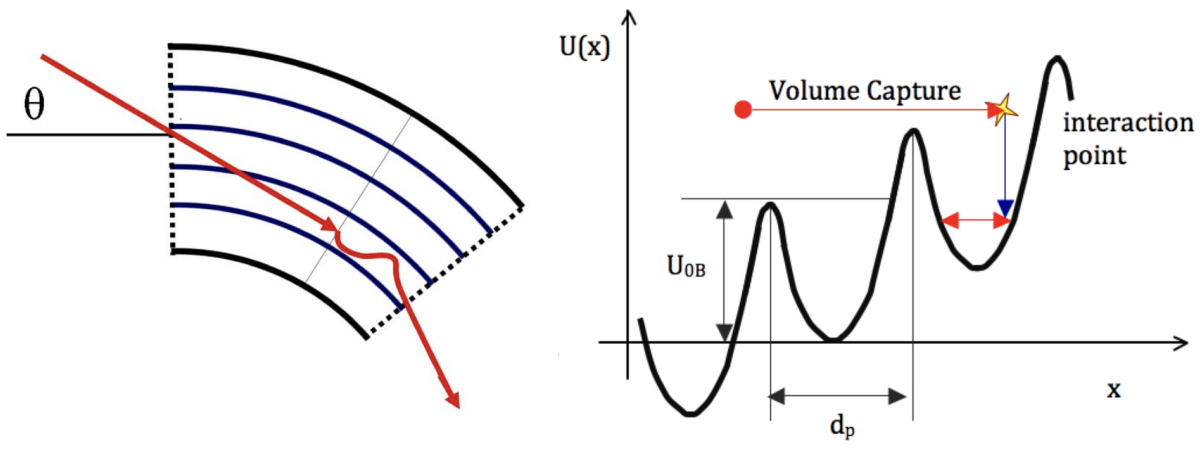

Figure 3.12: Volume Capture (graphics from [14]).

Volume reflection occurs when particles are reflected by interacting with the averaged potential of the crystal structures. It affects particles that impinge on bent crystals with an incident angle in the range between the critical angle and bending angle of the crystal [14].

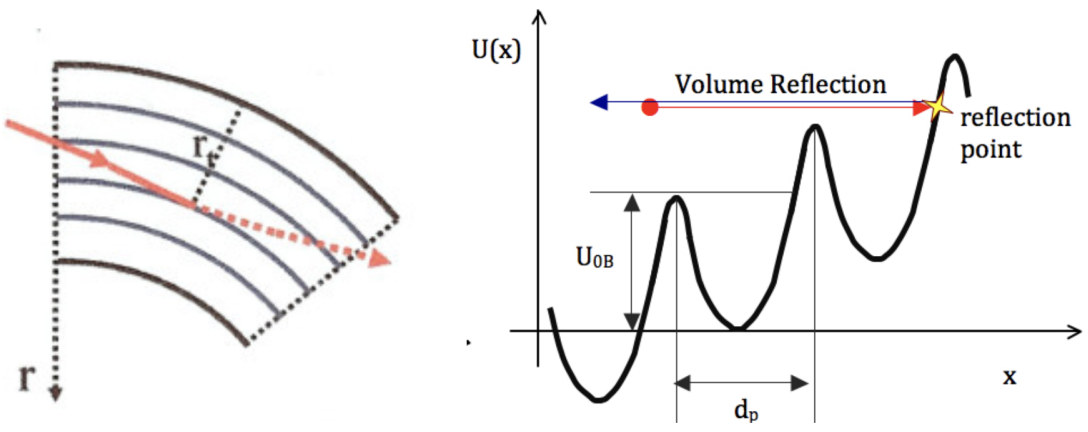

Figure 3.13: Volume Reflection (graphics from [14]).

Amorphous scattering implies that the particles treat the crystal as an amorphous material and do not interact with crystalline structures. In amorphous materials, the energy released by ionization as a result of a large number of random scatters may lead to particle absorption [12].

To have as many channeled particles as possible, it is important to choose the right parameters of the crystal, based on the critical values.

Such bent crystals can be produced with the use of a holder that clamps the crystal in order to induce a secondary curvature on the plane selected for particle steering. It can be divided into two categories that use different technologies. One of them is the manufacturing of strip crystals (ST), shown in Figure 3.14, silicon strips are mechanically bent at their extremities. As a result, an anticlastic curvature is induced along the (110) planes, which are used for channeling. The other one is the production of quasi-mosaic crystals (QM), which are thick silicon tiles clamped by a holder along their axis. This generates a quasi-mosaic curvature along the (111) planes [16].

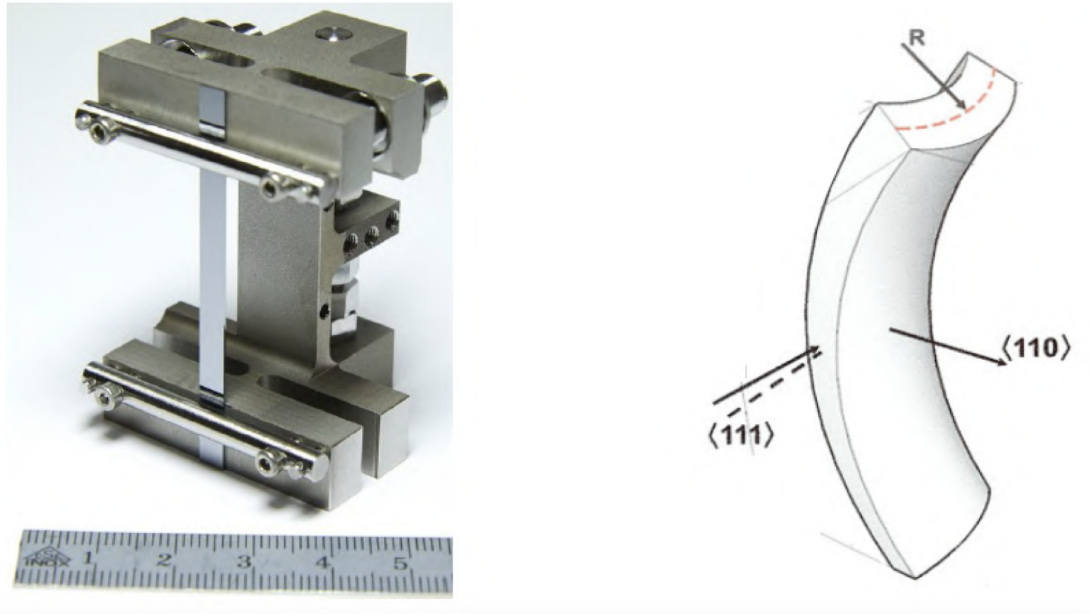

Figure 3.14: On the left side, is a picture of a strip crystal for the Large Hadron Collider with its titanium holder. On the right side, is a scheme of how the bending of the crystal is mechanically produced (graphics from [16]).

## 3.3 Beam focusing

To focus the beam while it passes through the crystal, the output of the crystal must be properly profiled. Such a crystal is then supposed to behave like a focusing lens. To profile the exit of the crystal, it is cut with a section of a cylinder to make a solid like the one shown in Figure 3.15, where a beam of particles enters the crystal from the left side and is channeled inside the crystal, and after exiting it, is focused at a distance F [17].

Crystal length:

$$L = R * \theta_b, \tag{3.10}$$

where L - length of the crystal, R - crystal bending radius, and $\theta_b$ - beam deflection angle.

The radius of the cutting cylinder:

$$r = \frac{1}{2}\sqrt{F^2 + R^2}, \tag{3.11}$$

where r - cylinder radius, F - focal length, and R - crystal bending radius.

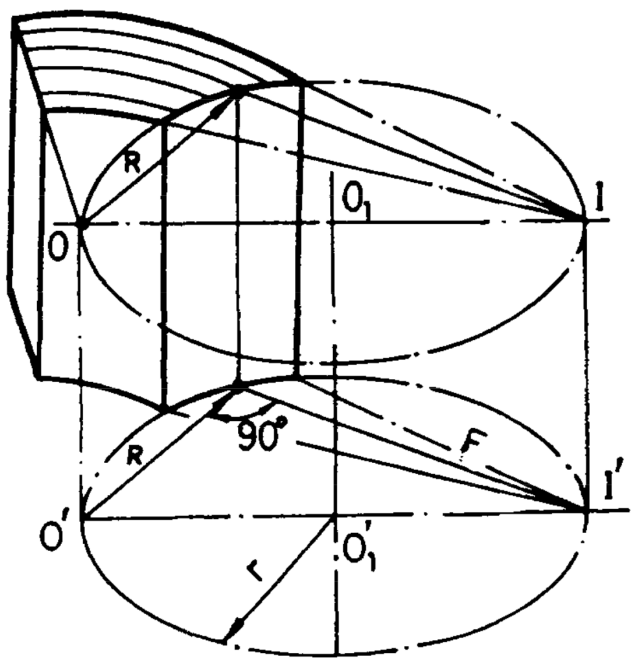

Figure 3.15: The trajectory of the electron beam after passing through the crystal lens (graphics from [17]).

## 3.4 Applications of bent crystals in high energy physics

Bent crystals have found diverse and innovative applications in the field of high-energy physics, enabling improved beam manipulation, extraction, and more collision opportunities in particle accelerators. Some examples are given below.

Bent crystals have proven to be promising tools for improving the performance of beam collimation at the Large Hadron Collider (LHC), the world's largest particle accelerator at the European Organization for Nuclear Research (CERN). Collimation is a crucial process that involves removing off-center particles from the beam to prevent damage to accelerator components. Bent crystals can serve as efficient beam-cleaning devices by deflecting halo particles away from the core beam, providing a more efficient means of beam cleaning compared to conventional collimation methods [14]. Crystal collimation techniques are now a part of the standard procedure used for ion beams in Run 3 and are being studied for use with proton beams [16, 18]. They are also being considered for the future HL-LHC operation.

The utilization of bent crystals in the Super Proton Synchrotron (SPS) at CERN has been investigated for beam collimation and proton extraction. Initial experiments focusing on beam halo collimation demonstrated promising results using bent crystals. Subsequent research has explored the extraction of protons, achieving an extraction efficiency of approximately 10%. These combined findings highlight the versatility and potential of bent crystals in enhancing both beam halo collimation and extraction processes within the SPS [19, 20].

Bent crystals have also sparked innovative ideas for achieving fixed-target experiments based on crystal-assisted beam halo splitting. Studies have proven that it can be operated safely without affecting the availability of the LHC for regular beam-beam collisions. While this concept is still under exploration, it showcases the potential of bent crystals to broaden the collision process in high-energy physics [21, 22].

In conclusion, bent crystals have emerged as versatile tools with a wide range of applications in high-energy physics. Their unique ability to channel particles and create tailored deflection forces has enabled significant advancements in beam manipulation and extraction. From collimation at the LHC to beam extraction at the SPS, innovative ideas on steering the beam to fixed-target collision points, and many more. These applications highlight the growing importance of bent crystals in shaping the future of particle accelerator technology and experimental research in the field of high-energy physics. There is also the possibility of using bent crystals in other fields, a proposal for one such application is presented in this study.

# 4. Numerical simulations

In order to be able to perform simulations, a program was prepared in the Geant4 environment. It is based on an existing example called *channeling* provided by default with the Geant4 code [23].

## 4.1 Geant4

The programming environment used to perform the calculations is Geant4 which is a toolkit for the simulation of the passage of particles through matter. Its areas of application include high-energy, nuclear, and accelerator physics, as well as studies in medical and space science. It offers a variety of features, such as tracking, geometry, physics simulations, and hits. The provided physical processes cover a broad spectrum, including electromagnetic, hadronic, and optical processes, a large set of long-lived particles, materials, and elements, over a wide range of energies. The toolkit results from a worldwide collaboration of physicists and software engineers. It was created using software engineering and object-oriented technology and implemented in the C++ programming language [24, 25, 26].

The version of Geant4 on which the simulation was prepared is geant4-v11.0.1.

## 4.2 Channeling example in Geant4

The *channeling* example provided by default with the Geant4 code demonstrates a simulation of the channeling process in a bent silicon crystal. The particles used are 400 GeV/c protons released at a distance of -10.5 m from the crystal with a divergence of 13.36 $\mu$rad and 11.25 $\mu$rad in the horizontal and vertical plane, respectively. The geometry also includes three Si detectors placed at distances of -9.998 m, -0.320 m, and 10.756 m from the bent crystal to perform measurements of incoming and outgoing angles of protons interacting with the bent Si crystal. All geometry is placed in a vacuum. Conventional Geant4 physics is complemented in this example by channeling and volume reflection physics. Extra information about the interplane potential, nuclei, and electron density of the Si crystal are given in the additional files provided with the example. The details about the used model [23] are given in the following paper [15].

The most important parameters of the discussed example are controlled by a macro file (see Figure 4.1), e.g.:

- /xtal/setBR XXX 0. 0. m - changes crystal bending to XXX meters;
- /xtal/setSize 1.0 70. XXX mm - changes crystal length to XXX millimeter;
- /xtal/setEC data/Si220pl - selects the (110) Si crystal plane of channeling [23].

The output from the *channeling* example is a ROOT file [27] named *ExExhCh.root* with the data object *ExExChTree* that contains:

- angXin - particle horizontal angle at the crystal entry;
- angYin - particle vertical angle at the crystal entry;

- posXin - particle horizontal position at the crystal entry;
- posYin - particle vertical position at the crystal entry;
- angXout - particle horizontal angle at the crystal exit;
- angYout - particle vertical angle at the crystal exit [23].

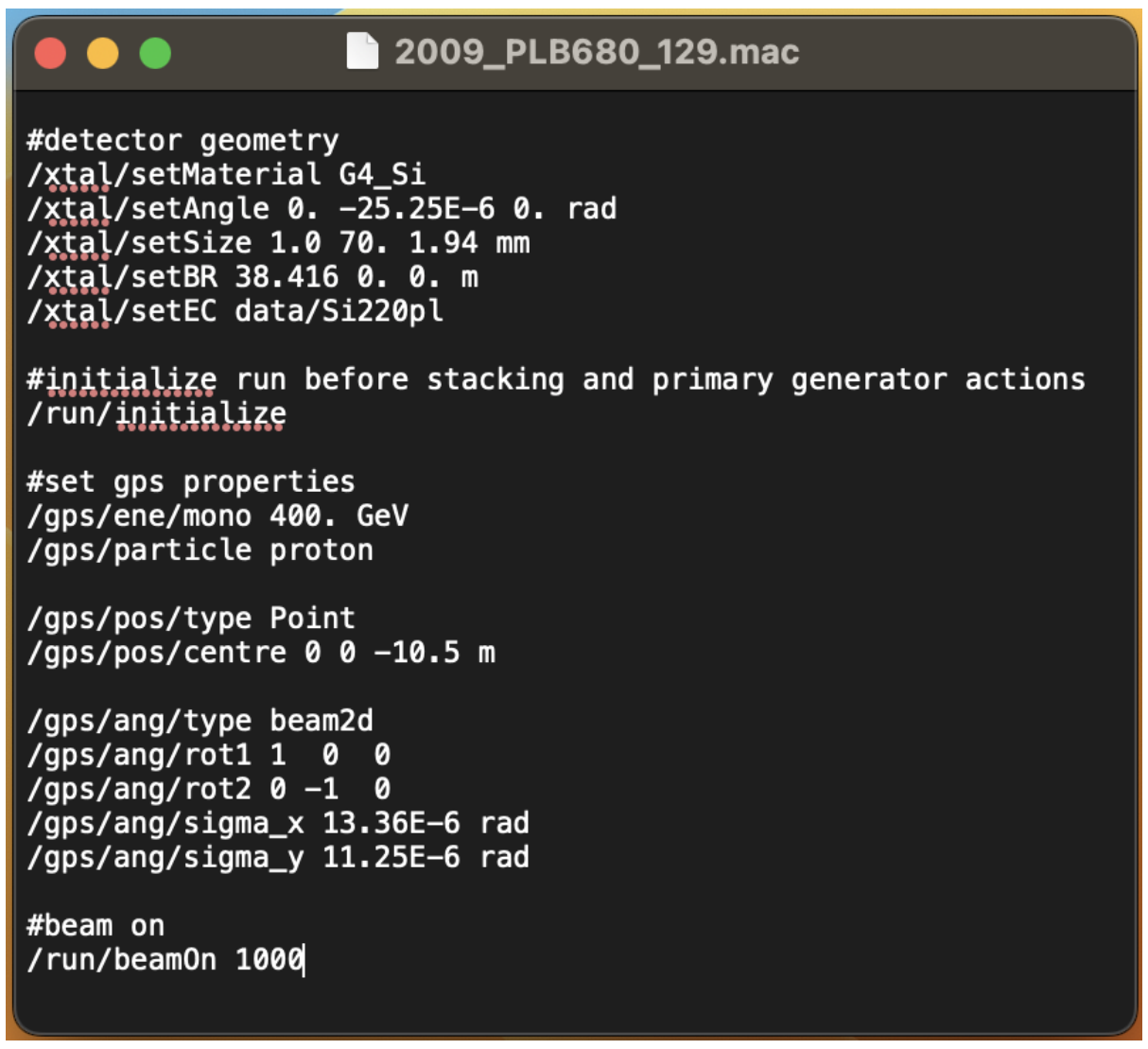

2009_PLB680_129.mac

```
#detector geometry
/xtal/setMaterial G4_Si
/xtal/setAngle 0. -25.25E-6 0. rad
/xtal/setSize 1.0 70. 1.94 mm
/xtal/setBR 38.416 0. 0. m
/xtal/setEC data/Si220pl

#initialize run before stacking and primary generator actions
/run/initialize

#set gps properties
/gps/ene/mono 400. GeV
/gps/particle proton

/gps/pos/type Point
/gps/pos/centre 0 0 -10.5 m

/gps/ang/type beam2d
/gps/ang/rot1 1  0  0
/gps/ang/rot2 0 -1  0
/gps/ang/sigma_x 13.36E-6 rad
/gps/ang/sigma_y 11.25E-6 rad

#beam on
/run/beamOn 1000
```

Figure 4.1: Macro "2009_PLB680_129.mac" from the channeling example in Geant4.

## 4.3 Added features

In order to obtain results from the profiled crystal, the simulation code was modified and additional features were added. The changed code is attached at the end of the thesis, as an appendix. The macro file has been expanded so that more parameters can be controlled and it is also possible to get visualization.

Added features are:

- an ability to geometrically profile the exit of the bent crystal with a cylinder;
- five more detectors behind the crystal which can detect angular distribution and positions, to be able to check the trajectory of the beam more precisely;
- water phantom which serves as a detector to check the energy deposition of the focused beam inside a tissue-like material;
- macro has been expanded to include more options for controlling the beam and the geometry of the simulation.

## 4.4 Description of new macro files

Four new macro files that control the modified simulation are called: parameters.mac, run.mac, vis.mac, and init_vis.mac.

The variable values that can be modified from the macro "parameters.mac" are as follows:

- **/xtal/setMaterial G4_Si** - selects the material that the crystal is made from, in this case, silicon;
- **/xtal/setAngle 0. -0.0015 0. rad** - changes the rotation of the crystal, so the incoming angle of the beam approaching the crystal, in this case, it is rotated by -0.0015 rad;
- **/xtal/setSize 2.2 10.0 12. mm** - changes the size of the crystal rectangular solid in the x, y, and z directions. In this instance, 2.2 mm is the width, 10.0 mm is the height, and 12.0 mm is the length of the crystal;
- **/xtal/setCylinder 2061.55 2000.0 497.99 mm** - switches the radius of the cylinder that cuts the exit of the crystal rectangular solid, and also changes the translation distances of the center of the cylinder with respect to the x and z axes. In this case, 2061.55 mm is the radius, 2000.0 mm is the translation vector on the x-axis, and 497.99 mm is the translation vector on the z-axis;
- **/xtal/setPosition 0. 0. 0. m** - the first parameter sets the position of the detectors on the x-axis, and the second parameter changes the position of the profiled crystal on the x-axis. So in such a case, all of the detectors and the crystal are situated in the 0.0 m position on the x-axis. The shift values for the detectors and the crystal refer to the center of these solids.
- **/xtal/setPhantom 0.3 0.3 0.3 m** - sets the width, height, and length of the water phantom. In this instance, all of those values are equal to 0.3 m;
- **/xtal/setPhantomPos 0.5 0. 1.0 m** - changes the position of the water phantom in all three directions. In this case, it is moved to the side due to the value of 0.5 m on the x-axis, and it is placed at the distance of 0.85 m behind the crystal, due to the value of 1.0 m on the z-axis. Water phantom shift values refer to the center of the solid;
- **/xtal/setBR 4000.0 0. 0. mm** - changes the bending radius of the crystal planes, in this situation, the value of the crystal bending radius is 4000.0 mm. The other two values of 0., are not applied in the simulation;
- **/xtal/setEC data/Si220pl** - selects the (110) Si crystal plane of channeling;
- **/xtal/setDetector 38.0 38.0 0.64 mm** - changes the width, height, and length of detectors. The detectors are made of vacuum as the rest of the world for the simulation;
- **/xtal/setDetectorPos1 0.01 0.5 1.0 m** - switches the distance at which the detectors are placed behind the crystal, part one refers to the first three detectors;
- **/xtal/setDetectorPos2 1.5 2.0 2.5 m** - changes the distance at which the detectors are placed behind the crystal, part two refers to the fourth, fifth, and sixth detector;
- **/gps/ene/mono 200. MeV** - selects the energy of the particles;
- **/gps/particle e-** - selects the type of the particles, in this case, it is an electron;
- **/gps/pos/type Beam** - selects the type of the particle bunch;
- **/gps/pos/centre 0 0 -10.5 m** - selects the starting point for the beam;

- **/gps/pos/rot1 1 0 0** and **/gps/pos/rot2 0 1 0** - defines a rotation matrix for the source of the particles;
- **/gps/pos/halfx 1. mm** - selects a half-width of the beam in the x-axis, in this case, the beam width is equal to 2 mm in the x-direction;
- **/gps/pos/halfy 0. mm** - defines a half-width of the beam in y-axis;
- **/gps/pos/sigma_x 0. mm** - changes the beam Gaussian distribution in x-axis;
- **/gps/pos/sigma_y 0. mm** - changes the beam Gaussian distribution in y-axis;
- **/gps/ang/type beam2d** - selects the type of angular beam distribution;
- **/gps/ang/rot1 1 0 0** and **/gps/ang/rot2 0 -1 0** - defines an angular rotation matrix for the source of the particles;
- **/gps/ang/sigma_x 0.0 rad** - changes the angular beam Gaussian distribution in x-axis;
- **/gps/ang/sigma_y 0.0 rad** - changes the angular beam Gaussian distribution in y-axis.

```
parameters.mac

#detector geometry
/xtal/setMaterial G4_Si
/xtal/setAngle 0. -0.0015 0. rad
/xtal/setSize 2.2 10.0 12. mm
/xtal/setCylinder 2061.55 2000.0 497.99 mm
/xtal/setPosition 0. 0. 0. m
/xtal/setPhantom 0.3 0.3 0.3 m
/xtal/setPhantomPos 0.5 0. 1.0 m
/xtal/setBR 4000.0 0. 0. mm
/xtal/setEC data/Si220pl
/xtal/setDetector 38.0 38.0 0.64 mm
/xtal/setDetectorPos1 0.01 0.5 1.0 m
/xtal/setDetectorPos2 1.5 2.0 2.5 m

#initialize run before stacking and primary generator actions
/run/initialize

#set gps properties
/gps/ene/mono 200. MeV
/gps/particle e-

/gps/pos/type Beam
/gps/pos/centre 0 0 -10.5 m
/gps/pos/rot1 1 0 0
/gps/pos/rot2 0 1 0
/gps/pos/halfx 1. mm
/gps/pos/halfy 0. mm
/gps/pos/sigma_x 0. mm
/gps/pos/sigma_y 0. mm

/gps/ang/type beam2d
/gps/ang/rot1 1  0  0
/gps/ang/rot2 0 -1  0
/gps/ang/sigma_x 0.0 rad
/gps/ang/sigma_y 0.0 rad
```

Figure 4.2: Macro "parameters.mac" from the modified channeling example in Geant4.

More possibilities for modifying the beam parameters can be found in the section on built-in commands in Geant4 available in the "Book For Application Developers" [28].

The "run.mac" macro takes the simulation guidelines from the "parameters.mac" macro and runs the simulation for 1000 particles. As a result, it creates an ExExhCh.root file with the saved output.

The macros "vis.mac" and "init_vis.mac" are necessary to invoke the visualization for the simulation. The "vis.mac" macro is a basic macro for creating visualization in Geant4. The "init_vis.mac" macro initializes the visualization and takes the setup for display from the "parameters.mac" file.

run.mac

```
#simulation set-up
/control/execute parameters.mac

#beam on
/run/beamOn 1000
```

Figure 4.3: Macro "run.mac" from the modified channeling example in Geant4.

vis.mac

```
## Visualisation
/vis/open OGL 1000 1000 # opens a visualisation driver
/vis/verbose errors
/vis/drawVolume # adds detector geometry
/vis/viewer/flush # executing the visualisation

/vis/geometry/set/visibility all 10

# set camera parameters
/vis/viewer/set/viewpointThetaPhi 5 60
# /vis/viewer/set/viewpointVector -1 0 0
/vis/viewer/set/lightsVector -1 0 0
/vis/viewer/zoom 8
/vis/viewer/panTo 0. 0.
# Specify style (surface, wireframe, auxiliary edges,...)
/vis/viewer/set/style wireframe
/vis/viewer/set/auxiliaryEdge true
/vis/viewer/set/lineSegmentsPerCircle 100
# declaring data added to the visualisation
/vis/scene/add/trajectories
/vis/scene/add/hits
/tracking/storeTrajectory 1
/vis/scene/add/trajectories smooth
# To superimpose all of the events from a given run:
/vis/scene/endOfEventAction accumulate

/vis/viewer/update
```

Figure 4.4: Macro "vis.mac" from the modified channeling example in Geant4.

init_vis.mac

```
# Set some default verbose
/control/verbose 2
/control/saveHistory
/run/verbose 2

#simulation set-up
/control/execute parameters.mac

/control/execute vis.mac
```

Figure 4.5: Macro "init_vis.mac" from the modified channeling example in Geant4.

It is also possible to run simulations for a larger number of events from visualization mode by executing the command "/control/execute run.mac", where the number of events will be specified. A disadvantage of performing simulations for a large number of particles from the visualization level is a significant increase in computational time.

## 4.5 Description of the simulation output

The output from the modified channeling example is the ExExhCh.root file with two TTrees (ROOT data structure). One of the TTrees is the ExExChTree. It consists of *leaves* that contain the following data:

- angXin - particle horizontal angle at the crystal entry, the same as in the basic channeling example;
- angYin - particle vertical angle at the crystal entry, the same as in the basic channeling example;
- posXin - particle horizontal position at the crystal entry, as in the basic channeling example;
- posYin - particle vertical position at the crystal entry, as in the basic channeling example;
- angXout1st - particle horizontal angle at the crystal exit, in this case, at the distance 0.01 m behind the crystal;
- angYout1st - particle vertical angle at the crystal exit, in this case, at the distance 0.01 m behind the crystal;
- posXout1st - particle horizontal position at the crystal exit, in this case, at the distance 0.01 m behind the crystal;
- posYout1st - particle vertical position at the crystal exit, in this case, at the distance 0.01 m behind the crystal;
- angXout2nd - particle horizontal angle at the crystal exit, in this case, at the distance 0.5 m behind the crystal;
- angYout2nd - particle vertical angle at the crystal exit, in this case, at the distance 0.5 m behind the crystal;
- posXout2nd - particle horizontal position at the crystal exit, in this case, at the distance 0.5 m behind the crystal;
- posYout2nd - particle vertical position at the crystal exit, in this case, at the distance 0.5 m behind the crystal;
- angXout3rd - particle horizontal angle at the crystal exit, in this case, at the distance 1.0 m behind the crystal;
- angYout3rd - particle vertical angle at the crystal exit, in this case, at the distance 1.0 m behind the crystal;
- posXout3rd - particle horizontal position at the crystal exit, in this case, at the distance 1.0 m behind the crystal;
- posYout3rd - particle vertical position at the crystal exit, in this case, at the distance 1.0 m behind the crystal;
- angXout4th - particle horizontal angle at the crystal exit, in this case, at the distance 1.5 m behind the crystal;
- angYout4th - particle vertical angle at the crystal exit, in this case, at the distance 1.5 m behind the crystal;
- posXout4th - particle horizontal position at the crystal exit, in this case, at the distance 1.5 m behind the crystal;
- posYout4th - particle vertical position at the crystal exit, in this case, at the distance 1.5 m behind

the crystal;
- angXout5th - particle horizontal angle at the crystal exit, in this case, at the distance 2.0 m behind the crystal;
- angYout5th - particle vertical angle at the crystal exit, in this case, at the distance 2.0 m behind the crystal;
- posXout5th - particle horizontal position at the crystal exit, in this case, at the distance 2.0 m behind the crystal;
- posYout5th - particle vertical position at the crystal exit, in this case, at the distance 2.0 m behind the crystal;
- angXout6th - particle horizontal angle at the crystal exit, in this case, at the distance 2.5 m behind the crystal;
- angYout6th - particle vertical angle at the crystal exit, in this case, at the distance 2.5 m behind the crystal;
- posXout6th - particle horizontal position at the crystal exit, in this case, at the distance 2.5 m behind the crystal;
- posYout6th - particle vertical position at the crystal exit, in this case, at the distance 2.5 m behind the crystal.

The other one of the TTrees is the PhantomDeposition which consists of leaves:
- Edep - energy deposition inside the water phantom;
- posX - position of particles inside the water phantom in the x-axis;
- posY - position of particles inside the water phantom in the y-axis;
- posZ - position of particles inside the water phantom in the z-axis.

Due to the fact that the detectors and the water phantom, which also serves as a detector, can overlap, only one of these two recording options can be used during a single simulation.

In order to be able to record trajectory data, the position of the detectors must be set to zero on the x-axis, and the position of the phantom must be moved to the side of the simulation setup, for example, 0.5 m on the x-axis. This is the layout that is proposed as default.

To collect data from the water phantom, the phantom should be set to 0.0 m on the x-axis, and the detectors should be moved to the side of the simulation arrangement, for example, 0.5 m on the x-axis.

## 4.6 Profiling of the crystal exit

Crystal profiling is performed through the process of cutting a part of the crystal body by a cylinder. The radius of the cylinder is related with the length of the crystal, as in Equation 3.11. Due to the chosen deflection angle of 1000 $\mu$rad, the ratio between the two values is of the order of magnitude $10^3$. For this reason, the section of the cylinder cutting the exit from the crystal is almost a straight line. The important part of crystal profiling is the translation of the center of the cylinder in relation to the outlet of the crystal, a schematic of such a translation is shown in Figure 4.6, where $d_x$ and $d_z$ are the shift values in the horizontal and longitudinal directions, respectively. The formulas for calculating these two parameters are as follows:

$$d_z = r \cdot sin\alpha \tag{4.1}$$

$$d_x = r \cdot sin\beta \tag{4.2}$$

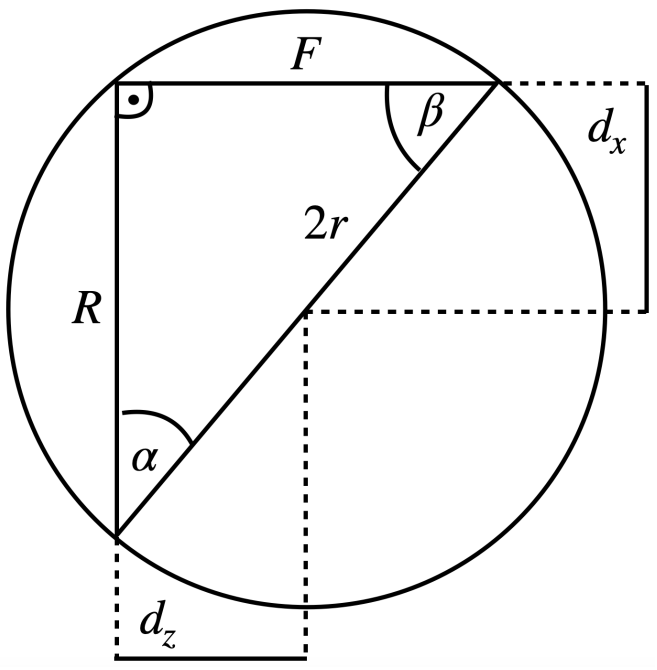

Figure 4.6: A scheme for determining the distance of translation of the cylinder with relation to the center of exit from the crystal, where R - the radius of the crystal, F - the focal length, 2r - the diameter of the cylinder, $d_x$ - the shift of the center of the cylinder in the x-plane, and $d_z$ - the shift of the center of the cylinder in the z-plane. That is the view from the top in Figure 3.15.

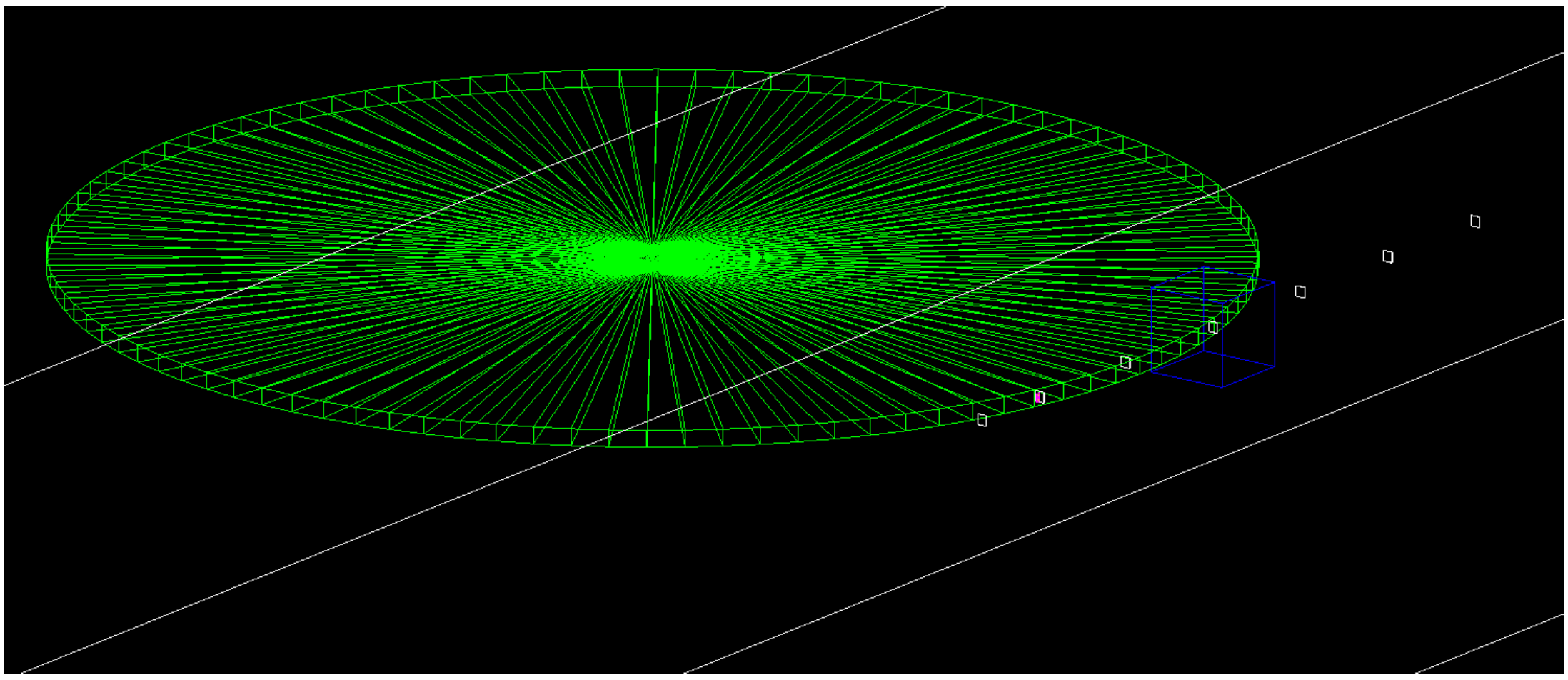

Figure 4.7: A visualization of the process of profiling the crystal exit, for the value of bending radius equal to 4000 mm, where the profiled crystal is pink, non-profiled crystal is yellow, profiling cylinder is green, detectors are white, and water phantom is blue.

An example of crystal exit profiling is visualized in Figures 4.7 and 4.8, where the profiled crystal is pink, non-profiled crystal is yellow, profiling cylinder is green, detectors are white, and water phantom is blue. Such a display is prepared only to visualize the solids that form the profiled crystal and how these solids overlap.

When preparing a set of parameters to simulate a crystal, the yellow solid that represents the bent crystal without profiling is 1.5 times the length of the profiled crystal in the middle of its width. So in this example, the crystal has a bending radius equal to 4000 mm, which means that the crystal's length is 4 mm (in the center of the profiled output), so the length of the yellow block is 12 mm. The value of the cylinder's radius is 2061.55, and the shift of its center, relative to the crystal's exit center, is +2000.0 mm in the x-axis, and +497.99 mm in the z-axis. With a combination of such parameters, it is possible to obtain a simulation of the profiled crystal in Geant4. The result is presented in Figure 4.9.

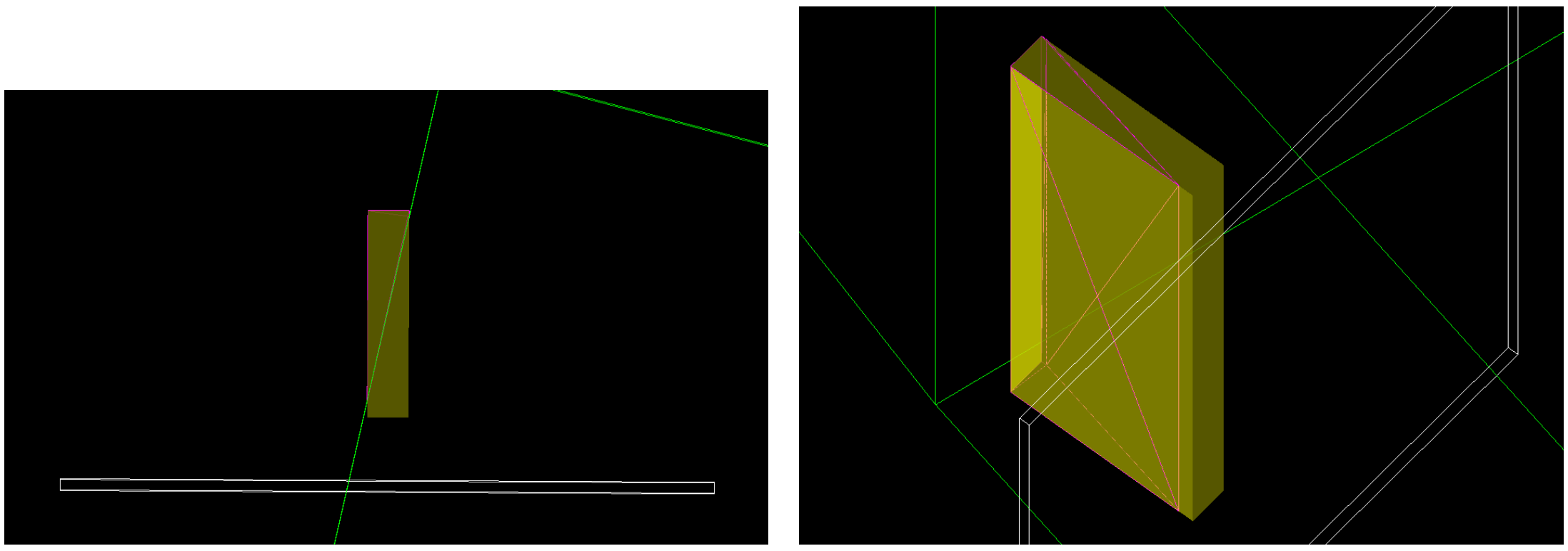

(a) view from above

(b) view from the side

Figure 4.8: A close-up of the process of profiling the crystal exit, for the value of bending radius equal to 4000 mm, where the profiled crystal is pink, the non-profiled crystal is yellow, and the profiling cylinder is green.

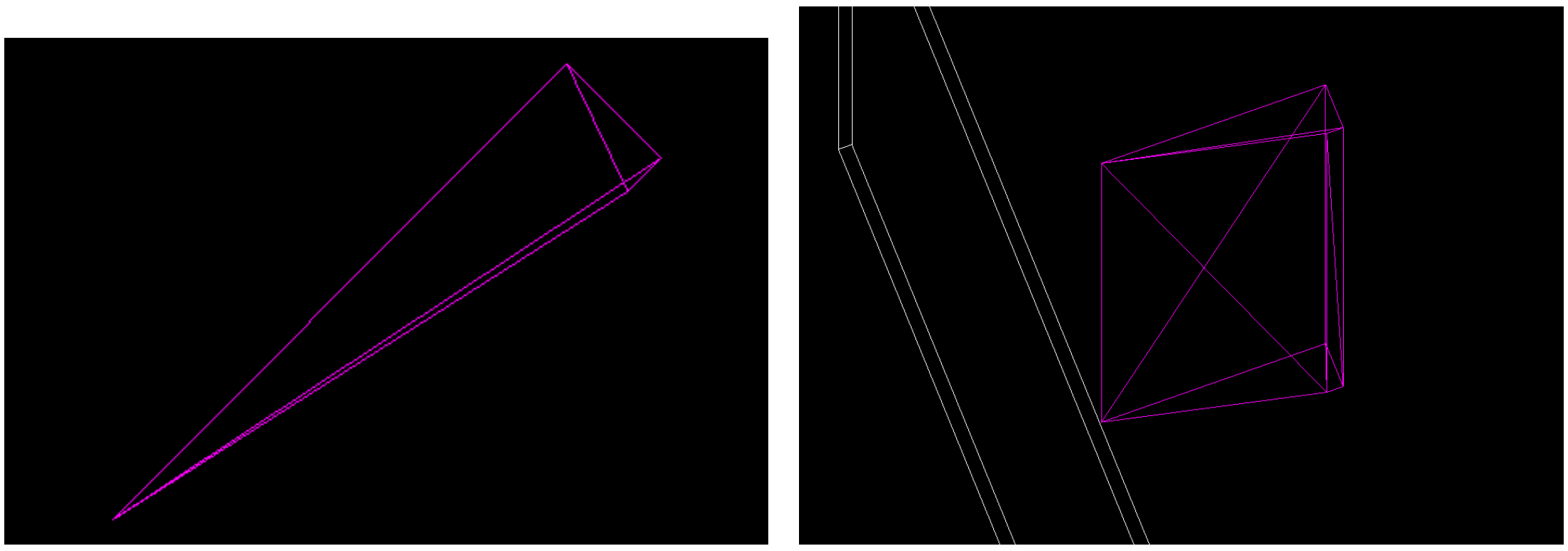

(a) view from above

(b) view from the side

Figure 4.9: A visualization of the profiled crystal, for the value of bending radius equal to 4000 mm.

In the case of a regular simulation, only the profiled crystal is visible, without the solids forming it, as shown in Figure 4.9. However, if one wants to visualize both solids forming the profiled crystal, as shown in Figures 4.7 and 4.8, then in the file src/DetectorConstruction.cc (A) it is needed to activate the lines of code related to:

- creating G4LogicalVolume for the non-profiled crystal and cylinder, i.e. lines 213-218,
- positioning these solids in the simulation space (G4PVPlacement), i.e. lines 284-298,
- and giving them attributes for visualization (G4VisAttributes), i.e. lines 356-362.

## 4.7 Data analysis and management

Analysis of the data obtained from Geant4 was performed in the Python [29] programming environment. Python library *ROOT* allows working seamlessly with root files, where the simulation output is stored.

Zenodo [30] is a CERN-operated platform, that preserves and shares research outputs like datasets, software, and preprints. It supports open access and is part of the European OpenAIRE initiative, aiding researchers in storing, sharing, and citing data reliably. The simulation code prepared for the purpose of this study was added to Zenodo and can be found at Digital Object Identifier (DOI) [31]: 10.5281/zenodo.8226536.

# 5. Results

This chapter presents and describes the results obtained from the prepared simulations. It shows the impact that properly bent and profiled silicon crystal can have on negatively charged particles when the channeling process occurs.

## 5.1 Distribution of the deflection angle

At first, it was checked whether the channeling phenomenon simulated in Geant4 gave the expected results from the simulation. In this case, a silicon non-profiled bent crystal was used. The parameters of the crystal are listed in Table 5.2. The material (Silicon) and crystal lattice (110) are the only ones available in Geant4. The value of the deflection angle (1000 $\mu$rad) was an arbitrary choice. The deflection angle was not extremely large, but at the same time, it was large enough that the electrons that were not channeled could be stopped by potentially adding a shield in front of a patient. The bending radius must be larger than the critical bending radius (Equation 3.8). The length and the bending radius of a crystal are directly related (Equation 3.9). In the next part, those two parameters were optimized to achieve maximum efficiency.

In every simulation prepared for this study, the energy of the electron beam is equal to 200 MeV, as listed in Table 5.1. This is typical energy of VHEE beams, already used in some previous studies (e.g. in [9]), and within reach of modern, compact electron linacs developed for potential use in readiotherpay (e.g. [8]).

Table 5.1: Parameters defining the beam.

| | |
|---|---|
| **Particle Type** | Electron |
| **Energy** | 200 MeV |

Table 5.2: Parameters defining the crystal.

| | |
|---|---|
| **Material** | Silicon Si |
| **Crystal Lattice** | 110 |
| **Critical Channeling Angle** | 447 $\mu$rad |
| **Critical Bending Radius** | 0.35 mm |
| **Deflection Angle** | 1000 $\mu$rad |
| **Bending Radius** | 4000 mm |
| **Length** | 4 mm |

The original channeling example in Geant4 was prepared for a proton beam with an energy of 400 GeV, which is deflected by a small angle of 50 $\mu$rad. It was not *a priori* known whether the available Geant4 model can perform a reasonable reconstruction of the channeling process using particles with the opposite charge to the proton (the electron), an energy that is smaller by three orders of magnitude, and a much larger beam deflection angle of 1000 $\mu$rad. The following part shows the verification of results for such a modification.

Figure 5.1 shows a two-dimensional histogram of the distribution of the deflection angle of particles that passed through a 4 mm long crystal, depending on the incoming angle. The presented result is from the detector positioned at a distance of 1 m behind the crystal. The distribution of the incoming beam was Gaussian and the spread of incoming angles ranged from -1500 to 1500 $\mu$rad. The beam was flat and only had a spread on the horizontal plane. The channeled particles are in part a) on the plot, this is the outcome that is the most crucial to this study. Particles deflected by an angle of 1000 $\mu$rad are the most numerous and there is also an area surrounding the pick in the range from 500 to 1500 $\mu$rad. Section b) contains dechanneled particles, so those that at some point stopped to be channeled. On the other hand, section c) includes the volume captured particles, so those that at some point started to be channeled, even though when they approached the crystal at first they did not meet the conditions suitable for the occurrence of channeling. Segment d) refers to the phenomenon of volume reflection, those are the particles that were reflected from the potential well of the crystal lattice. Sections e) are the result of particle scattering, which treated the crystal as amorphous matter.

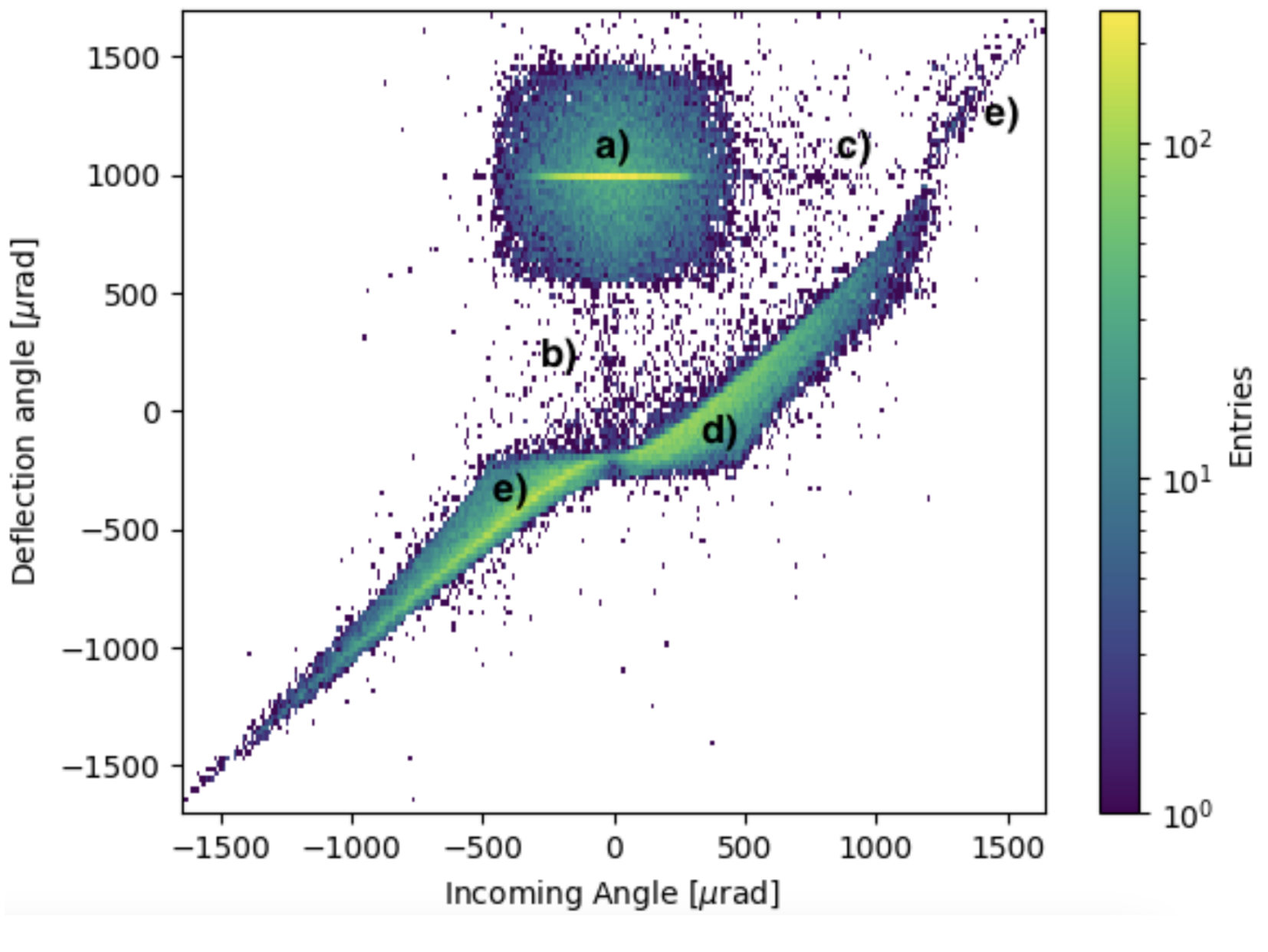

Figure 5.1: Distribution of the deflection angle of particles with energy 200 MeV that passed through a 4 mm long crystal, depending on the incoming angle; a) channeled particles, b) dechanneling, c) volume capture, d) volume reflection, e) amorphous scattering.

The result of the channeling phenomenon from the simulation is as expected, which means that it is the right software to carry out the further steps of this study. Similar plots can be found in papers describing the channeling of high-energy beams, e.g. in [32, 33, 34].

## 5.2 Optimization of the crystal length

Optimization of crystal parameters, such as length and bending radius, was performed for a number of different crystal lengths in the range of [0.004, 40] mm with channeling efficiency used as a figure of merit. The efficiency stands for the ratio of channeled particles to all electrons entering the crystal. In this case, a single-point source beam without any radial spread was used. The amount of channeled particles was checked in three ranges of angle: [950, 1050] $\mu$rad, [750, 1250] $\mu$rad, and [500, 1500] $\mu$rad. The angular spread in the range of [500, 1500] $\mu$rad translates into 1 mm of position spread at a distance of 1 m behind the crystal. With tumor sizes typically being in the order of centimeters (cm) [35], 1 mm of position spread is less than a typical tumor size so there is no need to limit the range for channeled particles and the entire range of [500, 1500] $\mu$rad of channeled electrons can be considered useful.

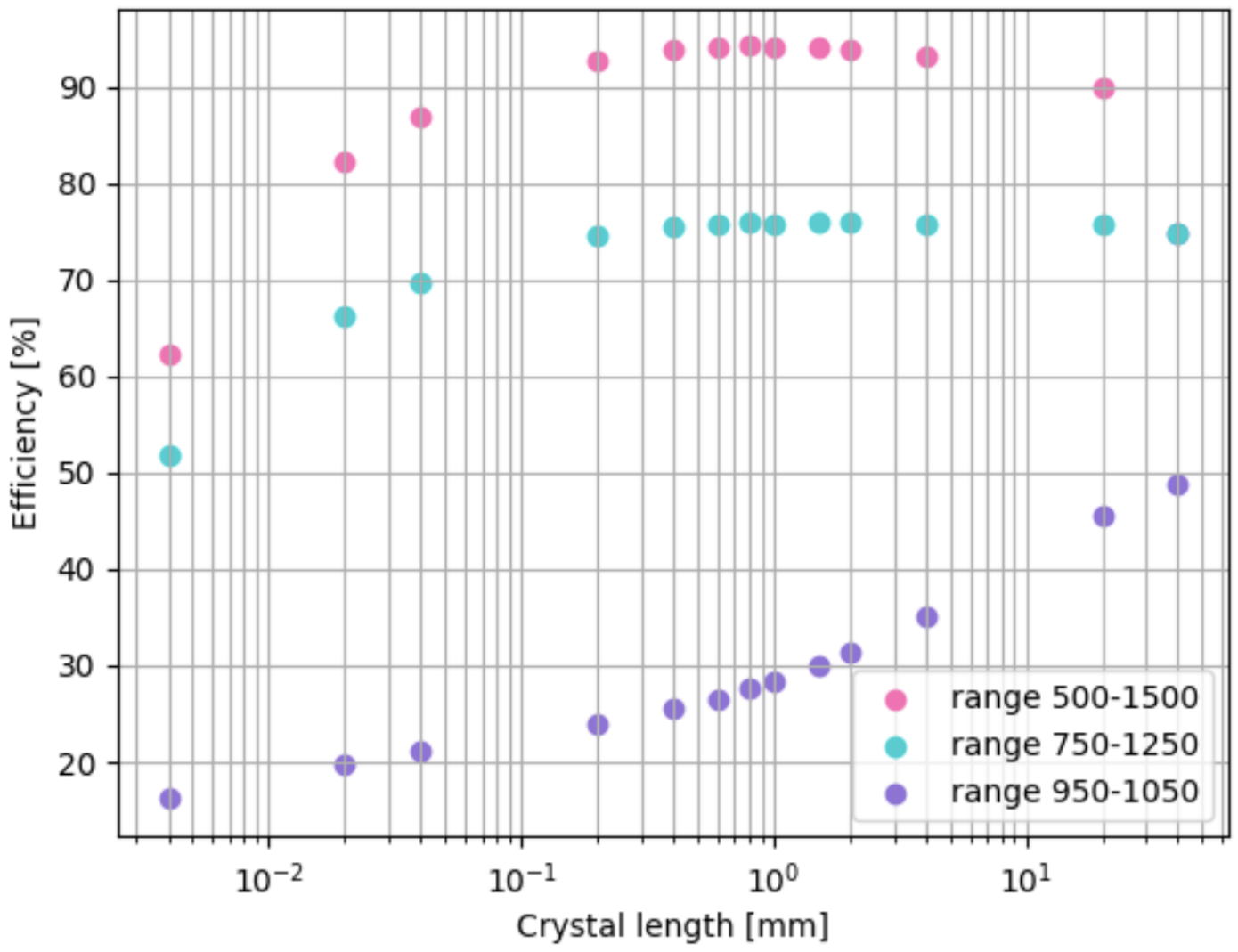

Figure 5.2: Efficiency of electron beam channeling as a function of crystal length. The ranges are the same as in Figure 5.3. Statistical errors, calculated as the standard error of the binomial distribution, are barely visible due to high statistics.

Due to the fact that the wider range of channeled particles is taken under consideration, more electrons can be counted into the efficiency parameter. For the range of [500, 1500] $\mu$rad the highest efficiency score was achieved for silicon crystals with the length of a couple of millimeters. The efficiency was at around 94$\%$. When the selected range was smaller, of [750, 1250] $\mu$rad, the distribution of the efficiency parameter was similar, but the highest score value was around 76$\%$. When the narrowest range of [950, 1050] $\mu$rad was considered, the efficiency increased as the crystal lengthened. In addition, the longer the crystal, the greater the possibility of losing particles due to the longer path they have to overcome inside the crystal, where they may encounter various interactions with the material.

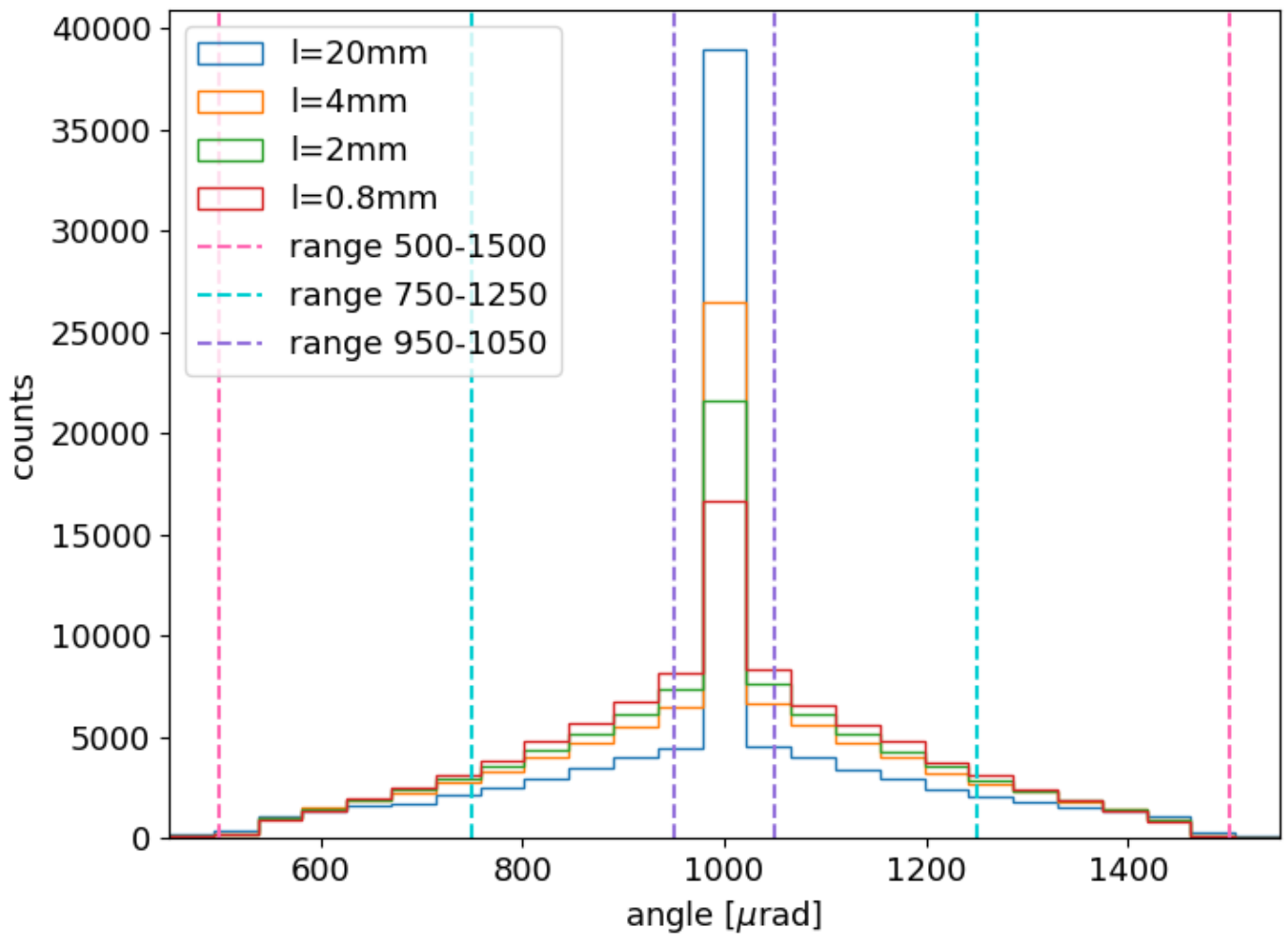

Figure 5.3: Angular distribution of channeled electrons for four crystal lengths of 0.8 mm, 2 mm, 4 mm, and 20 mm.

In summary, the most effective crystal length is considered to be a few millimeters long. For such crystals, the largest number of electrons can be channeled and the crystals are not too long, so there is little chance of losing particles that would have to travel too long inside the crystal and be exposed to additional interactions with it.

## 5.3 Beam focusing

In order to focus the beam the exit must be profiled as explained in chapter 3, in the section 3.3. The profiling is strictly geometric. This means that the output from the crystal is cut with a cylinder that has appropriately adjusted parameters depending on the bending radius of the crystal and the focal length. The focal length was chosen to be 1 m, so the only parameter that changes, depending on the crystal length, is the radius of the cutting cylinder, which is defined in Equation 3.11. Examples of the cylinder radius lengths are listed in Table 5.3.

Table 5.3: Radius lengths of the cylinder profiling the output of the crystal, depending on the bent crystal length.

| **Crystal length** | **Cylinder radius lengths** |
|---|---|
| 2 mm | 1118.03 mm |
| 4 mm | 2061.55 mm |
| 20 mm | 10012.49 mm |

As an example, a bent crystal with a length of 4 mm was chosen. The focusing ability was checked for the horizontal direction, for two types of pencil beams: one with a 2-mm-wide uniform distribution and one with a Gaussian distribution of a standard deviation of 2 mm. The bending radius of the crystal was 4000 mm, and the radius of the cutting cylinder was 2061.55 mm. The result of focusing the beam by profiling the exit of the crystal is presented in Figure 5.4. It shows the distribution of the beam position before entering the profiled crystal and after passing through the crystal at distances 0.5 m, 1 m, 1.5 m, 2 m, and 2.5 m respectively. The focusing point is at 1 m, it has a narrow peak, and at the base, the beam is spread in the width of about 1 mm.

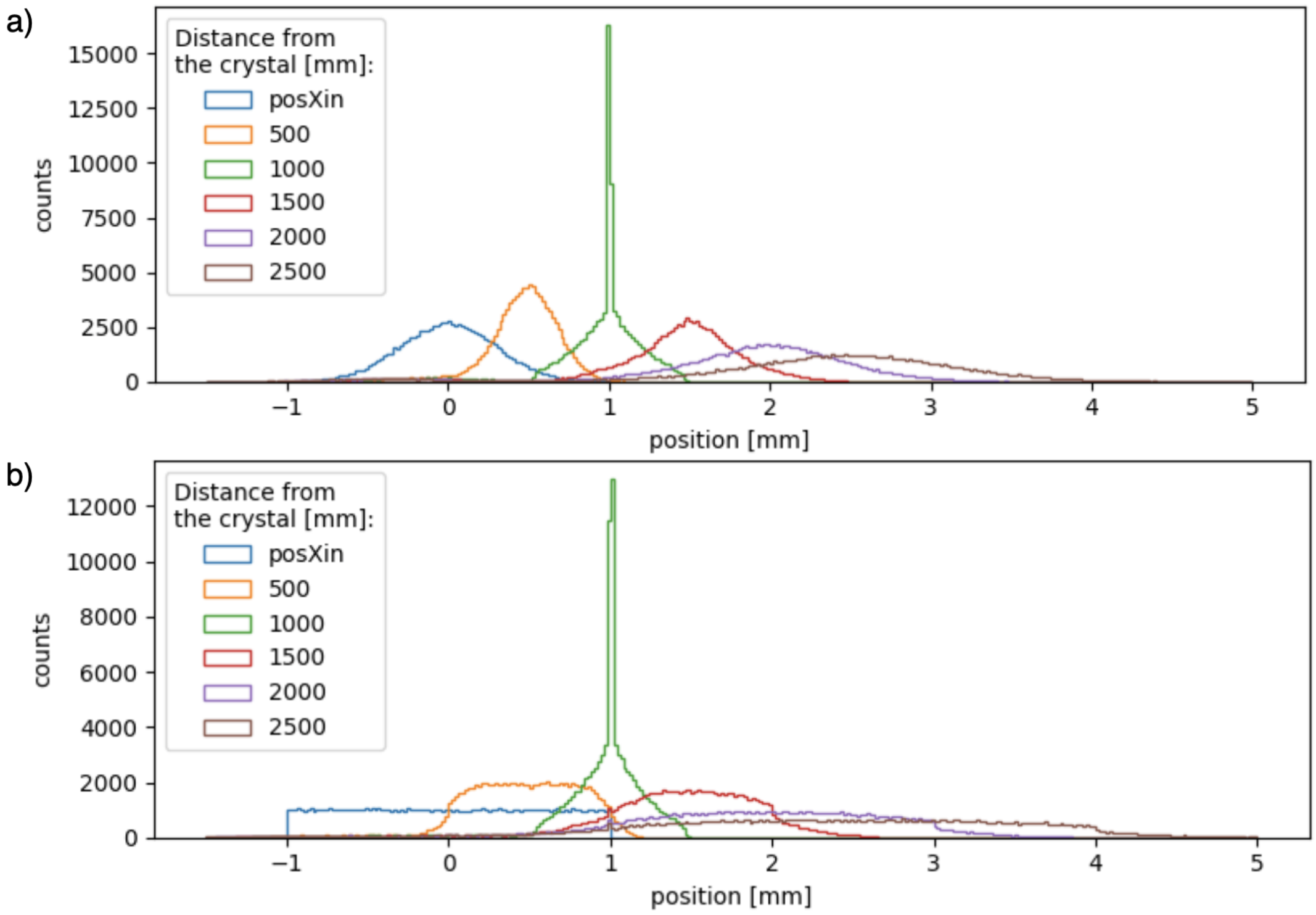

Figure 5.4: Distribution of the beam position before entering the profiled crystal with bending radius 4000mm, and after passing through the crystal at distances 0.5m, 1m, 1.5m, 2m, and 2.5m respectively, a) for a beam of Gaussian distribution, and b) for a beam of uniform distribution. Bin width is 20 $\mu$m.

To numerically represent the focusing of the beam, a range from which the particles were considered, was determined by the mean value +/- 3 times the standard deviation, and then the standard deviation of this range was calculated. The error for this parameter was calculated as the standard error of the mean (SEM). The obtained values are listed in Table 5.4. The values of the width of the beam at 1 m behind the crystal are similar to each other. At the same time, if we consider the ratio of the focused beam to the width of the beam approaching the crystal, a greater difference occurs in the case of a uniform beam, because the number of particles on the edges of the beam was much larger, but still the quality of the focused beam at a distance of 1 m is comparable to that of a Gaussian beam. The smaller the out/in ratio parameter, the greater the difference between the width of the incoming beam and the width of the focused beam. In the case where the beam was uniform, two approaches were taken, in the first (std) the same parameters were calculated as in the case for the beam with Gaussian distribution, and in the second (full) the entire width of the incoming beam was taken into account.

Table 5.4: Parameters of focusing the beam for a profiled crystal with bending radius 4000 mm.

| Beam distribution | Width type | Width in [mm] | Width at 1m [mm] | Ratio out/in |
|---|---|---|---|---|
| Gaussian | std, std | 0.2973 ± 0.0009 | 0.2313 ± 0.0007 | 0.778 |
| Uniform | std, std | 0.5776 ± 0.0018 | 0.2250 ± 0.0007 | 0.389 |
| Uniform | full, std | 2.0 | 0.2250 ± 0.0007 | 0.113 |

The graphs and parameters presented in this section prove that focusing the beam using a crystal with a properly profiled output is possible and such a crystal can be used as a focusing lens.

## 5.4 Energy deposition inside the water phantom

To check what the distribution of energy deposited inside the human body would be, a water phantom was added to the simulation since water is one of the closest materials that can be compared to tissue. The aim is to make the energy deposition curve a few mm wide and with a maximum at depth of around 15 cm into the phantom.

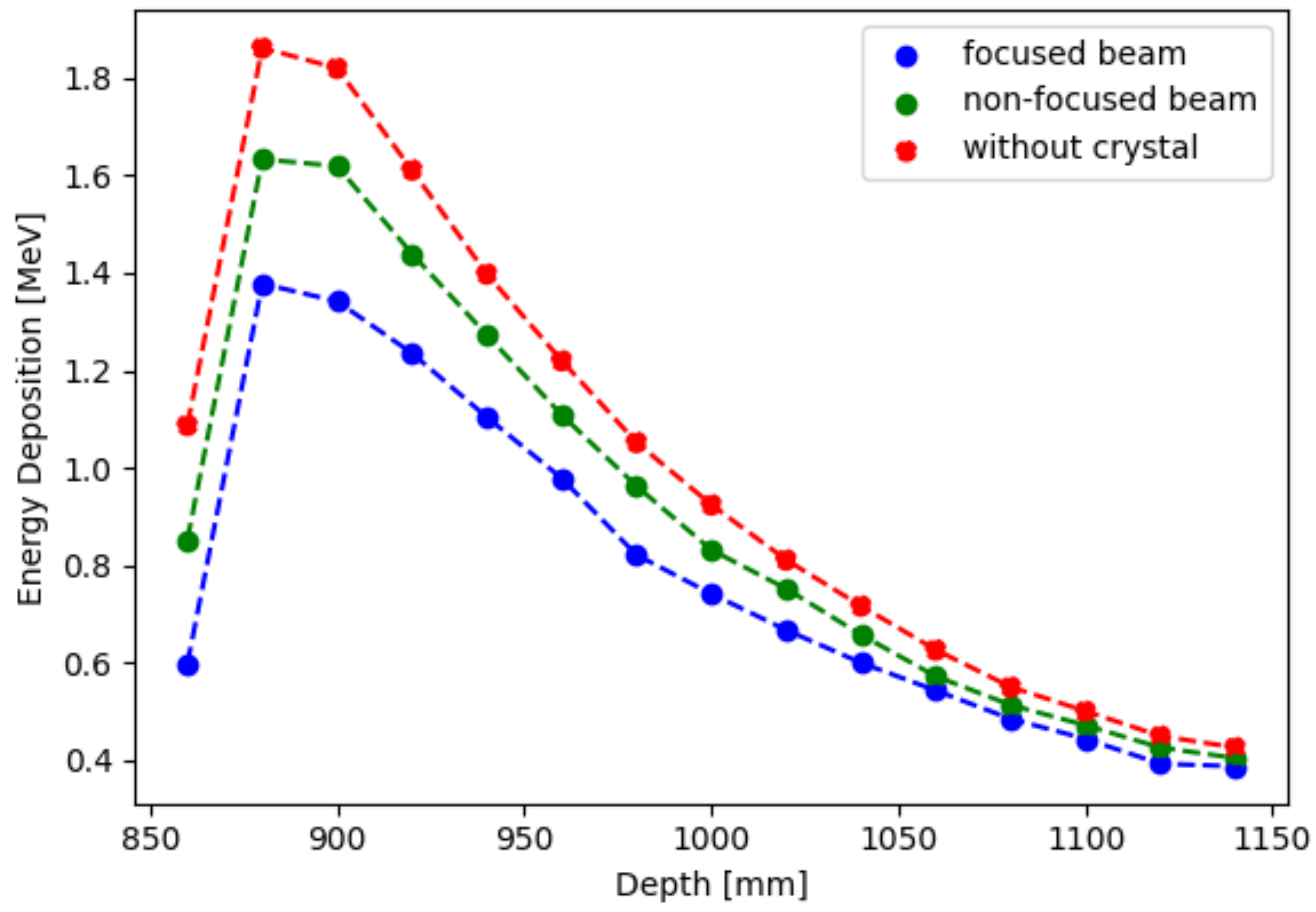

Figure 5.5: An example result of the averaged energy deposition as a function of phantom depth for an electron beam with an energy of 200 MeV, a crystal length of 4mm, and a cylinder radius length of 2061.55 mm. The phantom is placed at a distance of 850 mm behind the crystal. Statistical errors, calculated as the standard error of the mean (SEM), are barely visible due to high statistics.

Figure 5.5 represents the obtained energy deposition. Unfortunately, the energy deposition curve is far from expected. It resembles the shape of the deposition energy for 20 MeV electrons (Figure 2.1) or for VHEE, for which the focusing strength is not strong enough (Figure 2.5). The strong focusing is considered to be for focused beams with the f-number equal to f/3.8 or lower value [1]. The strength of focusing (f-number (F/D)) in this case is not enough and is equivalent to f/500. The reason for the weak focusing is that the width of the beam entering the crystal cannot be greater than about 2 mm, this is due to the way the exit from the crystal is profiled (Figure 4.9). At the same time, the focal length of 1 m is probably too long and should be shortened. In the Ph.D. thesis [1], to which the obtained result is compared, the width of the beam was around 20 cm, and in the case of the strongest focusing (f/1.2), the focus point was located at a distance of about 26 cm from the magnet lens.

In conclusion, works on finding an optimal set of crystal parameters such as bending radius, length of crystal, and focal length are ongoing. Focusing strength can be increased by reducing the focal length and increasing the crystal transverse dimension.

Alternatively, there is a possibility of preparing an array of focusing crystals to achieve something similar to a Fresnel lens. It would allow broadening the size of the beam approaching crystal lenses and result in stronger focusing, so the beam could penetrate deeper into the water phantom with a more narrow energy deposition curve. An example of such a Fresnel lens is shown in Figure 5.6.

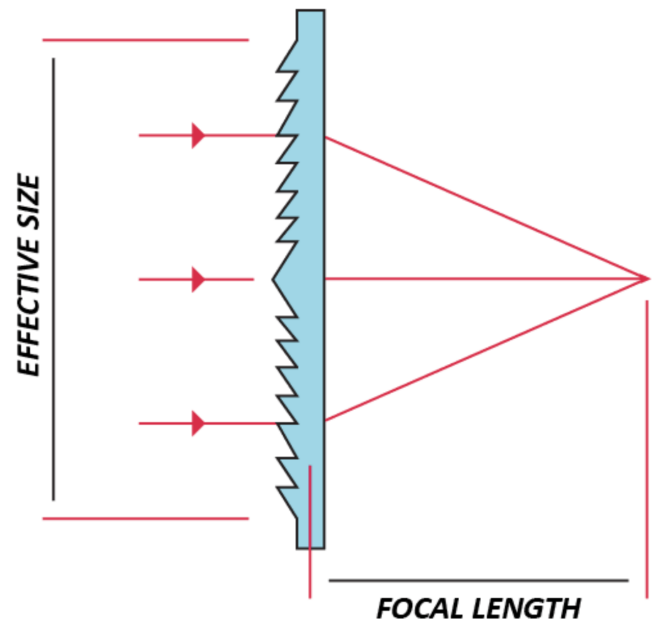

Figure 5.6: An example of a focusing Fresnel lens (graphics from [36]).

# 6. Conclusions

The aim of this thesis was to investigate the feasibility of focusing a very high-energy electron beam using crystal lenses for application in radiotherapy treatment.

For this purpose, a Geant4 simulation model has been developed by extending the available *channeling* example. Features implemented into the code cover: shaping the bent Si crystal for the purpose of providing focusing; five more detectors placed behind the crystal which can detect beam distribution, as a way to be able to check the trajectory of the beam more precisely; water phantom which serves as a detector to check the energy deposition of the focused beam inside a tissue-like material; and new expanded macros to include more options for controlling the beam and the geometry of the simulation. **The simulation model is well prepared for the development and optimization of crystal lenses that could be potentially used to focus high-energy electron beams.**

**Beam focusing by using bent crystals with profiled exits was demonstrated.** Crystals with lengths in the order of a few millimeters proved to be the most optimal set of parameters for deflecting the beam by an angle of 1000 $\mu$rad. The results obtained from two types of beams, with a Gaussian distribution and with a uniform distribution are similar. The focusing is stronger in the case of uniform beam distribution because there are significantly more particles at positions further from the middle of the beam approaching the crystal.

The energy deposition curve inside a water phantom is unfortunately far from expected. The obtained result resembles an outcome that can be acquired by a non-focused or weakly focused VHEE beam. The conclusion is that the focusing strength of the proposed crystal is not strong enough. The main reasons might be the limitation of the beam width approaching the crystal lens, which is around 2 mm at the maximum, and the focal length of 1 m might not be the optimal choice, it should be shorter.

The search for optimal crystal parameters, including the bending radius, crystal length, and focal length, is still ongoing. By shortening the focal length and lengthening the crystal transverse dimension, the focusing strength can be improved. As an alternative, it is possible to create something resembling a Fresnel lens by assembling a collection of focusing crystals.

# List of Figures

# List of Tables

The following appendices contain the source code of the modified parts of the Geant4 *channeling* example used for this thesis.

# List of Appendices

# A. src/DetectorConstruction.cc

```
//
// ********************************************************************
// * License and Disclaimer                                           *
// *                                                                  *
// * The  Geant4 software  is  copyright of the Copyright Holders  of *
// * the Geant4 Collaboration.  It is provided  under  the terms  and *
// * conditions of the Geant4 Software License,  included in the file *
// * LICENSE and available at  http://cern.ch/geant4/license .  These *
// * include a list of copyright holders.                             *
// *                                                                  *
// * Neither the authors of this software system, nor their employing *
// * institutes,nor the agencies providing financial support for this *
// * work  make  any representation or  warranty, express or implied, *
// * regarding  this  software system or assume any liability for its *
// * use.  Please see the license in the file  LICENSE  and URL above *
// * for the full disclaimer and the limitation of liability.         *
// *                                                                  *
// * This  code  implementation is the result of  the  scientific and *
// * technical work of the GEANT4 collaboration.                      *
// * By using,  copying,  modifying or  distributing the software (or *
// * any work based  on the software)  you  agree  to acknowledge its *
// * use  in  resulting  scientific  publications,  and indicate your *
// * acceptance of all terms of the Geant4 Software license.          *
// ********************************************************************
//
//

#include "DetectorConstruction.hh"

#include "G4Material.hh"
#include "G4NistManager.hh"

#include "G4Box.hh"
#include "G4Tubs.hh"
#include "G4SubtractionSolid.hh"
#include "G4IntersectionSolid.hh"

#include "G4LogicalVolume.hh"
#include "G4PVPlacement.hh"
#include "G4SystemOfUnits.hh"
#include "G4PhysicalConstants.hh"

#include "G4GeometryManager.hh"
#include "G4PhysicalVolumeStore.hh"
#include "G4LogicalVolumeStore.hh"
#include "G4SolidStore.hh"

#include "G4VisAttributes.hh"
#include "G4Colour.hh"

#include "G4CrystalExtension.hh"
#include "G4ExtendedMaterial.hh"
#include "G4LogicalCrystalVolume.hh"

#include "G4ChannelingMaterialData.hh"
#include "G4ChannelingOptrMultiParticleChangeCrossSection.hh"

#include "SensitiveDetector.hh"

#include "G4SDManager.hh"

#include "G4PSDoseDeposit.hh"

//....oooOO0OOooo........oooOO0OOooo........oooOO0OOooo........oooOO0OOooo......

DetectorConstruction::DetectorConstruction():
fECfileName("Si220pl"),
fMaterialName("G4_Si"),
fSizes(G4ThreeVector(2.2*CLHEP::mm,              // crystal's size in x-axis
                     15.0*CLHEP::mm,             // crystal's size in y-axis
                     12.0 * CLHEP::mm)),         // crystal's size in z-axis
fCylinder(G4ThreeVector(2061.55*CLHEP::mm,       // cylinder's radius
                        2000.0*CLHEP::mm,        // cylinder's shift in x-axis
                        497.99*CLHEP::mm)),      // cylinder's shift in z-axis
fPosition(G4ThreeVector(0.*CLHEP::m,             // detector's shift in x-axis
                        0.*CLHEP::m,             // crystal's shift in x-axis
                        0.*CLHEP::m)),
fPhantom(G4ThreeVector(0.3*CLHEP::m,             // phantom's size in x-axis
                       0.3*CLHEP::m,             // phantom's size in y-axis
                       0.3*CLHEP::m)),           // phantom's size in z-axis
```

```
fPhantomPos(G4ThreeVector(0.5*CLHEP::m,          // phantom's placement in x-axis
                          0.*CLHEP::m,           // phantom's placement in y-axis
                          1.0*CLHEP::m)),        // phantom's placement in z-axis
fBR(G4ThreeVector(4000.* CLHEP::mm,0.,0.)),      // crystal's bending radius
fAngles(G4ThreeVector(0.,-0.0015,0.)),           // incoming angle
fDetector(G4ThreeVector(38.0*CLHEP::mm,          // detector's size in x-axis
                        38.0*CLHEP::mm,          // detector's size in y-axis
                        0.64*CLHEP::mm)),        // detector's size in z-axis
fDetectorPos1(G4ThreeVector(0.01*CLHEP::m,       // 1st detector behind the crystal placement in z-axis
                            0.5*CLHEP::m,        // 2nd detector behind the crystal placement in z-axis
                            1.0*CLHEP::m)),      // 3rd detector behind the crystal placement in z-axis
fDetectorPos2(G4ThreeVector(1.5*CLHEP::m,        // 4th detector behind the crystal placement in z-axis
                            2.0*CLHEP::m,        // 5th detector behind the crystal placement in z-axis
                            2.5*CLHEP::m)){      // 6th detector behind the crystal placement in z-axis
    fMessenger = new DetectorConstructionMessenger(this);
}

//....oooOO0OOooo........oooOO0OOooo........oooOO0OOooo........oooOO0OOooo.....

DetectorConstruction::~DetectorConstruction(){;}

//....oooOO0OOooo........oooOO0OOooo........oooOO0OOooo........oooOO0OOooo.....

void DetectorConstruction::DefineMaterials(){;}

//....oooOO0OOooo........oooOO0OOooo........oooOO0OOooo........oooOO0OOooo......

G4VPhysicalVolume* DetectorConstruction::Construct(){

    //** World **//
    G4Material* worldMaterial =
      G4NistManager::Instance()->FindOrBuildMaterial("G4_Galactic");

    G4double worldSizeXY =  2. * CLHEP::meter;
    G4double worldSizeZ  = 22. * CLHEP::meter;

    G4Box* worldSolid = new G4Box("world.solid",
                                  worldSizeXY/2.,
                                  worldSizeXY/2.,
                                  worldSizeZ/2.);

    G4LogicalVolume* worldLogic = new G4LogicalVolume(worldSolid,
                                                      worldMaterial,
                                                      "world.logic");

    G4PVPlacement* worldPhysical = new G4PVPlacement(0,
                                                     G4ThreeVector(),
                                                     worldLogic,
                                                     "world.physic",
                                                     0,
                                                     false,
                                                     0);

    //** Detectors instantiation **//
    G4ThreeVector fDetectorSizes(G4ThreeVector(fDetector.x(),
                                               fDetector.y(),
                                               fDetector.z()));

    G4double fDetectorDistance[8] = {
        -9.998 * CLHEP::m,
        -0.320 * CLHEP::m,
        fDetectorPos1.x(),
        fDetectorPos1.y(),
        fDetectorPos1.z(),
        fDetectorPos2.x(),
        fDetectorPos2.y(),
        fDetectorPos2.z(),
    };


    G4Box* ssdSolid = new G4Box("ssd.solid",
                                fDetectorSizes.x()/2.,
                                fDetectorSizes.y()/2.,
                                fDetectorSizes.z()/2.);

    G4Material* detectorMaterial =
      G4NistManager::Instance()->FindOrBuildMaterial("G4_Galactic");
    G4LogicalVolume* ssdLogic =
      new G4LogicalVolume(ssdSolid,
                          detectorMaterial,
                          "ssd.logic");

    for(size_t i1=0;i1<8;i1++){
        new G4PVPlacement(0,
                          G4ThreeVector(fPosition.x(),
                                        0.,
                                        fDetectorDistance[i1]),
                          ssdLogic,
                          "ssd.physic",
```

```
                        worldLogic,
                        false,
                        i1);

}

//** Crystal solid parameters - Start **//
// G4Box* crystalSolid = new G4Box("crystal.solid",
//                              fSizes.x()/2.,
//                              fSizes.y()/2.,
//                              fSizes.z()/2.);

//** Crystal solid parameters - End **//


//** Crystal solid parameters with profiled exit - Start**//
G4Box* crystalSolidT = new G4Box("crystal.solid",
                               fSizes.x()/2.,
                               fSizes.y()/2.,
                               fSizes.z()/2.);

G4VSolid* crystalCut = new G4Tubs("crystal.cut",
                                0.*CLHEP::mm,
                                fCylinder.x(),
                                30. * CLHEP::mm,
                                0.,
                                2*CLHEP::pi);

G4ThreeVector translateVector(fCylinder.y(),0.*CLHEP::mm,fCylinder.z()); //cylinders length 4mm - the good focusing

G4ThreeVector rotationVector(CLHEP::halfpi,0.,0.); //cylinder - the good focusing

G4RotationMatrix* rotCut = new G4RotationMatrix;
if(fAngles.x()!=0. || rotationVector.x()!=0.){
  rotCut->rotateX(rotationVector.x());
}
if(fAngles.y()!=0. || rotationVector.y()!=0.){
  rotCut->rotateY(rotationVector.y());
}
if(fAngles.z()!=0. || rotationVector.z()!=0.){
  rotCut->rotateZ(rotationVector.z());
}

// G4LogicalVolume* tubsLogic = new G4LogicalVolume(crystalSolidT,
//                                               worldMaterial,
//                                               "world.logicTubs");
// G4LogicalVolume* cutLogic = new G4LogicalVolume(crystalCut,
//                                              worldMaterial,
//                                              "world.logicCut");

G4VSolid* crystalSolid = new G4SubtractionSolid("crystal.solid",
                                              crystalSolidT,
                                              crystalCut,
                                              rotCut,
                                              translateVector);

//** Crystal solid parameters with profiled exit - Start**//



//** Crystal Definition Start **//
G4Material* mat =
  G4NistManager::Instance()->FindOrBuildMaterial("G4_Si");
G4ExtendedMaterial* Crystal =
  new G4ExtendedMaterial("crystal.material",mat);

Crystal->RegisterExtension(std::unique_ptr<G4CrystalExtension>(
        new G4CrystalExtension(Crystal)));
G4CrystalExtension* crystalExtension =
  (G4CrystalExtension*)Crystal->RetrieveExtension("crystal");
crystalExtension->SetUnitCell(
    new G4CrystalUnitCell(5.43 * CLHEP::angstrom,
                          5.43 * CLHEP::angstrom,
                          5.43 * CLHEP::angstrom,
                          CLHEP::halfpi,
                          CLHEP::halfpi,
                          CLHEP::halfpi,
                          227));

Crystal->RegisterExtension(std::unique_ptr<G4ChannelingMaterialData>(
        new G4ChannelingMaterialData("channeling")));
G4ChannelingMaterialData* crystalChannelingData =
  (G4ChannelingMaterialData*)Crystal->RetrieveExtension("channeling");
crystalChannelingData->SetFilename(fECfileName);

if(fBR!=G4ThreeVector()){
  crystalChannelingData->SetBR(fBR.x());
}
```

```
    G4LogicalCrystalVolume* crystalLogic =
      new G4LogicalCrystalVolume(crystalSolid,
                                 Crystal,
                                 "crystal.logic");
    crystalLogic->SetVerbose(1);

    G4RotationMatrix* rot = new G4RotationMatrix;
    if(fAngles.x()!=0.){
      rot->rotateX(fAngles.x());
    }
    if(fAngles.y()!=0.){
      rot->rotateY(fAngles.y());
    }
    if(fAngles.z()!=0.){
      rot->rotateZ(fAngles.z());
    }

    new G4PVPlacement(rot,
                      G4ThreeVector(fPosition.y(), 0., 0.),
                      crystalLogic,
                      "crystal.physic",
                      worldLogic,
                      false,
                      0);

    // new G4PVPlacement(rot,
    //                   G4ThreeVector(),
    //                   tubsLogic,
    //                   "tubs.physic",
    //                   worldLogic,
    //                   false,
    //                   0);

    // new G4PVPlacement(rotCut,
    //                   translateVector,
    //                   cutLogic,
    //                   "cut.physic",
    //                   worldLogic,
    //                   false,
    //                   0);

    //** Crystal Definition - End **//


    //** Water Phantom - Start **//

    G4Material* phantomMaterial =
      G4NistManager::Instance()->FindOrBuildMaterial("G4_WATER");

    G4Box* phantomSolid = new G4Box("phantom.solid",
                                    fPhantom.x()/2.,
                                    fPhantom.y()/2.,
                                    fPhantom.z()/2.);

    G4LogicalVolume* phantomLogic = new G4LogicalVolume(phantomSolid,
                                                        phantomMaterial,
                                                        "phantom.logic");

    new G4PVPlacement(0,
                      G4ThreeVector(fPhantomPos.x(),
                                    fPhantomPos.y(),
                                    fPhantomPos.z()),
                      phantomLogic,
                      "phantom.physic",
                      worldLogic,
                      false,
                      0);


    //** Water Phantom - End **//


#ifndef G4MULTITHREADED
    G4ChannelingOptrMultiParticleChangeCrossSection* testMany =
    new G4ChannelingOptrMultiParticleChangeCrossSection();
    testMany->AttachTo(crystalLogic);
    G4cout << " Attaching biasing operator " << testMany->GetName()
    << " to logical volume " << crystalLogic->GetName()
    << G4endl;

    G4VSensitiveDetector* telescope = new SensitiveDetector("/telescope");
    G4SDManager::GetSDMpointer()->AddNewDetector(telescope);
    for(unsigned int i1=0;i1<3;i1++){
        ssdLogic->SetSensitiveDetector(telescope);
    }
#endif

    // Visualization
    G4VisAttributes* crisVis = new G4VisAttributes(G4Colour::Magenta());
```

```
    crystalLogic->SetVisAttributes(crisVis);

    G4VisAttributes* worldVis = new G4VisAttributes(G4Colour::White());
    worldLogic->SetVisAttributes(worldVis);

    G4VisAttributes* phantomVis = new G4VisAttributes(G4Colour::Blue());
    phantomLogic->SetVisAttributes(phantomVis);

    // G4VisAttributes* crisCut = new G4VisAttributes(G4Colour::Green());
    // //crisCut->SetForceSolid(true);
    // cutLogic->SetVisAttributes(crisCut);

    // G4VisAttributes* crisTubs = new G4VisAttributes(G4Colour(1.0, 1.0, 0.0, 0.6));
    // crisTubs->SetForceSolid(true);
    // tubsLogic->SetVisAttributes(crisTubs);

    return worldPhysical;
}

//....oooOO0OOooo........oooOO0OOooo........oooOO0OOooo........oooOO0OOooo......

#ifdef G4MULTITHREADED
void DetectorConstruction::ConstructSDandField(){
    G4LogicalVolume* crystalLogic =
      G4LogicalVolumeStore::GetInstance()->GetVolume("crystal.logic");
    G4ChannelingOptrMultiParticleChangeCrossSection* testMany =
    new G4ChannelingOptrMultiParticleChangeCrossSection();
    testMany->AttachTo(crystalLogic);
    G4cout << " Attaching biasing operator " << testMany->GetName()
    << " to logical volume " << crystalLogic->GetName()
    << G4endl;

    G4LogicalVolume* ssdLogic =
      G4LogicalVolumeStore::GetInstance()->GetVolume("ssd.logic");
    G4VSensitiveDetector* telescope = new SensitiveDetector("/telescope");
    G4SDManager::GetSDMpointer()->AddNewDetector(telescope);
    for(unsigned int i1=0;i1<8;i1++){
        ssdLogic->SetSensitiveDetector(telescope);

    G4LogicalVolume* phantomLogic =
      G4LogicalVolumeStore::GetInstance()->GetVolume("phantom.logic");
    G4VSensitiveDetector* dose = new SensitiveDetector("/dose",true);
    phantomLogic->SetSensitiveDetector(dose);

    }
}
#else
void DetectorConstruction::ConstructSDandField(){;}
#endif

//....oooOO0OOooo........oooOO0OOooo........oooOO0OOooo........oooOO0OOooo......
```

# B. include/DetectorConstruction.hh

```
//
// ********************************************************************
// * License and Disclaimer                                           *
// *                                                                  *
// * The  Geant4 software  is  copyright of the Copyright Holders  of *
// * the Geant4 Collaboration.  It is provided  under  the terms  and *
// * conditions of the Geant4 Software License,  included in the file *
// * LICENSE and available at  http://cern.ch/geant4/license .  These *
// * include a list of copyright holders.                             *
// *                                                                  *
// * Neither the authors of this software system, nor their employing *
// * institutes,nor the agencies providing financial support for this *
// * work  make  any representation or  warranty, express or implied, *
// * regarding  this  software system or assume any liability for its *
// * use.  Please see the license in the file  LICENSE  and URL above *
// * for the full disclaimer and the limitation of liability.         *
// *                                                                  *
// * This  code  implementation is the result of  the  scientific and *
// * technical work of the GEANT4 collaboration.                      *
// * By using,  copying,  modifying or  distributing the software (or *
// * any work based  on the software)  you  agree  to acknowledge its *
// * use  in  resulting  scientific  publications,  and indicate your *
// * acceptance of all terms of the Geant4 Software license.          *
// ********************************************************************
// ----------------------------------------------------------------
//

#ifndef DetectorConstruction_h
#define DetectorConstruction_h 1
#endif

#include "G4VUserDetectorConstruction.hh"

#include "G4LogicalVolume.hh"
#include "G4VPhysicalVolume.hh"
#include "G4RunManager.hh"
#include "DetectorConstructionMessenger.hh"

#include "globals.hh"


//....oooOO0OOooo........oooOO0OOooo........oooOO0OOooo........oooOO0OOooo......

class DetectorConstruction : public G4VUserDetectorConstruction
{
public:

    DetectorConstruction();
    ~DetectorConstruction();

    void DefineMaterials();
    G4VPhysicalVolume* Construct();

private:
    void ConstructSDandField();

private:
    DetectorConstructionMessenger* fMessenger;

public:
    G4String GetEC() {return fECfileName;}
    void SetEC(G4String aString) {fECfileName = aString;}

    G4String GetMaterial() {return fMaterialName;}
    void SetMaterial(G4String aString) {fMaterialName = aString;}

    G4ThreeVector GetSizes() {return fSizes;}
    void SetSizes(G4ThreeVector a3vec) {fSizes = a3vec;}

    G4ThreeVector GetCylinder() {return fCylinder;}
    void SetCylinder(G4ThreeVector a3vec) {fCylinder = a3vec;}

    G4ThreeVector GetPosition() {return fPosition;}
    void SetPosition(G4ThreeVector a3vec) {fPosition = a3vec;}

    G4ThreeVector GetPhantom() {return fPhantom;}
    void SetPhantom(G4ThreeVector a3vec) {fPhantom = a3vec;}

    G4ThreeVector GetPhantomPos() {return fPhantomPos;}
    void SetPhantomPos(G4ThreeVector a3vec) {fPhantomPos = a3vec;}
```

```
    G4ThreeVector GetBR() {return fBR;}
    void SetBR(G4ThreeVector a3vec) {fBR = a3vec;}

    G4ThreeVector GetAngles() {return fAngles;}
    void SetAngles(G4ThreeVector a3vec) {fAngles = a3vec;}

    G4ThreeVector GetDetector() {return fDetector;}
    void SetDetector(G4ThreeVector a3vec) {fDetector = a3vec;}

    G4ThreeVector GetDetectorPos1() {return fDetectorPos1;}
    void SetDetectorPos1(G4ThreeVector a3vec) {fDetectorPos1 = a3vec;}

    G4ThreeVector GetDetectorPos2() {return fDetectorPos2;}
    void SetDetectorPos2(G4ThreeVector a3vec) {fDetectorPos2 = a3vec;}

private:
    G4String fECfileName;
    G4String fECOfileName;
    G4String fMaterialName;
    G4ThreeVector fSizes;
    G4ThreeVector fCylinder;
    G4ThreeVector fPosition;
    G4ThreeVector fPhantom;
    G4ThreeVector fPhantomPos;
    G4ThreeVector fBR;
    G4ThreeVector fAngles;
    G4ThreeVector fDetector;
    G4ThreeVector fDetectorPos1;
    G4ThreeVector fDetectorPos2;
};
```

# C. src/DetectorConstructionMessenger.cc

```
//
// ********************************************************************
// * License and Disclaimer                                           *
// *                                                                  *
// * The  Geant4 software  is  copyright of the Copyright Holders  of *
// * the Geant4 Collaboration.  It is provided  under  the terms  and *
// * conditions of the Geant4 Software License,  included in the file *
// * LICENSE and available at  http://cern.ch/geant4/license .  These *
// * include a list of copyright holders.                             *
// *                                                                  *
// * Neither the authors of this software system, nor their employing *
// * institutes,nor the agencies providing financial support for this *
// * work  make  any representation or  warranty, express or implied, *
// * regarding  this  software system or assume any liability for its *
// * use.  Please see the license in the file  LICENSE  and URL above *
// * for the full disclaimer and the limitation of liability.         *
// *                                                                  *
// * This  code  implementation is the result of  the  scientific and *
// * technical work of the GEANT4 collaboration.                      *
// * By using,  copying,  modifying or  distributing the software (or *
// * any work based  on the software)  you  agree  to acknowledge its *
// * use  in  resulting  scientific  publications,  and indicate your *
// * acceptance of all terms of the Geant4 Software license.          *
// ********************************************************************
//

#include "DetectorConstructionMessenger.hh"
#include "DetectorConstruction.hh"
#include "G4UIdirectory.hh"
#include "G4UIcmdWithADoubleAndUnit.hh"
#include "G4UIcmdWithABool.hh"
#include "G4UIcmdWithAString.hh"
#include "G4UIcmdWith3VectorAndUnit.hh"
#include "G4UIcmdWith3Vector.hh"
#include "G4UIcmdWithADouble.hh"
#include "G4RunManager.hh"

#include "G4ios.hh"

DetectorConstructionMessenger::
DetectorConstructionMessenger(
                              DetectorConstruction* mpga)
:fTarget(mpga){
    fMyXtalDirectory = new G4UIdirectory("/xtal/");
    fMyXtalDirectory->SetGuidance("Crystal␣setup␣control␣commands.");

    fXtalMaterialCmd = new G4UIcmdWithAString("/xtal/setMaterial",
                                              this);
    fXtalMaterialCmd->SetGuidance("Set␣crystal␣material.");
    fXtalMaterialCmd->SetParameterName("xMat",true);
    fXtalMaterialCmd->SetDefaultValue("G4_Si");

    fXtalSizeCmd = new G4UIcmdWith3VectorAndUnit("/xtal/setSize",this);
    fXtalSizeCmd->SetGuidance("Set␣crystal␣size.");
    fXtalSizeCmd->SetParameterName("xtalSizeX",
                                   "xtalSizeY",
                                   "xtalSizeZ",
                                   true);
    fXtalSizeCmd->SetDefaultValue(G4ThreeVector(6.,2.,6.));
    fXtalSizeCmd->SetDefaultUnit("mm");

    fXtalCylinderCmd = new G4UIcmdWith3VectorAndUnit("/xtal/setCylinder",this);
    fXtalCylinderCmd->SetGuidance("Set␣cylinder␣size.");
    fXtalCylinderCmd->SetParameterName("xtalCylinderX",
                                       "xtalCylinderY",
                                       "xtalCylinderZ",
                                       true);
    fXtalCylinderCmd->SetDefaultValue(G4ThreeVector(6.,2.,6.));
    fXtalCylinderCmd->SetDefaultUnit("mm");

    fXtalPositionCmd = new G4UIcmdWith3VectorAndUnit("/xtal/setPosition",this);
    fXtalPositionCmd->SetGuidance("Set␣the␣x-axis␣position␣of␣the␣detestors␣and␣the␣crystal␣(in␣both␣cases␣the␣default␣value␣is␣0).")
        ↪ ;
    fXtalPositionCmd->SetParameterName("xtalPositionX",
                                       "xtalPositionY",
                                       "xtalPositionZ",
                                       true);
    fXtalPositionCmd->SetDefaultValue(G4ThreeVector(0.,0.,0.));
    fXtalPositionCmd->SetDefaultUnit("m");
```

```
    fXtalPhantomCmd = new G4UIcmdWith3VectorAndUnit("/xtal/setPhantom",this);
    fXtalPhantomCmd->SetGuidance("Set phantom size.");
    fXtalPhantomCmd->SetParameterName("xtalPhantomX",
                                      "xtalPhantomY",
                                      "xtalPhantomZ",
                                      true);
    fXtalPhantomCmd->SetDefaultValue(G4ThreeVector(0.3,0.3,0.3));
    fXtalPhantomCmd->SetDefaultUnit("m");

    fXtalPhantomPosCmd = new G4UIcmdWith3VectorAndUnit("/xtal/setPhantomPos",this);
    fXtalPhantomPosCmd->SetGuidance("Set phantom position.");
    fXtalPhantomPosCmd->SetParameterName("xtalPhantomPosX",
                                         "xtalPhantomPosY",
                                         "xtalPhantomPosZ",
                                         true);
    fXtalPhantomPosCmd->SetDefaultValue(G4ThreeVector(0.5,0.,1.));
    fXtalPhantomPosCmd->SetDefaultUnit("m");

    fXtalBRCmd = new G4UIcmdWith3VectorAndUnit("/xtal/setBR",this);
    fXtalBRCmd->SetGuidance("Set crystal bending radius.");
    fXtalBRCmd->SetParameterName("xtalBRX",
                                 "xtalBRY",
                                 "xtalBRZ",
                                 true);
    fXtalBRCmd->SetDefaultValue(G4ThreeVector(0.,0.,0.));
    fXtalBRCmd->SetDefaultUnit("mm");

    fXtalAngleCmd = new G4UIcmdWith3VectorAndUnit("/xtal/setAngle",this);
    fXtalAngleCmd->SetGuidance("Set crystal angles.");
    fXtalAngleCmd->SetParameterName("xtalAngleX",
                                    "xtalAngleY",
                                    "xtalAngleZ",
                                    true);
    fXtalAngleCmd->SetDefaultValue(G4ThreeVector(0.,0.,0.));
    fXtalAngleCmd->SetDefaultUnit("rad");

    fXtalECCmd = new G4UIcmdWithAString("/xtal/setEC",
                                        this);
    fXtalECCmd->SetGuidance("Set crystal EC.");
    fXtalECCmd->SetParameterName("xEC",true);
    fXtalECCmd->SetDefaultValue("data/Si220pl");

    fXtalDetectorCmd = new G4UIcmdWith3VectorAndUnit("/xtal/setDetector",this);
    fXtalDetectorCmd->SetGuidance("Set detector size.");
    fXtalDetectorCmd->SetParameterName("xtalDetectorX",
                                       "xtalDetectorY",
                                       "xtalDetectorZ",
                                       true);
    fXtalDetectorCmd->SetDefaultValue(G4ThreeVector(38.,38.,0.64));
    fXtalDetectorCmd->SetDefaultUnit("mm");

    fXtalDetectorPos1Cmd = new G4UIcmdWith3VectorAndUnit("/xtal/setDetectorPos1",this);
    fXtalDetectorPos1Cmd->SetGuidance("Set detector position in z-axis part 1.");
    fXtalDetectorPos1Cmd->SetParameterName("xtalDetectorPos1X",
                                           "xtalDetectorPos1Y",
                                           "xtalDetectorPos1Z",
                                           true);
    fXtalDetectorPos1Cmd->SetDefaultValue(G4ThreeVector(0.01,0.5,1.));
    fXtalDetectorPos1Cmd->SetDefaultUnit("m");

    fXtalDetectorPos2Cmd = new G4UIcmdWith3VectorAndUnit("/xtal/setDetectorPos2",this);
    fXtalDetectorPos2Cmd->SetGuidance("Set detector position in z-axis part 2.");
    fXtalDetectorPos2Cmd->SetParameterName("xtalDetectorPos2X",
                                           "xtalDetectorPos2Y",
                                           "xtalDetectorPos2Z",
                                           true);
    fXtalDetectorPos2Cmd->SetDefaultValue(G4ThreeVector(1.5,2.,2.5));
    fXtalDetectorPos2Cmd->SetDefaultUnit("m");
}

//....oooOO0OOooo........oooOO0OOooo........oooOO0OOooo........oooOO0OOooo....

DetectorConstructionMessenger::
~DetectorConstructionMessenger(){
    delete fXtalMaterialCmd;
    delete fXtalSizeCmd;
    delete fXtalCylinderCmd;
    delete fXtalPositionCmd;
    delete fXtalPhantomCmd;
    delete fXtalPhantomPosCmd;
    delete fXtalAngleCmd;
    delete fXtalECCmd;
    delete fXtalBRCmd;
    delete fXtalDetectorCmd;
    delete fXtalDetectorPos1Cmd;
    delete fXtalDetectorPos2Cmd;
}

//....oooOO0OOooo........oooOO0OOooo........oooOO0OOooo........oooOO0OOooo....
```

```
void DetectorConstructionMessenger::SetNewValue(
                                                   G4UIcommand *command,
                                                   G4String newValue){
    if(command==fXtalMaterialCmd ){
        fTarget->SetMaterial(newValue);
    }
    if(command==fXtalSizeCmd ){
        fTarget->SetSizes(fXtalSizeCmd->GetNew3VectorValue(newValue));
    }
    if(command==fXtalCylinderCmd ){
        fTarget->SetCylinder(fXtalCylinderCmd->GetNew3VectorValue(newValue));
    }
    if(command==fXtalPositionCmd ){
        fTarget->SetPosition(fXtalPositionCmd->GetNew3VectorValue(newValue));
    }
    if(command==fXtalPhantomCmd ){
        fTarget->SetPhantom(fXtalPhantomCmd->GetNew3VectorValue(newValue));
    }
    if(command==fXtalPhantomPosCmd ){
        fTarget->SetPhantomPos(fXtalPhantomPosCmd->GetNew3VectorValue(newValue));
    }
    if(command==fXtalBRCmd ){
        fTarget->SetBR(fXtalBRCmd->GetNew3VectorValue(newValue));
    }
    if(command==fXtalAngleCmd ){
        fTarget->SetAngles(fXtalAngleCmd->GetNew3VectorValue(newValue));
    }
    if(command==fXtalECCmd ){
        fTarget->SetEC(newValue);
    }
    if(command==fXtalDetectorCmd ){
        fTarget->SetDetector(fXtalDetectorCmd->GetNew3VectorValue(newValue));
    }
    if(command==fXtalDetectorPos1Cmd ){
        fTarget->SetDetectorPos1(fXtalDetectorPos1Cmd->GetNew3VectorValue(newValue));
    }
    if(command==fXtalDetectorPos2Cmd ){
        fTarget->SetDetectorPos2(fXtalDetectorPos2Cmd->GetNew3VectorValue(newValue));
    }
}

//....oooOO0OOooo........oooOO0OOooo........oooOO0OOooo........oooOO0OOooo....

G4String DetectorConstructionMessenger::GetCurrentValue(
                                                         G4UIcommand * command){
    G4String cv;

    if( command==fXtalMaterialCmd ){
        cv = fTarget->GetMaterial();
    }
    if( command==fXtalSizeCmd ){
        cv = fXtalSizeCmd->ConvertToString(fTarget->GetSizes(),"mm");
    }
    if( command==fXtalCylinderCmd ){
        cv = fXtalCylinderCmd->ConvertToString(fTarget->GetCylinder(),"mm");
    }
    if( command==fXtalPositionCmd ){
        cv = fXtalPositionCmd->ConvertToString(fTarget->GetPosition(),"m");
    }
    if( command==fXtalPhantomCmd ){
        cv = fXtalPhantomCmd->ConvertToString(fTarget->GetPhantom(),"m");
    }
    if( command==fXtalPhantomPosCmd ){
        cv = fXtalPhantomPosCmd->ConvertToString(fTarget->GetPhantomPos(),"m");
    }
    if( command==fXtalBRCmd ){
        cv = fXtalBRCmd->ConvertToString(fTarget->GetBR(),"mm");
    }
    if( command==fXtalAngleCmd ){
        cv = fXtalAngleCmd->ConvertToString(fTarget->GetAngles(),"rad");
    }
    if( command==fXtalECCmd ){
        cv = fTarget->GetEC();
    }
    if( command==fXtalDetectorCmd ){
        cv = fXtalDetectorCmd->ConvertToString(fTarget->GetDetector(),"mm");
    }
    if( command==fXtalDetectorPos1Cmd ){
        cv = fXtalDetectorPos1Cmd->ConvertToString(fTarget->GetDetectorPos1(),"m");
    }
    if( command==fXtalDetectorPos2Cmd ){
        cv = fXtalDetectorPos2Cmd->ConvertToString(fTarget->GetDetectorPos2(),"m");
    }

    return cv;
}
//....oooOO0OOooo........oooOO0OOooo........oooOO0OOooo........oooOO0OOooo....
```

# D. include/DetectorConstructionMessenger.hh

```
//
// ********************************************************************
// * License and Disclaimer                                           *
// *                                                                  *
// * The  Geant4 software  is  copyright of the Copyright Holders  of *
// * the Geant4 Collaboration.  It is provided  under  the terms  and *
// * conditions of the Geant4 Software License,  included in the file *
// * LICENSE and available at  http://cern.ch/geant4/license .  These *
// * include a list of copyright holders.                             *
// *                                                                  *
// * Neither the authors of this software system, nor their employing *
// * institutes,nor the agencies providing financial support for this *
// * work  make  any representation or  warranty, express or implied, *
// * regarding  this  software system or assume any liability for its *
// * use.  Please see the license in the file  LICENSE  and URL above *
// * for the full disclaimer and the limitation of liability.         *
// *                                                                  *
// * This  code  implementation is the result of  the  scientific and *
// * technical work of the GEANT4 collaboration.                      *
// * By using,  copying,  modifying or  distributing the software (or *
// * any work based  on the software)  you  agree  to acknowledge its *
// * use  in  resulting  scientific  publications,  and indicate your *
// * acceptance of all terms of the Geant4 Software license.          *
// ********************************************************************
//
// ----------------------------------------------------------------
//

#ifndef DetectorConstructionMessenger_h
#define DetectorConstructionMessenger_h 1

class DetectorConstruction;
class G4UIdirectory;
class G4UIcmdWithADoubleAndUnit;
class G4UIcmdWithAIntAndUnit;
class G4UIcmdWith3VectorAndUnit;
class G4UIcmdWith3Vector;
class G4UIcmdWithABool;
class G4UIcmdWithAString;
class G4UIcmdWithADouble;

#include "G4UImessenger.hh"
#include "globals.hh"

class DetectorConstructionMessenger: public G4UImessenger
{
  public:
    DetectorConstructionMessenger(
                DetectorConstruction* mpga);
    ~DetectorConstructionMessenger();

    virtual void SetNewValue(G4UIcommand * command,G4String newValues);
    virtual G4String GetCurrentValue(G4UIcommand * command);

  private:
    DetectorConstruction * fTarget;

    G4UIdirectory* fMyXtalDirectory;

    G4UIcmdWithAString*  fXtalMaterialCmd;
    G4UIcmdWith3VectorAndUnit* fXtalSizeCmd;
    G4UIcmdWith3VectorAndUnit* fXtalCylinderCmd;
    G4UIcmdWith3VectorAndUnit* fXtalPositionCmd;
    G4UIcmdWith3VectorAndUnit* fXtalPhantomCmd;
    G4UIcmdWith3VectorAndUnit* fXtalPhantomPosCmd;
    G4UIcmdWith3VectorAndUnit* fXtalBRCmd;
    G4UIcmdWith3VectorAndUnit* fXtalAngleCmd;
    G4UIcmdWithAString*  fXtalECCmd;
    G4UIcmdWith3VectorAndUnit* fXtalDetectorCmd;
    G4UIcmdWith3VectorAndUnit* fXtalDetectorPos1Cmd;
    G4UIcmdWith3VectorAndUnit* fXtalDetectorPos2Cmd;

};

#endif
```

# E. src/SensitiveDetector.cc

```
//
// ********************************************************************
// * License and Disclaimer                                           *
// *                                                                  *
// * The  Geant4 software  is  copyright of the Copyright Holders  of *
// * the Geant4 Collaboration.  It is provided  under  the terms  and *
// * conditions of the Geant4 Software License,  included in the file *
// * LICENSE and available at  http://cern.ch/geant4/license .  These *
// * include a list of copyright holders.                             *
// *                                                                  *
// * Neither the authors of this software system, nor their employing *
// * institutes,nor the agencies providing financial support for this *
// * work  make  any representation or  warranty, express or implied, *
// * regarding  this  software system or assume any liability for its *
// * use.  Please see the license in the file  LICENSE  and URL above *
// * for the full disclaimer and the limitation of liability.         *
// *                                                                  *
// * This  code  implementation is the result of  the  scientific and *
// * technical work of the GEANT4 collaboration.                      *
// * By using,  copying,  modifying or  distributing the software (or *
// * any work based  on the software)  you  agree  to acknowledge its *
// * use  in  resulting  scientific  publications,  and indicate your *
// * acceptance of all terms of the Geant4 Software license.          *
// ********************************************************************
//

#include "SensitiveDetector.hh"
#include "SensitiveDetectorHit.hh"
#include "G4HCofThisEvent.hh"
#include "G4TouchableHistory.hh"
#include "G4Track.hh"
#include "G4Step.hh"
#include "G4SDManager.hh"
#include "G4Navigator.hh"
#include "G4ios.hh"
#include "G4AnalysisManager.hh"

SensitiveDetector::SensitiveDetector(G4String name, G4bool isPhantom, G4ThreeVector phantomPosition):
  G4VSensitiveDetector(name){
    G4String HCname;
    collectionName.insert(HCname="collection");
    fHCID = -1;
    fIsPhantom = isPhantom;
    fPhantomPosition = phantomPosition;
}

//....oooOO0OOooo........oooOO0OOooo........oooOO0OOooo........oooOO0OOooo....

SensitiveDetector::~SensitiveDetector(){
}

//....oooOO0OOooo........oooOO0OOooo........oooOO0OOooo........oooOO0OOooo....

void SensitiveDetector::Initialize(G4HCofThisEvent*HCE){
    fHitsCollection =
        new SensitiveDetectorHitsCollection(SensitiveDetectorName,
                                            collectionName[0]);
    if(fHCID<0){
        fHCID = G4SDManager::GetSDMpointer()->GetCollectionID(fHitsCollection);
    }
    HCE->AddHitsCollection(fHCID,fHitsCollection);
}

//....oooOO0OOooo........oooOO0OOooo........oooOO0OOooo........oooOO0OOooo....

G4bool SensitiveDetector::ProcessHits(G4Step*aStep,
                                      G4TouchableHistory*){

  // only care about primary particles (aka electrons)
  if(aStep->GetTrack()->GetTrackID()>1) return true;

  if(fIsPhantom == false){
    // G4cout<<"A step in a SD that is not a Phantom"<<G4endl;
    //what to do with no phantom
    G4StepPoint* postStepPoint = aStep->GetPostStepPoint();
    if(!(postStepPoint->GetStepStatus() == fGeomBoundary)) return true;

    G4StepPoint* preStepPoint = aStep->GetPreStepPoint();

    G4TouchableHistory* theTouchable =
```

```
        (G4TouchableHistory*)(preStepPoint->GetTouchable());
    G4VPhysicalVolume* thePhysical = theTouchable->GetVolume(0);
    G4int copyNo = thePhysical->GetCopyNo();

    G4ThreeVector worldPos = preStepPoint->GetPosition();

    SensitiveDetectorHit* aHit =
        new SensitiveDetectorHit(copyNo);
    aHit->SetLayerID(copyNo);
    aHit->SetWorldPos(worldPos);

    fHitsCollection->insert(aHit);
    return true;
  }else{
    //what to do with a phantom
    G4ThreeVector position = aStep->GetTrack()->GetPosition();
    G4double Edep = aStep->GetTotalEnergyDeposit();

    // G4VPhysicalVolume* physVolume = aStep->GetTrack()->GetVolume();
    // G4LogicalVolume* logVolume = physVoulme->GetLogicalVolume();
    // G4ThreeVector phantomPosition = physVolume->GetTranslation();

    G4AnalysisManager* analysisManager = G4AnalysisManager::Instance();
    analysisManager->FillNtupleDColumn(1,0, position.x() / CLHEP::mm);
    analysisManager->FillNtupleDColumn(1,1, position.y() / CLHEP::mm);
    analysisManager->FillNtupleDColumn(1,2, position.z() / CLHEP::mm);
    analysisManager->FillNtupleDColumn(1,3, Edep / CLHEP::MeV);
    analysisManager->AddNtupleRow(1);

    return true;
  }
}

//....oooOO0OOooo........oooOO0OOooo........oooOO0OOooo........oooOO0OOooo....

void SensitiveDetector::EndOfEvent(G4HCofThisEvent* /*HCE*/){
}

//....oooOO0OOooo........oooOO0OOooo........oooOO0OOooo........oooOO0OOooo....
```

# F. include/SensitiveDetector.hh

```
//
// ********************************************************************
// * License and Disclaimer                                           *
// *                                                                  *
// * The  Geant4 software  is  copyright of the Copyright Holders  of *
// * the Geant4 Collaboration.  It is provided  under  the terms  and *
// * conditions of the Geant4 Software License,  included in the file *
// * LICENSE and available at  http://cern.ch/geant4/license .  These *
// * include a list of copyright holders.                             *
// *                                                                  *
// * Neither the authors of this software system, nor their employing *
// * institutes,nor the agencies providing financial support for this *
// * work  make  any representation or  warranty, express or implied, *
// * regarding  this  software system or assume any liability for its *
// * use.  Please see the license in the file  LICENSE  and URL above *
// * for the full disclaimer and the limitation of liability.         *
// *                                                                  *
// * This  code  implementation is the result of  the  scientific and *
// * technical work of the GEANT4 collaboration.                      *
// * By using,  copying,  modifying or  distributing the software (or *
// * any work based  on the software)  you  agree  to acknowledge its *
// * use  in  resulting  scientific  publications,  and indicate your *
// * acceptance of all terms of the Geant4 Software license.          *
// ********************************************************************
// ------------------------------------------------------------------
//

#ifndef SensitiveDetector_h
#define SensitiveDetector_h 1

#include "G4VSensitiveDetector.hh"
#include "SensitiveDetectorHit.hh"
class G4Step;
class G4HCofThisEvent;
class G4TouchableHistory;

class SensitiveDetector : public G4VSensitiveDetector
{
  public:
  SensitiveDetector(G4String name, G4bool isPhantom=false, G4ThreeVector phantomPosition=G4ThreeVector(0,0,0));
      virtual ~SensitiveDetector();

      virtual void Initialize(G4HCofThisEvent*HCE);
      virtual G4bool ProcessHits(G4Step*aStep,G4TouchableHistory*ROhist);
      virtual void EndOfEvent(G4HCofThisEvent*HCE);

  private:
      SensitiveDetectorHitsCollection * fHitsCollection;
      G4int fHCID;
      G4bool fIsPhantom;
      G4ThreeVector fPhantomPosition;
};

#endif
```

# G. src/EventAction.cc

```
//
// ********************************************************************
// * License and Disclaimer                                           *
// *                                                                  *
// * The  Geant4 software  is  copyright of the Copyright Holders  of *
// * the Geant4 Collaboration.  It is provided  under  the terms  and *
// * conditions of the Geant4 Software License,  included in the file *
// * LICENSE and available at  http://cern.ch/geant4/license .  These *
// * include a list of copyright holders.                             *
// *                                                                  *
// * Neither the authors of this software system, nor their employing *
// * institutes,nor the agencies providing financial support for this *
// * work  make  any representation or  warranty, express or implied, *
// * regarding  this  software system or assume any liability for its *
// * use.  Please see the license in the file  LICENSE  and URL above *
// * for the full disclaimer and the limitation of liability.         *
// *                                                                  *
// * This  code  implementation is the result of  the  scientific and *
// * technical work of the GEANT4 collaboration.                      *
// * By using,  copying,  modifying or  distributing the software (or *
// * any work based  on the software)  you  agree  to acknowledge its *
// * use  in  resulting  scientific  publications,  and indicate your *
// * acceptance of all terms of the Geant4 Software license.          *
// ********************************************************************
//

#include "EventAction.hh"

#include "G4RunManager.hh"

#include "G4Event.hh"
#include "G4EventManager.hh"
#include "G4HCofThisEvent.hh"
#include "G4VHitsCollection.hh"
#include "G4TrajectoryContainer.hh"
#include "G4Trajectory.hh"
#include "G4VVisManager.hh"
#include "G4SDManager.hh"
#include "G4UImanager.hh"
#include "G4ios.hh"
#include "G4SystemOfUnits.hh"
#include "G4AnalysisManager.hh"

#include "SensitiveDetectorHit.hh"

EventAction::EventAction():
fSDHT_ID(-1){}

//....oooOO0OOooo........oooOO0OOooo........oooOO0OOooo........oooOO0OOooo....

EventAction::~EventAction(){;}

//....oooOO0OOooo........oooOO0OOooo........oooOO0OOooo........oooOO0OOooo....

void EventAction::BeginOfEventAction(const G4Event*){;}

//....oooOO0OOooo........oooOO0OOooo........oooOO0OOooo........oooOO0OOooo....

void EventAction::EndOfEventAction(const G4Event* evt){
    G4SDManager* SDman = G4SDManager::GetSDMpointer();


    G4ThreeVector ssd[8];
    ssd[0]= G4ThreeVector(0.,0.,0.);
    ssd[1]= G4ThreeVector(0.,0.,0.);
    ssd[2]= G4ThreeVector(0.,0.,0.);
    ssd[3]= G4ThreeVector(0.,0.,0.);
    ssd[4]= G4ThreeVector(0.,0.,0.);
    ssd[5]= G4ThreeVector(0.,0.,0.);
    ssd[6]= G4ThreeVector(0.,0.,0.);
    ssd[7]= G4ThreeVector(0.,0.,0.);

    if(fSDHT_ID == -1) {
        G4String sdName;
        if(SDman->FindSensitiveDetector(sdName="telescope",0)){
          fSDHT_ID = SDman->GetCollectionID(sdName="telescope/collection");
        }
    }
    // G4cout<<"[DEBUG] fSDHT_ID="<<fSDHT_ID<<G4endl;
```

```
SensitiveDetectorHitsCollection* sdht = 0;
G4HCofThisEvent *hce = evt->GetHCofThisEvent();

if(hce){
    if(fSDHT_ID != -1){
        G4VHitsCollection* aHCSD = hce->GetHC(fSDHT_ID);
        sdht = (SensitiveDetectorHitsCollection*)(aHCSD);
    }
}

int bTotalHits = 0;
if(sdht){
    int n_hit_sd = sdht->entries();
    for(int i2=0;i2<8;i2++){
        for(int i1=0;i1<n_hit_sd;i1++)
        {
            SensitiveDetectorHit* aHit = (*sdht)[i1];
            if(aHit->GetLayerID()==i2) {
                ssd[i2] = aHit->GetWorldPos();
                bTotalHits++;
            }
        }
    }
}

if(bTotalHits > 2){
    G4AnalysisManager* analysisManager = G4AnalysisManager::Instance();
    G4double angXin  = (ssd[1].x() - ssd[0].x()) / (ssd[1].z() - ssd[0].z());
    G4double angYin  = (ssd[1].y() - ssd[0].y()) / (ssd[1].z() - ssd[0].z());
    analysisManager->FillNtupleDColumn(0, 0, angXin * 1.E6 * CLHEP::rad);
    analysisManager->FillNtupleDColumn(0, 1, angYin * 1.E6 * CLHEP::rad);

    double posXin = ssd[1].x() - angXin * ssd[1].z();
    double posYin = ssd[1].y() - angYin * ssd[1].z();
    analysisManager->FillNtupleDColumn(0, 2, posXin / CLHEP::mm);
    analysisManager->FillNtupleDColumn(0, 3, posYin / CLHEP::mm);

    G4double angXout1st = (ssd[2].x() - posXin) / (ssd[2].z());
    G4double angYout1st = (ssd[2].y() - posYin) / (ssd[2].z());
    analysisManager->FillNtupleDColumn(0, 4, angXout1st * 1.E6 * CLHEP::rad);
    analysisManager->FillNtupleDColumn(0, 5, angYout1st * 1.E6 * CLHEP::rad);

    double posXout1st = ssd[2].x() - angXin * ssd[2].z();
    double posYout1st = ssd[2].y() - angYin * ssd[2].z();
    analysisManager->FillNtupleDColumn(0, 6, posXout1st / CLHEP::mm);
    analysisManager->FillNtupleDColumn(0, 7, posYout1st / CLHEP::mm);

    G4double angXout2nd = (ssd[3].x() - posXin) / (ssd[3].z());
    G4double angYout2nd = (ssd[3].y() - posYin) / (ssd[3].z());
    analysisManager->FillNtupleDColumn(0, 8, angXout2nd * 1.E6 * CLHEP::rad);
    analysisManager->FillNtupleDColumn(0, 9, angYout2nd * 1.E6 * CLHEP::rad);

    double posXout2nd = ssd[3].x() - angXin * ssd[3].z();
    double posYout2nd = ssd[3].y() - angYin * ssd[3].z();
    analysisManager->FillNtupleDColumn(0, 10, posXout2nd / CLHEP::mm);
    analysisManager->FillNtupleDColumn(0, 11, posYout2nd / CLHEP::mm);

    G4double angXout3rd = (ssd[4].x() - posXin) / (ssd[4].z());
    G4double angYout3rd = (ssd[4].y() - posYin) / (ssd[4].z());
    analysisManager->FillNtupleDColumn(0, 12, angXout3rd * 1.E6 * CLHEP::rad);
    analysisManager->FillNtupleDColumn(0, 13, angYout3rd * 1.E6 * CLHEP::rad);

    double posXout3rd = ssd[4].x() - angXin * ssd[4].z();
    double posYout3rd = ssd[4].y() - angYin * ssd[4].z();
    analysisManager->FillNtupleDColumn(0, 14, posXout3rd / CLHEP::mm);
    analysisManager->FillNtupleDColumn(0, 15, posYout3rd / CLHEP::mm);

    G4double angXout4th = (ssd[5].x() - posXin) / (ssd[5].z());
    G4double angYout4th = (ssd[5].y() - posYin) / (ssd[5].z());
    analysisManager->FillNtupleDColumn(0, 16, angXout4th * 1.E6 * CLHEP::rad);
    analysisManager->FillNtupleDColumn(0, 17, angYout4th * 1.E6 * CLHEP::rad);

    double posXout4th = ssd[5].x() - angXin * ssd[5].z();
    double posYout4th = ssd[5].y() - angYin * ssd[5].z();
    analysisManager->FillNtupleDColumn(0, 18, posXout4th / CLHEP::mm);
    analysisManager->FillNtupleDColumn(0, 19, posYout4th / CLHEP::mm);

    G4double angXout5th = (ssd[6].x() - posXin) / (ssd[6].z());
    G4double angYout5th = (ssd[6].y() - posYin) / (ssd[6].z());
    analysisManager->FillNtupleDColumn(0, 20, angXout5th * 1.E6 * CLHEP::rad);
    analysisManager->FillNtupleDColumn(0, 21, angYout5th * 1.E6 * CLHEP::rad);

    double posXout5th = ssd[6].x() - angXin * ssd[6].z();
    double posYout5th = ssd[6].y() - angYin * ssd[6].z();
    analysisManager->FillNtupleDColumn(0, 22, posXout5th / CLHEP::mm);
    analysisManager->FillNtupleDColumn(0, 23, posYout5th / CLHEP::mm);

    G4double angXout6th = (ssd[7].x() - posXin) / (ssd[7].z());
    G4double angYout6th = (ssd[7].y() - posYin) / (ssd[7].z());
```

```
        analysisManager->FillNtupleDColumn(0, 24, angXout6th * 1.E6 * CLHEP::rad);
        analysisManager->FillNtupleDColumn(0, 25, angYout6th * 1.E6 * CLHEP::rad);

        double posXout6th = ssd[7].x() - angXin * ssd[7].z();
        double posYout6th = ssd[7].y() - angYin * ssd[7].z();
        analysisManager->FillNtupleDColumn(0, 26, posXout6th / CLHEP::mm);
        analysisManager->FillNtupleDColumn(0, 27, posYout6th / CLHEP::mm);

        analysisManager->AddNtupleRow();
    }
}

//....oooOO0OOooo........oooOO0OOooo........oooOO0OOooo........oooOO0OOooo....
```

# H. src/RunAction.cc

```
//
// ********************************************************************
// * License and Disclaimer                                           *
// *                                                                  *
// * The  Geant4 software  is  copyright of the Copyright Holders  of *
// * the Geant4 Collaboration.  It is provided  under  the terms  and *
// * conditions of the Geant4 Software License,  included in the file *
// * LICENSE and available at  http://cern.ch/geant4/license .  These *
// * include a list of copyright holders.                             *
// *                                                                  *
// * Neither the authors of this software system, nor their employing *
// * institutes,nor the agencies providing financial support for this *
// * work  make  any representation or  warranty, express or implied, *
// * regarding  this  software system or assume any liability for its *
// * use.  Please see the license in the file  LICENSE  and URL above *
// * for the full disclaimer and the limitation of liability.         *
// *                                                                  *
// * This  code  implementation is the result of  the  scientific and *
// * technical work of the GEANT4 collaboration.                      *
// * By using,  copying,  modifying or  distributing the software (or *
// * any work based  on the software)  you  agree  to acknowledge its *
// * use  in  resulting  scientific  publications,  and indicate your *
// * acceptance of all terms of the Geant4 Software license.          *
// ********************************************************************
//

#include "RunAction.hh"

#include "G4Run.hh"
#include "G4RunManager.hh"
#include "G4UnitsTable.hh"
#include "G4SystemOfUnits.hh"
#include "G4AnalysisManager.hh"

//....oooOO0OOooo........oooOO0OOooo........oooOO0OOooo........oooOO0OOooo......

RunAction::RunAction(): G4UserRunAction(){
    G4RunManager::GetRunManager()->SetPrintProgress(10);

    G4AnalysisManager* analysisManager = G4AnalysisManager::Instance();

    //** Set defaults **//
    analysisManager->SetVerboseLevel(1);
    analysisManager->SetFirstHistoId(1);

    //** Creating ntuple **//
    analysisManager->CreateNtuple("ExExChTree", "Angles and Positions");
    analysisManager->CreateNtupleDColumn("angXin");
    analysisManager->CreateNtupleDColumn("angYin");
    analysisManager->CreateNtupleDColumn("posXin");
    analysisManager->CreateNtupleDColumn("posYin");
    analysisManager->CreateNtupleDColumn("angXout1st");
    analysisManager->CreateNtupleDColumn("angYout1st");
    analysisManager->CreateNtupleDColumn("posXout1st");
    analysisManager->CreateNtupleDColumn("posYout1st");
    analysisManager->CreateNtupleDColumn("angXout2nd");
    analysisManager->CreateNtupleDColumn("angYout2nd");
    analysisManager->CreateNtupleDColumn("posXout2nd");
    analysisManager->CreateNtupleDColumn("posYout2nd");
    analysisManager->CreateNtupleDColumn("angXout3rd");
    analysisManager->CreateNtupleDColumn("angYout3rd");
    analysisManager->CreateNtupleDColumn("posXout3rd");
    analysisManager->CreateNtupleDColumn("posYout3rd");
    analysisManager->CreateNtupleDColumn("angXout4th");
    analysisManager->CreateNtupleDColumn("angYout4th");
    analysisManager->CreateNtupleDColumn("posXout4th");
    analysisManager->CreateNtupleDColumn("posYout4th");
    analysisManager->CreateNtupleDColumn("angXout5th");
    analysisManager->CreateNtupleDColumn("angYout5th");
    analysisManager->CreateNtupleDColumn("posXout5th");
    analysisManager->CreateNtupleDColumn("posYout5th");
    analysisManager->CreateNtupleDColumn("angXout6th");
    analysisManager->CreateNtupleDColumn("angYout6th");
    analysisManager->CreateNtupleDColumn("posXout6th");
    analysisManager->CreateNtupleDColumn("posYout6th");
    analysisManager->FinishNtuple();

    //** Create new ntuple for energy deposition **//
    analysisManager->CreateNtuple("PhantomDeposition", "Energy deposition in the phantom");
    analysisManager->CreateNtupleDColumn("posX");
```

```
    analysisManager->CreateNtupleDColumn("posY");
    analysisManager->CreateNtupleDColumn("posZ");
    analysisManager->CreateNtupleDColumn("Edep");
    analysisManager->FinishNtuple();


}

//....oooOO0OOooo........oooOO0OOooo........oooOO0OOooo........oooOO0OOooo......

RunAction::~RunAction(){
}

//....oooOO0OOooo........oooOO0OOooo........oooOO0OOooo........oooOO0OOooo......

void RunAction::BeginOfRunAction(const G4Run*){
    G4AnalysisManager* analysisManager = G4AnalysisManager::Instance();
    G4String fileName = "ExExCh.root";
    analysisManager->OpenFile(fileName);
    G4cout << "Using " << analysisManager->GetType() << G4endl;
}

//....oooOO0OOooo........oooOO0OOooo........oooOO0OOooo........oooOO0OOooo......

void RunAction::EndOfRunAction(const G4Run*)
{
    G4AnalysisManager* analysisManager = G4AnalysisManager::Instance();
    analysisManager->Write();
    analysisManager->CloseFile();

}

//....oooOO0OOooo........oooOO0OOooo........oooOO0OOooo........oooOO0OOooo......
```